\documentclass[online]{aa}  

\usepackage{graphicx}
\usepackage{amsmath}
\usepackage{txfonts}
\usepackage{caption}

\usepackage{pdflscape}
\usepackage{enumitem,amssymb}
\newlist{todolist}{itemize}{2}
\setlist[todolist]{label=$\square$}
\usepackage{pifont}

\nolinenumbers

\usepackage{varioref}
\usepackage{multirow}
\usepackage{multicol}
\usepackage{txfonts}
\usepackage{booktabs}
\usepackage{xcolor}
\usepackage[nonewpage]{imakeidx}
\usepackage{idxlayout}

\usepackage[pdftex]{hyperref}
\RequirePackage{color}
\definecolor{darkblue}{cmyk}{55,17,0,0}
\hypersetup{colorlinks=true,
  linkcolor=darkblue,
  linkbordercolor=darkblue,
  citecolor=darkblue,
  urlcolor=black
  }

\newcommand{\modest}{\mbox{MODEST}}

\newcommand{\Percquietsun}{68.2\,\%}               
\newcommand{\Perccoldumbra}{0.7\,\%}               
\newcommand{\Percbadpixels}{0.07\,\%}              

\newcommand{\PercColdumbraWRTtotalumbra}{11.7\,\%}

\newcommand{\numinvs}{942}                         
\newcommand{\numars}{117}                          
\newcommand{\numpixels}{1.4$\times10^{8}$}         
\newcommand{\numdatapoints}{6.3$\times10^{10}$}    

\graphicspath{{modest_fig2tab/}}

\newcommand{\kms}[1]{{km s$^{-1}$#1}}

  \titlerunning{The MODEST catalog: depth-dependent spatially coupled inversions of sunspots}
\title{The MODEST catalog of depth-dependent spatially coupled inversions of sunspots observed by Hinode/SOT-SP}
   
   \authorrunning{{Castellanos~Dur\'an} et al.}

\begin{document}

   \author{J.~S. Castellanos~Dur\'an\inst{1,2\thanks{\hbox{\email{castellanos@mps.mpg.de}}}}, N. Milanovic\inst{1}, A. Korpi-Lagg\inst{1,3}, B. L\"optien\inst{1}, M. van~Noort\inst{1}, \and S.~K. Solanki\inst{1,4}}
   \institute{Max Planck Institute for Solar System Research, Justus-von-Liebig-Weg 3, D-37077, G\"ottingen, Germany \label{inst1}
   \and
   Georg-August-Universit\"at G\"ottingen, Friedrich-Hund-Platz 1, D-37077, G\"ottingen,  Germany\label{inst2}\and
   Department of Computer Science, Aalto University, PO Box 15400, FI-00076 Aalto, Finland \label{inst3}\and
   School of Space Research, Kyung Hee University, Yongin, 446-101, Gyeonggi, Republic of Korea\label{inst4}  }

   \date{Submitted: January 19, 2024; accepted: February 22, 2024}

 
  \abstract{
We present a catalog that we named MODEST containing depth-dependent information on the atmospheric conditions inside sunspot groups of all types. The catalog is currently composed of \numinvs{} observations of \numars{} individual active regions with sunspots that cover all types of features observed in the solar photosphere. 
We use the SPINOR-2D code to perform  spatially coupled inversions of the Stokes profiles observed by Hinode/SOT-SP at high spatial resolution.  SPINOR-2D accounts for the unavoidable degradation of the spatial information due to the point spread function of the telescope. The sunspot sample focuses on complex sunspot groups, but simple sunspots are also part of the catalog for completeness. Sunspots were observed from 2006 to 2019, covering parts of solar cycles 23 and 24. The catalog is a living resource, as with time, more sunspot groups will be included.
}

   \keywords{Sunspots; Sun: photosphere; Sun: magnetic fields}
   \maketitle

\section{Introduction}\label{sec:intro-modest}

Sunspots are magnetic structures that are comparatively cool and hence dark in continuum images with respect to their surroundings and form the hearts of active regions \citep[for a review see e.g.,][]{Solanki2003, Borrero2011LRSP}. They play a central role in and are often used as tracers of solar magnetic activity. Although sunspots have been studied for over four centuries, many of their properties are still not well known or understood. In an effort to change this, we provide a new catalog of high resolution maps of physical parameters within the sunspots and in their surroundings.

The thermal, magnetic field and dynamic properties of magnetic features in the lower solar atmosphere, such as sunspots, pores, plage regions, etc., are encoded in the intensity and polarization properties of the solar spectrum. The polarization state of sunlight is fully described by the Stokes profiles, where the intensity of the light is represented by Stokes $I(\lambda)$, the linear polarization by Stokes $Q(\lambda)$ and $U(\lambda)$, and the circular polarization by Stokes $V(\lambda)$.

It is necessary to solve an \textit{inverse problem} to retrieve the conditions in the solar atmosphere from the measured Stokes parameters \citep[see][for a  review]{delToroIniesta2016LRSP}. These so-called inversions use as input the atomic data relevant to the transitions underlying the spectral lines in the solar spectrum. In a first step, a model of the solar atmosphere is formulated. The complexity of this model varies, depending on the type of available observations and the level of detail of the involved physical processes. Then the radiative transfer equation of polarized light is solved to produce synthetic Stokes profiles that are compared with the observations. The model is iteratively modified until the synthetic Stokes profiles match the observations. The $\chi^2$-merit function is usually used as a quantitative measure of the quality of the fit. 

The atmospheric conditions retrieved from an inversion depend on the simplifications entering into the atmospheric model underlying the inversions. The atmosphere recovered for a given model does not have to resemble the true stratification of the solar atmosphere at the analyzed location. Inversions are usually done by integrating the radiative transfer equation numerically and the propagation of numerical errors is intrinsic. Different physical parameters might leave very similar imprints in the observable (the polarization state and intensity of light), which can lead to ambiguities. It is customary to use the merit function $\chi^2$ as a metric of how well the atmospheric model resembles the observations. However, there is no guarantee that the $\chi^2$-hypersurface has a single minimum so that the final result can in some ill-posed cases depend on the initial guess. Taken together, the above points make inversions a fine art, requiring experience and thorough interpretation of the obtained results.

\begin{figure*}
    \centering
    \includegraphics[width=.95\textwidth]{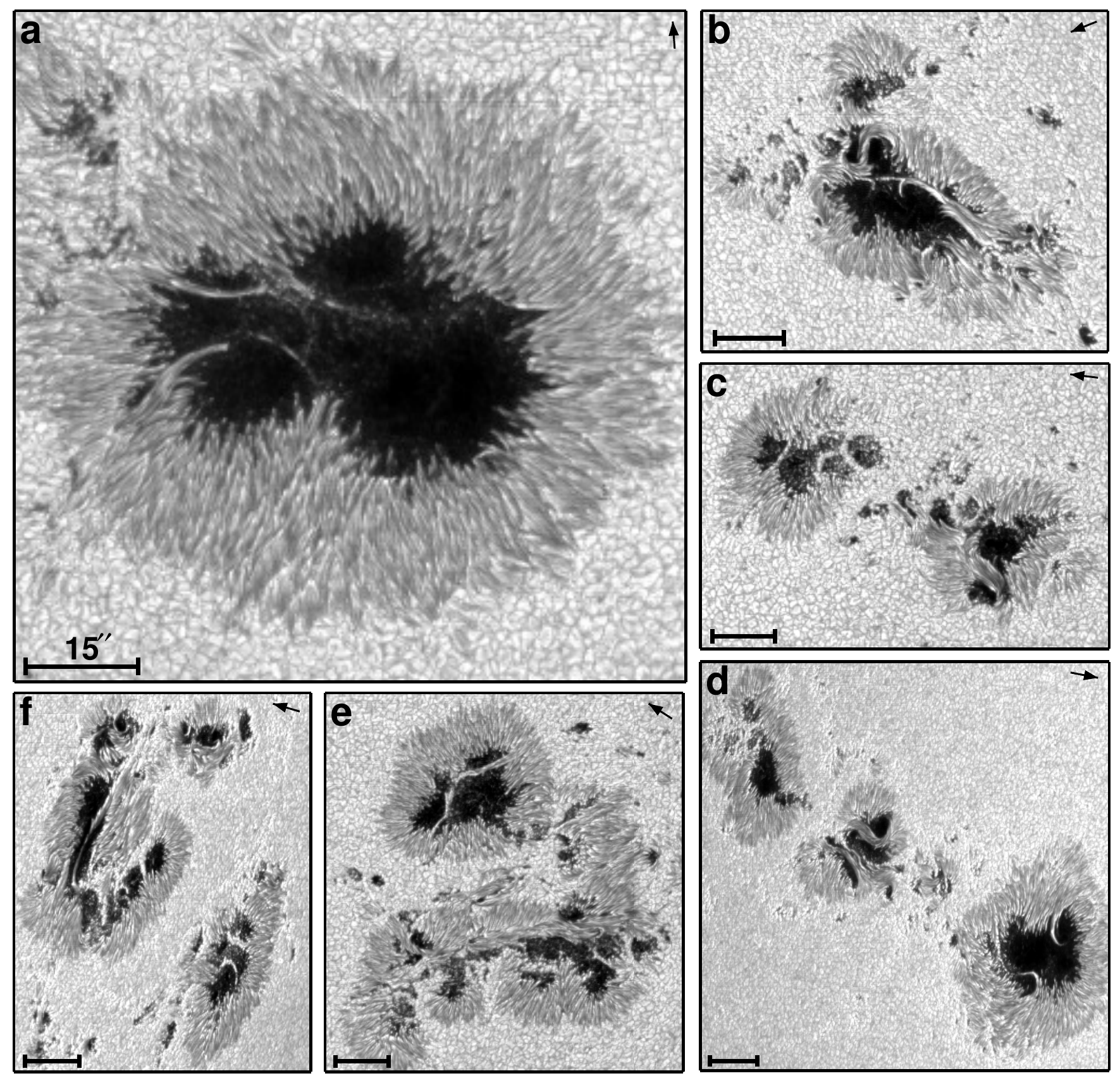}
    \caption{Best-fit continuum images of sunspot groups part of \modest{}.  Panels display continuum images of the sunspot groups AR\,11943 (2014 January 7; a), AR\,11726 (2013 April 23; b), AR\,12645 (2017 April 3; c), AR\,11295 (2011 September 24; d), AR\,12242 (2014 December 19; e), and AR\,12673 (2017 September 8; f).  Bars on the bottom-left have a length of 15\arcsec{}$\approx$10\,800\,km (no foreshortening correction applied). Arrows located on the top-right part of each image mark the direction of solar disk center.} 
    \label{fig:sunspotsmodest}
\end{figure*}

The plane-parallel Milne-Eddington (ME) approximation is the simplest atmospheric model. Inversion codes that apply this approximation are extremely useful when quick inversions are needed \citep[e.g.,][]{Borrero2011}. This model is commonly used because a ME-type atmosphere has an analytical solution \citep{Unno1956PASJ...ME, Rachkovsky962IzKry..ME, Rachkovsky1967IzKry...ME}. As a result, inversion codes that rely on ME-type atmospheres can be almost straightforwardly turned into pipelines to retrieve information on some atmospheric properties at the average height of formation of the spectral lines.  Milne-Eddington inversions can also be used to infer chromospheric magnetic fields and line-of-sight velocities for the \ion{He}{I} triplet at 10830\,\AA{} as these form in a thin layer \citep[e.g.,][]{Rueedi1995A&A...10830, Lagg2004, Lagg2009, Sowmya2022A&A...Supersonicflows}. Milne-Eddington inversions are a fast and convenient tool to find the \textit{average} atmospheric parameters over the height range where the spectral line is formed when inverting large datasets. In particular for data taken by the Spectro-Polarimeter \citep[SP;][]{Ichimoto2008SoPh} onboard the Japanese Hinode solar mission \citep{Kosugi2007}, the ME-inversion is a part of the standard data processing pipeline (Level-2 data product).  The code used by the Hinode consortium is the Milne-Eddington gRid Linear Inversion Network \citep[MERLIN;][]{SkumanichLites1987ApJ...Merlin}. MERLIN inversions of Hinode data are provided by the \citet{merlin2inversion...hinode}.

Different parts of an absorption line are formed at different heights in the solar atmosphere. The core of the line forms in higher layers than its wings. Any variation or gradient in the atmosphere where the line is formed can leave an imprint on its shape and polarization. For example, gradients in the velocity and the magnetic field result in asymmetric or even complex Stokes profiles \citep{Solanki1988A&A...fluxtubes, SolankiMotavon1993A&A}, which are ubiquitous \citep{Solanki1984A&A...fluxtubes, Sigwarth1999A&A...magelements}. Unfortunately, by construction, the ME-inversions fail to model the observed Stokes profiles when such gradients along the line-of-sight exist, since they can only reproduce simple profiles without asymmetries; a significant shortcoming given the strong asymmetries almost invariably present outside sunspot umbrae.

\begin{table*}[htbp]
\centering
\caption{Atomic information of the observed spectral lines.  }\label{tab:atomicinfo-modest}\index{abundance}\index{Land\'e factor}\index{Land\'e factor!effective}\index{ionization potential}
\begin{tabular}{lccp{0.1cm}ccclrccccc}
\toprule[1.5pt] 
\multirow{2}{*}{element}  & wavelength &lower  & & upper & \multirow{2}{*}{$g_{\rm lower}$}& \multirow{2}{*}{$g_{\rm upper}$}& \multirow{2}{*}{$g_{\rm eff}$} & \multirow{2}{*}{$\log(g_l^{\star}f)$}   & energy & \multirow{2}{*}{abundance} & 1st ionization& 2nd ionization   \\ 
  & \multicolumn{1}{c}{(\AA)}      & level & & level& &&& & lower [eV]& & potential [eV] & potential [ev]     \\ 
\midrule[1.5pt] 
\ion{Fe}{I}  & 6301.5012  &$z\,^5\!P_2^{\rm o}$& - &$e\,^5\!D_2$ &1.83&1.5 & 1.67 &-0.745&3.654&\multirow{2}{*}{7.50}&\multirow{2}{*}{7.9024}&\multirow{2}{*}{16.1879}\\[-3mm]
\ion{Fe}{I}  & 6302.4936  &$z\,^5\!P_1^{\rm o}$& - &$e\,^5\!D_0$ &2.50& --
& 2.50 &-1.203&3.687& \\[-4mm]
\bottomrule[1.5pt] 
\end{tabular} 
\end{table*}

Depth-dependent codes are commonly used in solar physics to model stratified atmospheres (e.g., SIR \citep{RuizCobo1992ApJ...SIR}, NICOLE \citep{Socas-Navarro1998ApJ...NICOLEI,Socas-Navarro2015A&A...NICOLEII}, SPINOR \citep{Solanki1987PhDT,Frutiger2000}, SNAPI \citep{Milic2018A&A...SNAPI}, among others). 
 Such height-stratified inversions are able to retrieve variations of the physical conditions with optical depth \citep[for a review see e.g.,][]{delaCruzRodriguezVanNoort2017SSRv..review}.
However, depending on the photospheric features of interest, the model used to interpret the observations might need to be adjusted. For example, the formation height of the same spectral line from quiet Sun to sunspots can differ by hundreds of kilometers \citep[see e.g., Fig.~2 of][]{Smitha2021...Ti22micron}. This large difference requires a change in the grid locations where the atmospheric parameters are determined. 

Typically, it is difficult to obtain excellent fits to all pixels inside the field of view (FOV) if a single atmospheric model is used. It has been customary that different features, such as the quiet sun, plage, penumbra, and umbra, were modeled with different settings for the atmospheric parameters \citep[e.g,][]{Collados1994A&A...spots, BellotRubio2003A&A...evershed}. Depth-dependent spatially coupled inversions were proposed to solve this as well as other problems
\citep[][hereafter coupled inversions]{vanNoort2012A&A,vanNoort2013A&A}. They account for the parasitic light coming from adjacent pixels due to the unavoidable blurring caused by the point spread function (PSF) of the telescope. 
The coupled inversions applied to Hinode/SOT-SP data have been used to study different features of the solar photosphere such as, among others, the umbral dots \citep{Riethmueller2013A&A...UD}, the Wilson depression \citep{Loeptien2020A&A...wilson}, the umbra-penumbra boundary \citep{Loeptien2020A&A...umbrapenumbra}, light bridges \citep{Lagg2014A&A...LB, CastellanosDuran2020}, penumbral filaments \citep{vanNoort2013A&A,Tiwari2013}, counter Evershed flows \citep{ Siu-Tapia2017A&A,Siutapia2019, CastellanosDuran2023...ejectionCEFs}.

Recently, the concept behind the coupled inversions was generalized by \citet{delaCruzRodriguez2019AA...coupled} to account for observations taken by different facilities that may have different spatial, or spectral resolutions, or that are rotated relative to each other \citep[see, e.g.,][for a multi-observatory dataset where such inversions could be applied]{RouppevanderVoort2020A&A...obssstiris}.  To our knowledge, there is only one existing photospheric catalog of depth-dependent inversions, which mainly focuses on simple sunspots and covers $\sim$50 Hinode/SOT-SP scans \citep[see e.g.,][]{Loptien2018A&A}.

In this work, we present the Max-Planck Open Database of Elaborate inversions of SunspoTs or \modest{} for short. The \modest{} catalog of sunspot groups covers different types of active regions (ARs) with sunspots, including the most complex ones. Figure~\ref{fig:sunspotsmodest} shows examples of different ARs that are part of \modest{}. For many ARs, \modest{} also provides some (limited) temporal coverage as they cross the solar disk. \modest{} is one of the first catalogs of stratified sunspots inversions where the same atmospheric model was used for all observations while being able to fit all types of features observed on the Sun with one set of free atmospheric parameters.  The \modest{} catalog will enable statistical studies of depth-dependent conditions inside a large variety of sunspots.

\section{Data and sunspot sample} \label{sec:obs-modest}

\subsection{Data calibration}
The \modest{} catalog uses spectropolarimetric data taken by the Spectropolarimeter \citep[SP;][]{Ichimoto2008SoPh} attached to the Solar Optical Telescope \citep[SOT;][]{Tsuneta2008} onboard Hinode \citep{Kosugi2007}. Hinode/SOT-SP performs high spectral and spatial resolution spectropolarimetric observations of the solar photosphere in the line pair \ion{Fe}{I}~6301.5\,\AA{} and 6302.5\,\AA{}. The magnetic sensitivity of these lines, given by the effective Land\'e factor ($g_{\rm eff}$), is 1.67 and 2.5 respectively \citep[cf.][]{SolankiStenflo1985A&A...geff}. Table~\ref{tab:atomicinfo-modest} summarizes the atomic parameters of these transitions.

Hinode/SOT-SP has a spectral sampling of 21.5\,m\AA{}, while the spatial resolution varies depending on the observing mode, with the ``fast'' mode having 0\farcs297$\times$0\farcs32 pixel size and the ``normal'' mode 0\farcs149$\times$0\farcs16. The fast mode decreases the time needed to scan a given area on the solar surface, to the detriment of the spatial resolution with respect to the normal mode. Depending on the science case, either mode can be more suitable, and therefore the sample of sunspots presented here contains observations taken in both modes. The data were calibrated using the standard \texttt{sp\_prep} routines that are part of the Solarsoft package and were reduced using the nominal Hinode/SOT-SP pipeline \citep{Lites2013SoPh}. In addition, all data were checked for continuum polarization that is sometimes observed in Stokes $Q(\lambda)$, $U(\lambda)$ and $V(\lambda)$ \citep[see Fig.~1 in][]{Okamoto2018ApJ}. In case such an offset was found in the polarization, we removed the offset by requiring the continuum polarization to be zero. No additional correction to account for gray spectral stray light was applied (see Appendix~\ref{sec:greystrayModest}).  In addition, for some particularly large Hinode/SOT-SP scans, we did not invert the full scan, but we rather manually centered the sunspot group and cut the scan in the spatial plane (See Sect.~\ref{sec:tiling}).

\subsection{Sample of sunspots}

Since coupled inversions are computationally expensive, only a small subset of all Hinode observations could be included in the \modest{} database, so that we need to select. We made use of the Lockheed Martin Solar and Astrophysics Laboratory (LMSAL) database\footnote{\url{http://sot.lmsal.com/data/sot/level2d/}} Hinode/SOT-SP to select the sample of sunspot groups to invert. This database contains ME-inversions of all Hinode/SOT-SP observations, which were performed with the MERLIN code. However, this database does not specify whether individual scans include a specific AR or not. On the date we started selecting our sample (2019 July 27), more than 22\,000 scans were available.

We started by creating an initial list of all ARs with sunspots that were visible on the Sun from the date when Hinode started observing until 2019 July 27. Here we used the Solar Regions Summaries\footnote{\url{ftp://ftp.swpc.noaa.gov/pub/warehouse}} made by the Space Weather Prediction Center. These are daily reports about all visible ARs on the Sun. For each visible AR with sunspots, the SRS reports contain the NOAA number, the estimated area, the latitude and longitude, the modified Z\"urich class \citep{McIntosh1990SoPh...ClassARs}, magnetic sunspot class \citep{Hale1919ApJ...Bcycle, Kunzel1960AN...DeltaSpots,Kunzel1965AN....spots}, among others. If an AR was observed for more than one day, it occurs more than once on the initial list. To see which of these ARs were observed by Hinode/SOT-SP, we used the SOT planning files\footnote{\url{https://sot.lmsal.com/operations/timeline/}}. We kept only ARs that were visible on dates that appeared in the SOT planning files or that had a NOAA number that was mentioned in those. Finally, we also excluded all regions that had an estimated sunspot area smaller than 100\,MSH, where MSH stands for a millionth of the solar hemisphere, i.e., $1\mu$SH$=(6\farcs3)^2$.  This filtering reduced the number of scans that we further investigated from more than 22\,000 to about 9\,100.

These conditions could not completely exclude other types of Hinode/SOT-SP observations that were performed on the same day as an observation of a suitable active region, however. These scans were excluded in the subsequent filtering process. Sunspots can be identified easily in maps of the magnetic field strength. Hence, to select only scans including sunspots, we made use of maps of the magnetic field strength provided by the MERLIN inversions in the LMSAL Level-2 database. With the help of the ME-inversions, we checked for each scan in our sample if it had at least one connected region, where the inferred magnetic field strength was greater than 2\,kG, and where the area of the region was larger or equal to (10\arcsec{})$^2$.

This approach did not work well for wide scans that included the limb. Sometimes the MERLIN inversion assigns very strong magnetic fields to the parts of the FOV outside the limb (usually 5\,kG, which is the upper limit in the MERLIN code), which would be falsely classified as sunspots by our approach. Consequently, we removed scans including the limb from the sample, by checking if the center of the connected region is too close to or even outside the limb (if its central pixel lies at a radius greater than the disk radius minus 50\arcsec{}, which corresponds to $\mu\approx 0.35$). After applying these conditions, we were left with a sample of about 4500 ``suitable'' scans. Unfortunately, some of these scans still contained off-limb regions, since the coordinates of the center of the FOV in the Level-2 data header were not always accurate \citep[cf.][]{Fouhey2022..SPcoords} or the scans were too wide. 

Initially, \modest{} was conceived to study statistically superstrong magnetic fields observed in sunspots \citep[cf.][]{Okamoto2018ApJ, CastellanosDuran2020}. For this reason, we sorted the ``suitable'' scans of sunspot groups in descending order, according to the number of pixels inside the connected regions that had magnetic field strength of exactly 5\,kG, which is the maximum value allowed of the magnetic field strength in the MERLIN inversions. From this point on, we inspected each of the remaining scans manually before deciding whether to invert it or not. This step reduced the number of scans by another factor of $\sim$4.

 \begin{figure*}[htbp]
 \begin{center}
 \includegraphics[width=.49\textwidth]{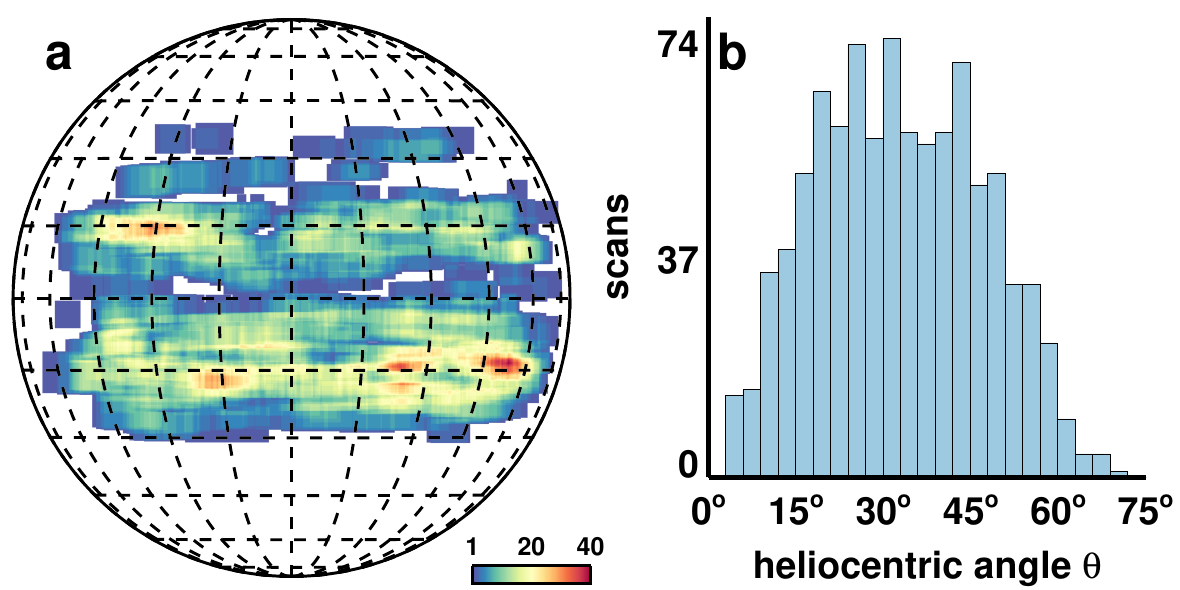}\includegraphics[width=.49\textwidth]{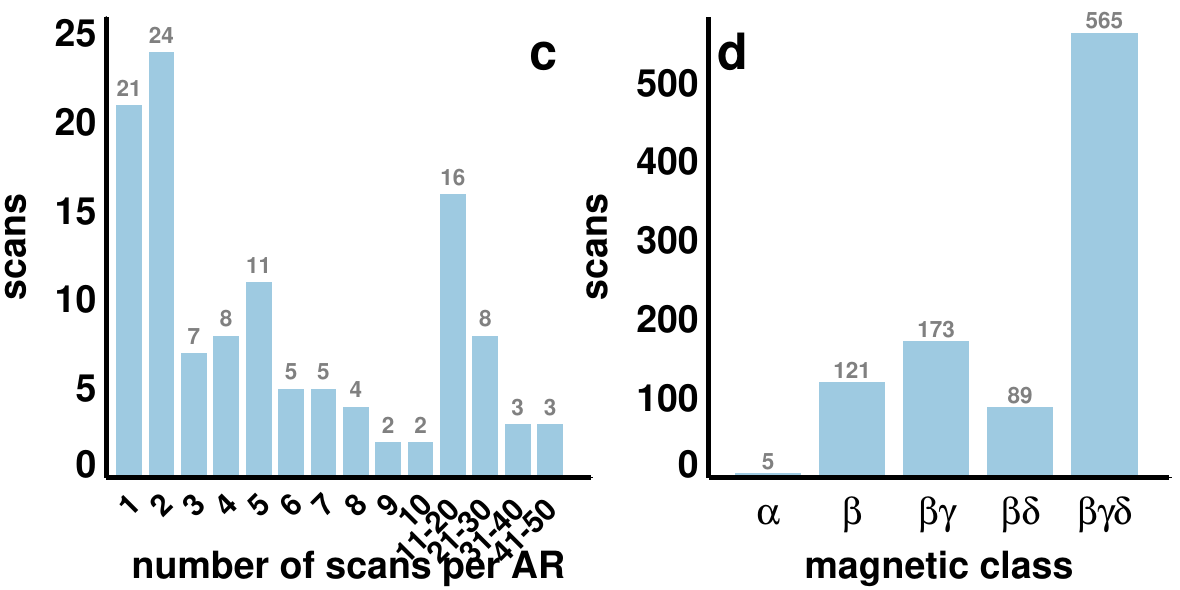} 
 \caption{Characteristics of the sample of sunspots in the archive. (a) Location of all scans on the solar disk. The color gives the number of scans observed at roughly the same location on the disk. (b) Heliocentric angle of the center of each field of view. (c) Number of scans of a given sunspot group (or part of it). (d) Magnetic classification of the sunspot group, with each scan being counted separately. }\label{fig:sMODESTSample}
 \end{center}
 \end{figure*}

As we were originally interested in complex sunspots with very strong magnetic fields, in particular in bipolar light bridges, the current \modest{} sample focuses on complex ARs with sunspots, in particular $\delta$-groups. To increase the usability of our catalog also for other studies involving sunspots, we later added scans of $\alpha$ and $\beta$ spots, and we will continue to do so in the future. Figure~\ref{fig:sMODESTSample} shows a summary of the current sample of sunspot groups that are part of \modest{}. As of now, \modest{} consists of \numinvs{} scans of \numars{} individual ARs with sunspots (either whole or part of them), and most of the inverted scans contain multiple sunspots. Panel (\ref{fig:sMODESTSample}a) displays the density of all scans projected onto the solar disk. Panel (\ref{fig:sMODESTSample}b) shows the heliocentric angle of the center of each scan. Multiple scans were inverted for many ARs to provide information on the temporal evolution of at least some of the ARs. Panel (\ref{fig:sMODESTSample}c) compiles the number of scans that cover a given AR, or part of it. Panel (\ref{fig:sMODESTSample}d) arranges all sunspot groups in terms of their magnetic class.

The current \modest{}'s sample covers ARs located roughly within $\pm60^{\circ}$ of longitude and $\pm25^{\circ}$ in latitude on the solar disk. Table~\ref{tab:sample-ARS-MODEST} summarizes the sample of the scans inverted so far. Table~\ref{tab:sample-ARS-MODEST} lists information about the inverted scans: the OBS\_ID, NOAA number, date, time, type of the Hinode scan (either fast with a pixel size $\sim$0\farcs32 or normal $\sim$0\farcs16), coordinates of the FOV, $\mu-$value (cosine of the heliocentric angle), number of tiles that the FOV was divided into (see below), size of the FOV, and area, Z\"urich classification and magnetic classification of the AR.

\section{Inversion approach}\label{sec:inversions-modest}

\subsection{Spatially coupled inversions}\label{sec:modestcoupledinversions}

In this work, we applied the so-called spatially coupled inversions  \citep{vanNoort2012A&A, vanNoort2013A&A}, which make use of the knowledge of the telescope PSF to remove its effects at the same time as obtaining information on the solar atmosphere from the measured Stokes profiles. The recorded images at the entrance slit of the SP are the result of the convolution of the undisturbed solar image with the PSF of the aplanatic 50-cm $f/9$ Gregorian telescope. 
The PSF includes the effects of the spider holding the M2 mirror and the central obscuration \citep[see Fig.~10 in][]{vanNoort2012A&A}. 
As a result of this convolution, the information of a given pixel is distributed over its surroundings. To account for these effects, \citet{vanNoort2012A&A} proposed to include the effects of the PSF in the inversion procedure. 

The inversions are performed using the SPINOR code \citep{Frutiger2000}.  SPINOR fits the spectropolarimetric observations with a stratified model of the solar atmosphere. 
SPINOR uses the STOPRO routines that solve the radiative transfer equation of polarized light \citep{Solanki1987PhDT}, and employs the Levenberg-Marquardt minimization algorithm \citep{Levenberg1994...minimization,  Marquardt1963...Algorithm} to fit the synthetic Stokes profiles to the observed ones. 
The module for spatially coupled inversions in the SPINOR code can be activated when the PSF of the instrument is known (SPINOR-2D inversions).

Table~\ref{tab:atomicinfo-modest} summarizes the atomic parameters used in the inversion procedure, where $g_{\rm lower}$ and $g_{\rm upper}$ are the Land\'e factors of the lower and upper energy levels of the considered transitions, and $\log(g_l^{\star}f)$ is the  weighted oscillator strength of the line. 
The Land\'e factors are calculated from the atomic configuration assuming $LS$-coupling. The effective (triplet) Land\'e factors are shown for completeness but are not used during the inversion.

\begin{figure*}[htbp]
    \centering

    \includegraphics[width=0.34\textwidth]{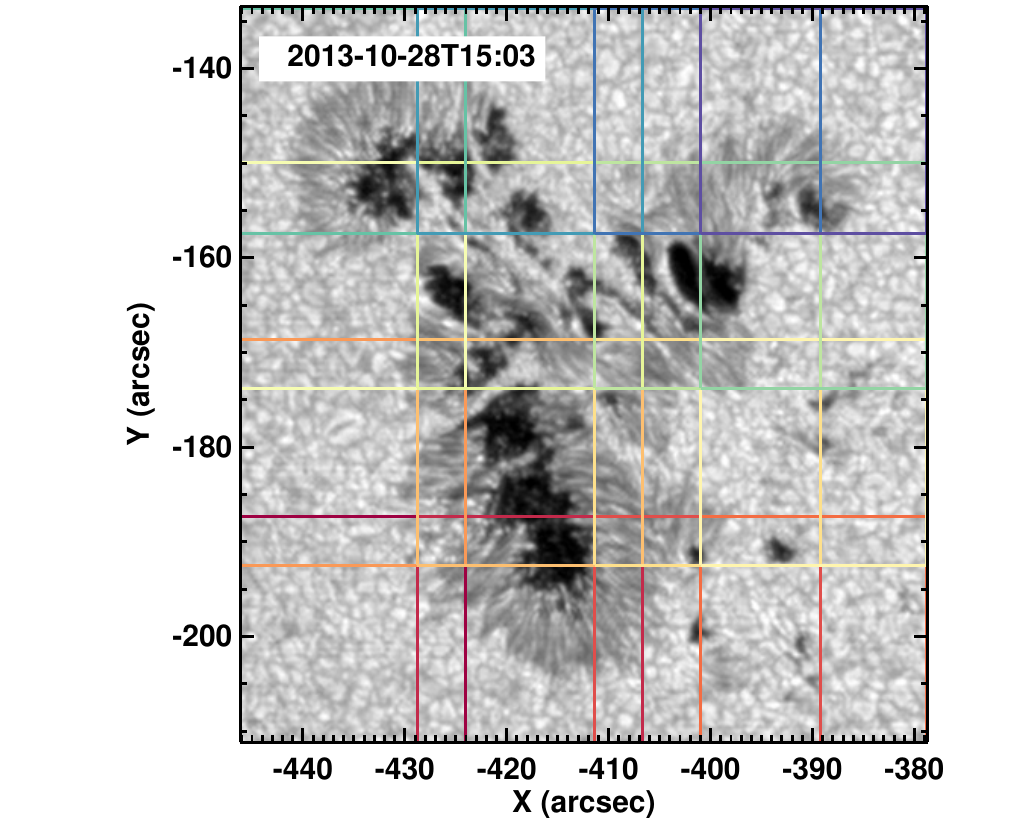}\hspace{5pt}\includegraphics[width=0.34\textwidth]{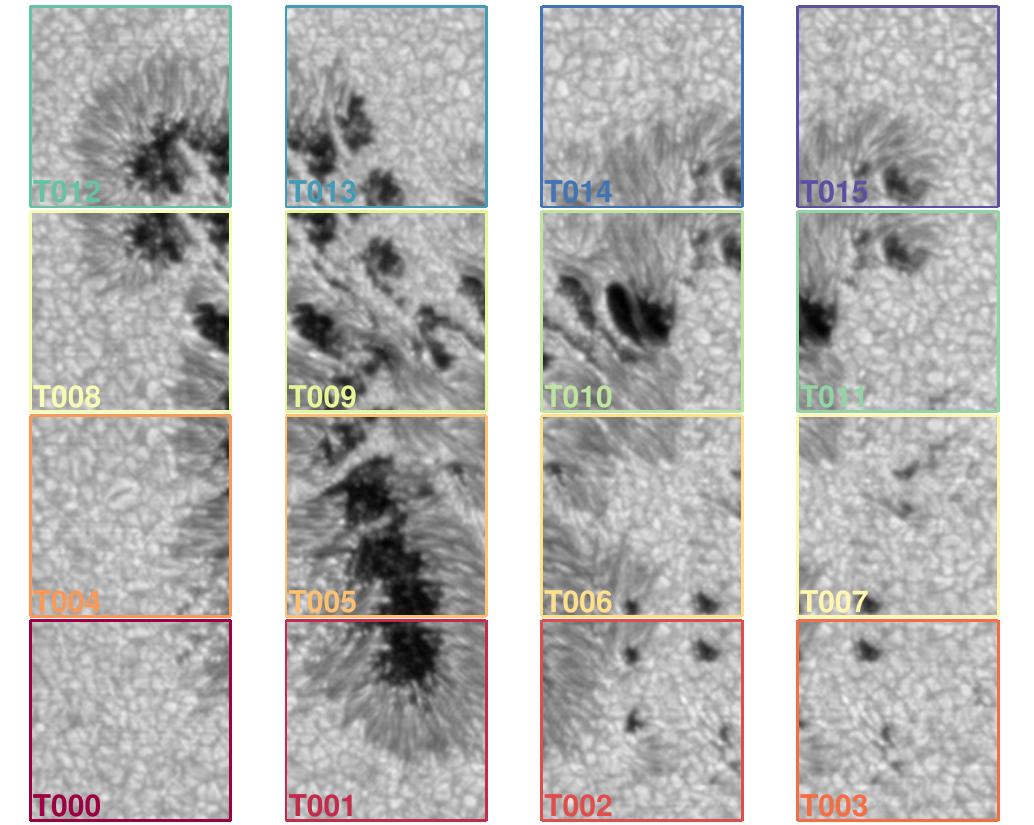}\\[1pt]
    \includegraphics[width=0.34\textwidth]{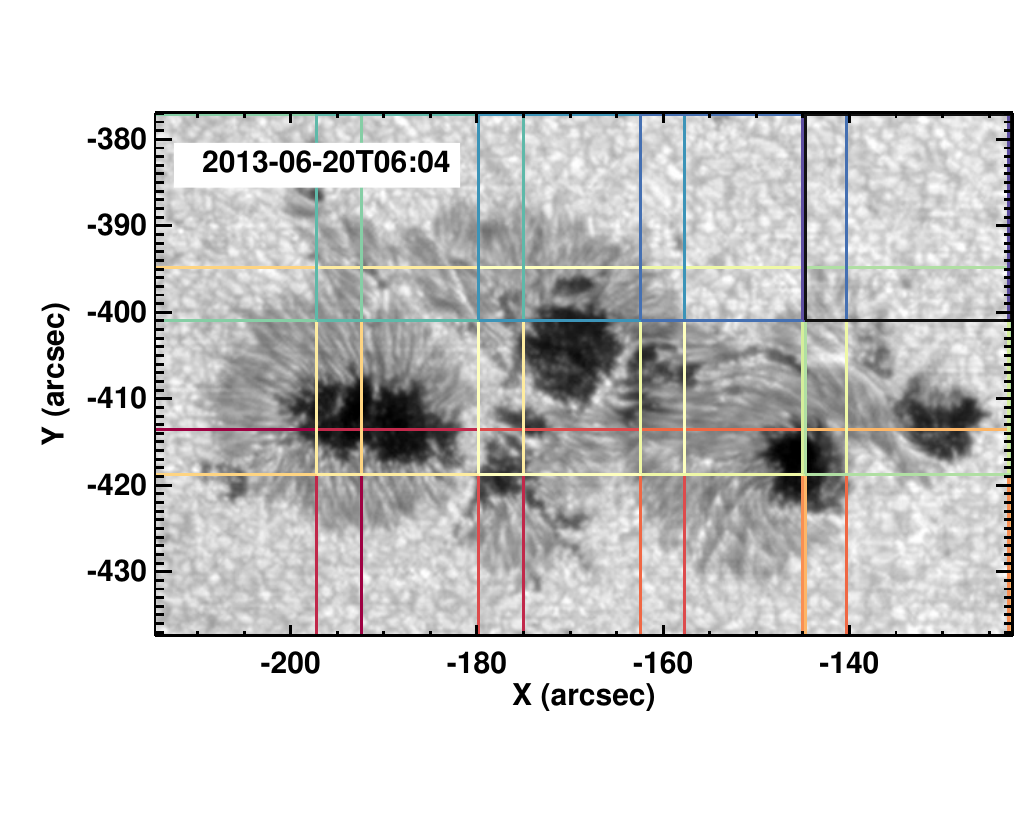}\hspace{5pt}\includegraphics[width=0.34\textwidth]{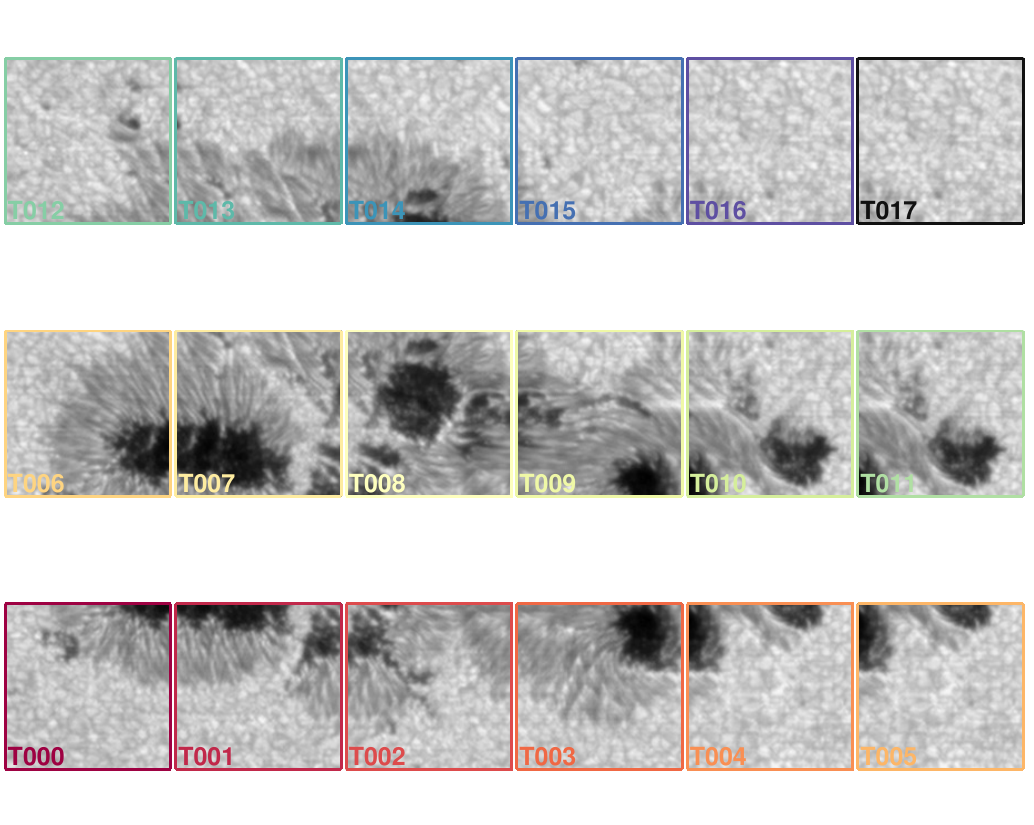}\\[1pt]
    \includegraphics[width=0.34\textwidth]{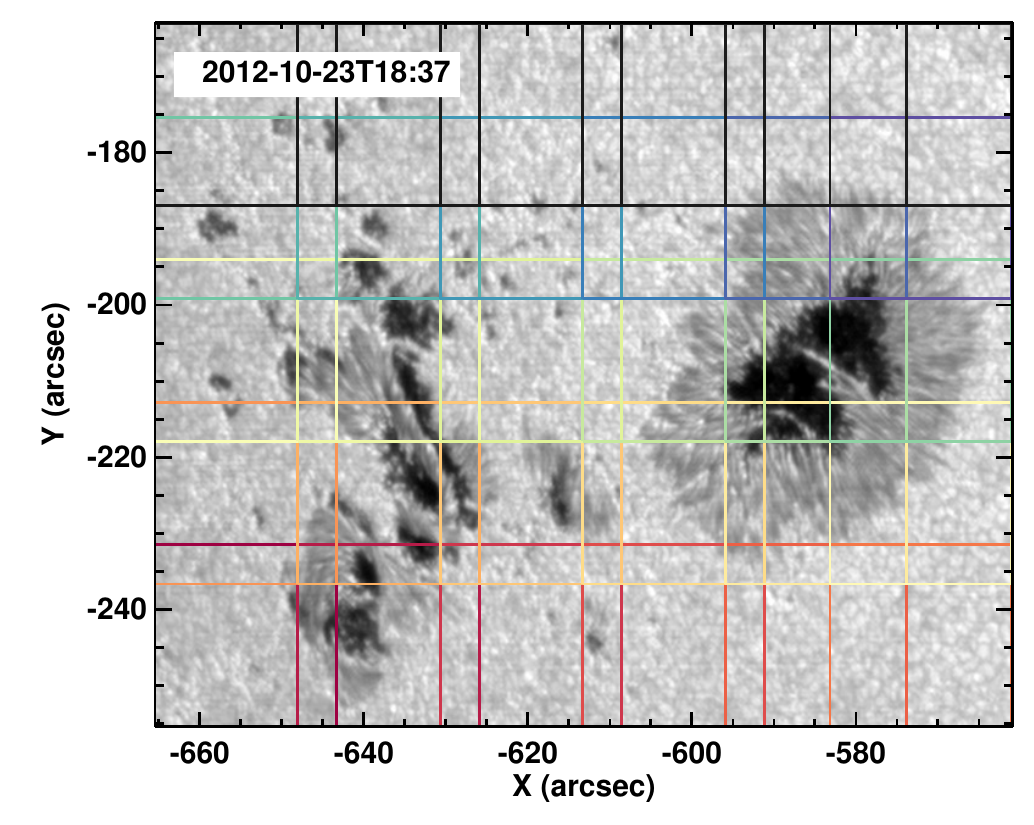}\hspace{5pt}\includegraphics[width=0.34\textwidth]{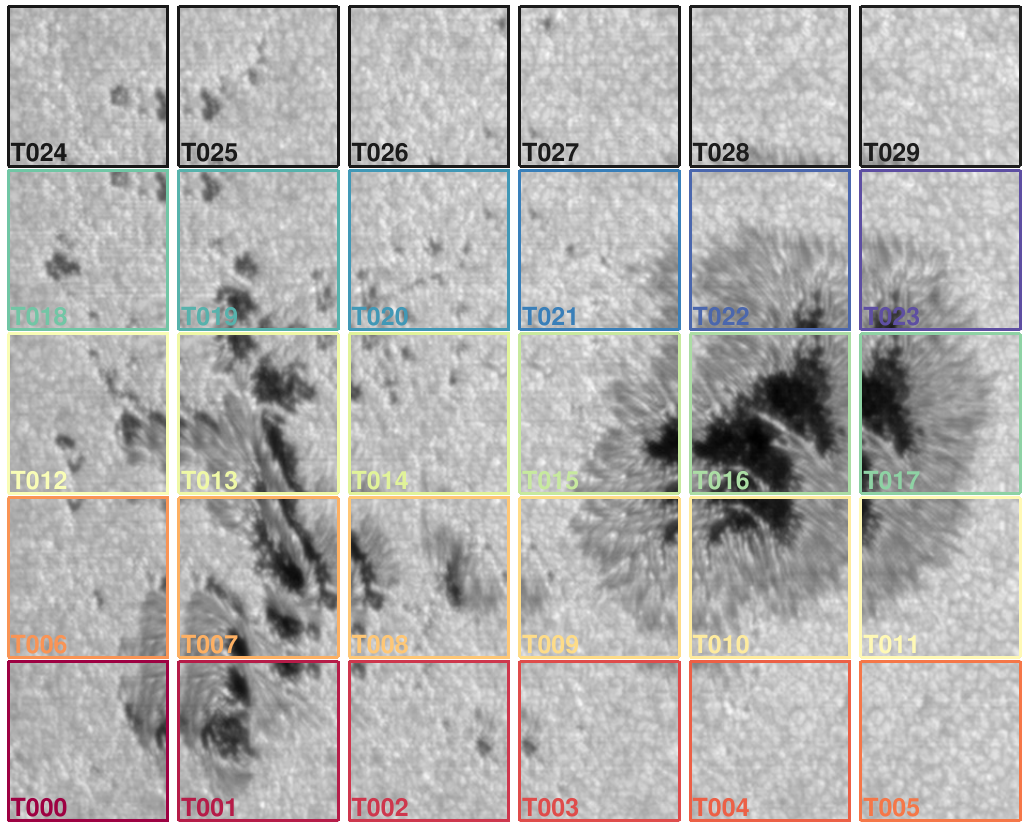}\\[1pt]
    \includegraphics[width=0.34\textwidth]{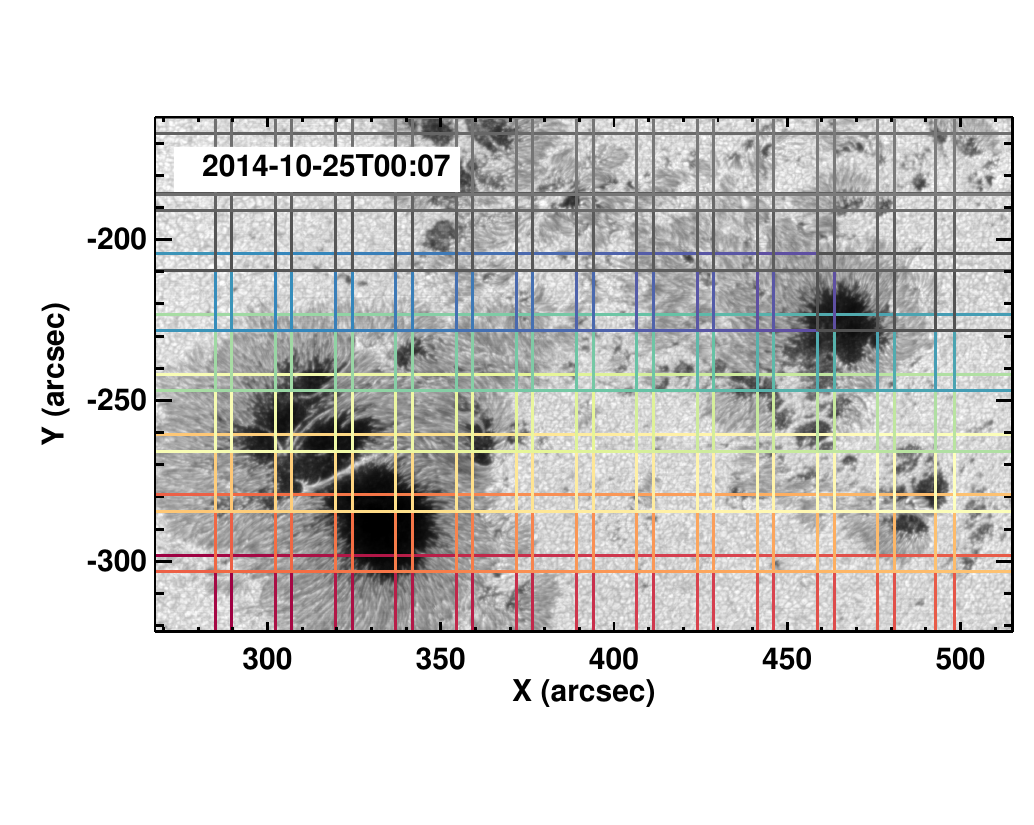}\hspace{5pt}\includegraphics[width=0.34\textwidth]{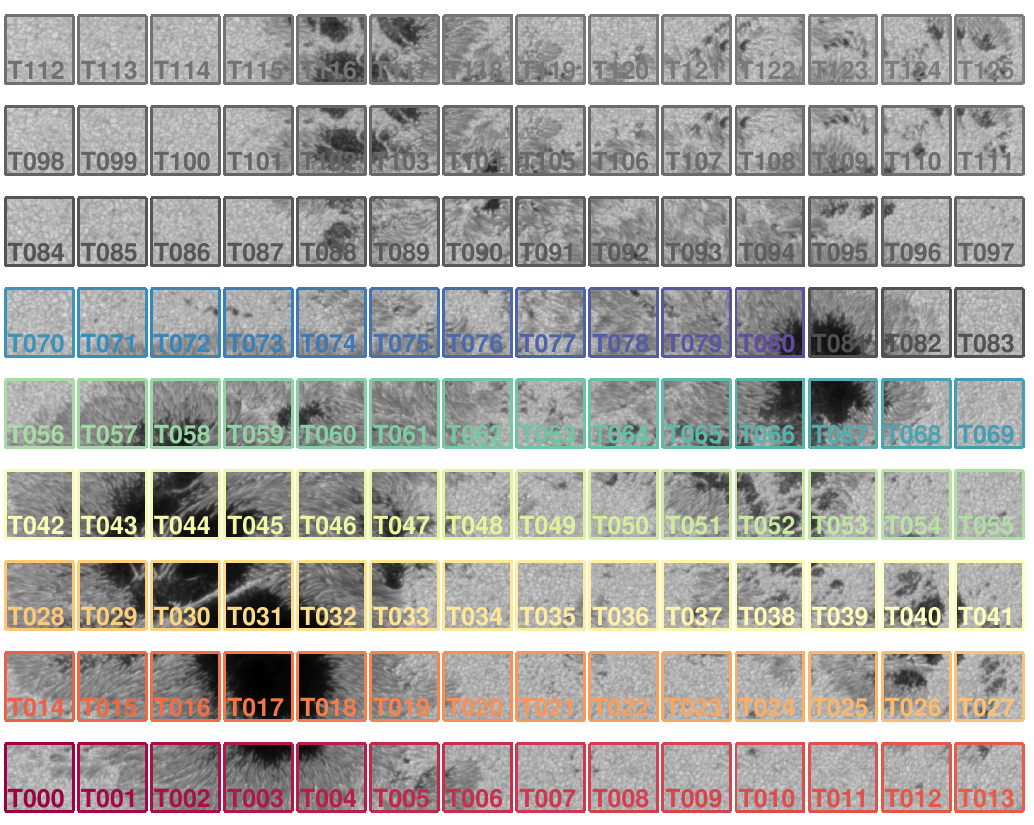}
    \caption{Illustration of the division of the FOV into tiles for the coupled inversions. From top to bottom, the continuum maps as observed by Hinode/SOT-SP of AR\,11882, AR\,11775, AR\,11560, and AR\,12192. Colored lines mark the location of the tiles in which the FOV is divided (left). Notice that the tiles are designed to overlap. Right column shows the individual continuum maps of each individual tile.}
    \label{fig:tiling_modest}
\end{figure*}

\subsection{Node position and free parameters}\label{sec:freeparameters}

The number of free parameters in the atmospheric model is related (1) to the number of nodes placed at different depths at which the parameters are determined,  and (2) to the physical quantities that play a role in the formation of the spectral line.  A larger number of free parameters generally improves the quality of the fit. This can come at the cost of the uniqueness of the solution, in particular, if some free parameters are not sufficiently independent of each other. 
For the \modest{} catalog, we use 16 free parameters to model a stratified solar photosphere. The temperature (T), magnetic field strength ($B$), inclination ($\gamma$), azimuth ($\varphi$) and line-of-sight velocity ($v_{\rm LOS}$) each contribute with three free parameters located at different optical depths.
The micro-turbulence ($v_{\rm micro}$) was modeled without depth-dependence (one free parameter) as the $\chi^2$-hypersurface is relatively insensitive to depth-dependent changes due to micro-turbulence velocities \citep{Frutiger2000}. 
We convolve the synthetic profile with the measured SP spectral transmission profile, which has a FWHM of 24.3 m\AA{} and extended wings, so no other parameter was necessary to account for the broadening of the \ion{Fe}{I} line pair.

\begin{figure*}
    \centering
    \includegraphics[width=.99\textwidth]{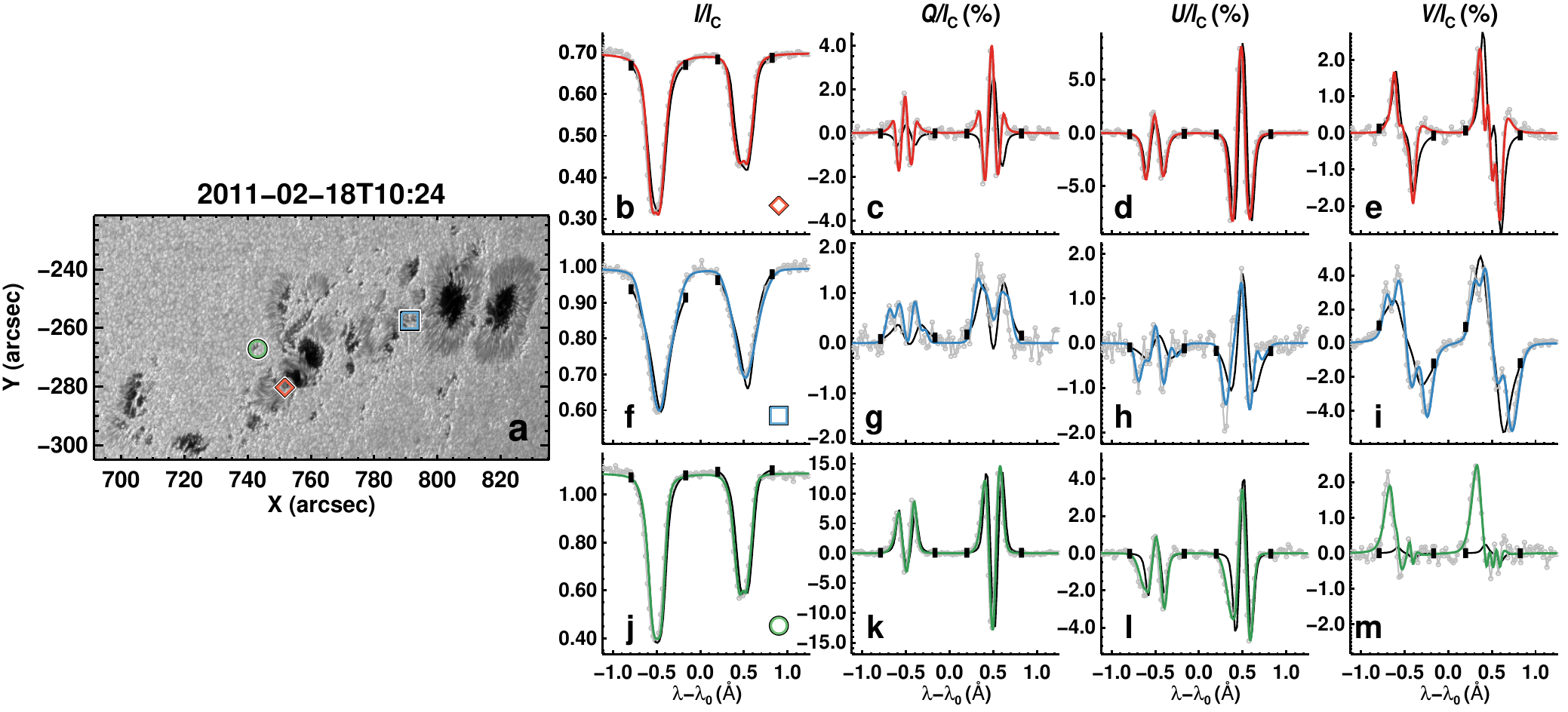}
    \caption{
    Fits obtained by the coupled inversions to the observed Stokes profiles. Panel (a) shows the continuum image of the AR\,11158 when it was located at $\mu\approx0.6$. Columns 2 to 5 show the observed Stokes profiles (gray dotted lines) and the retrieved fits obtained by the coupled inversions (colored lines), and the ME-inversion by MERLIN (black lines). Vertical black bars mark the spectral windows that MERLIN uses for the Level-2 inversions. The locations of the plotted profiles are marked on the continuum images by the colored symbols. $\lambda_0=6302$\,\AA{}.
    }
    \label{fig:modest_fits1}
\end{figure*}

\begin{figure*}
    \centering
    \includegraphics[width=.99\textwidth]{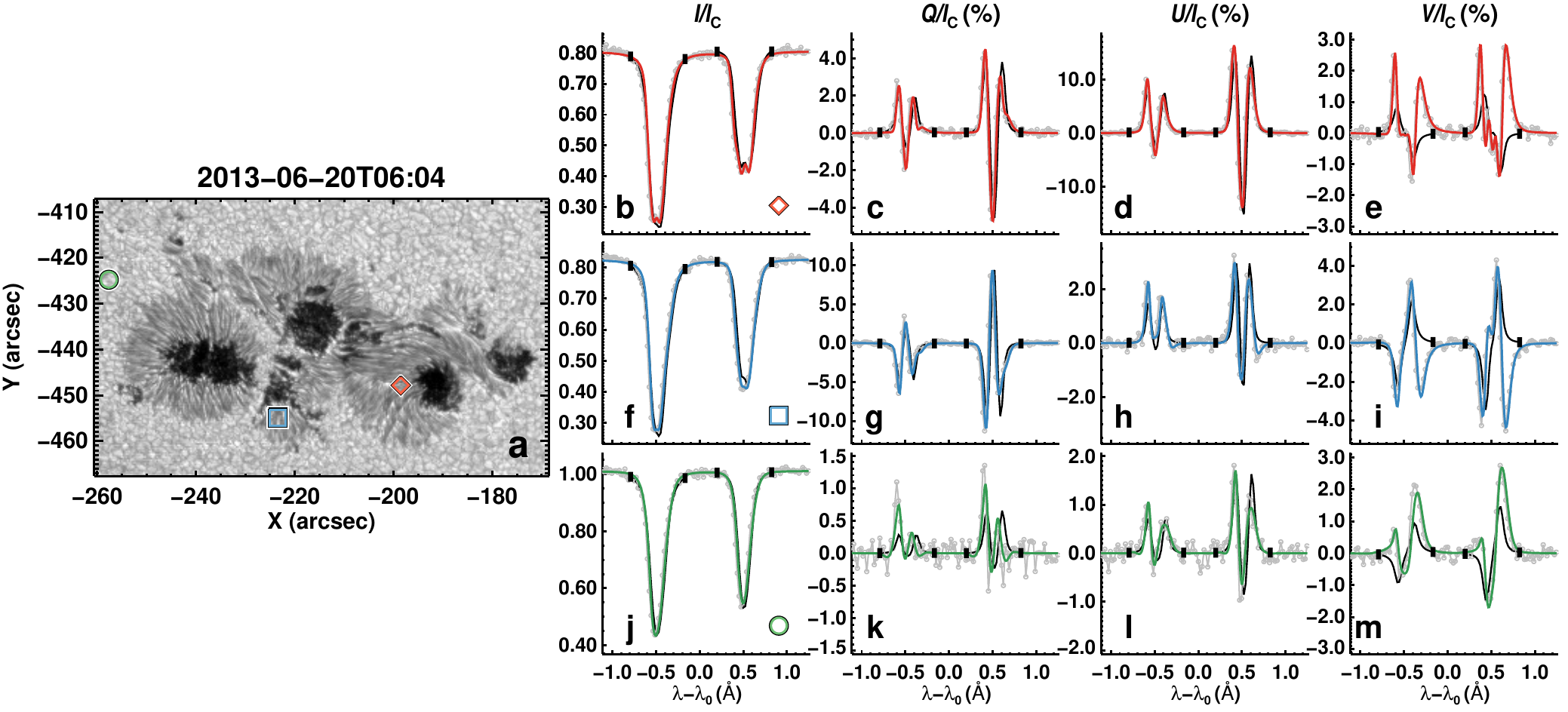}
    \caption{The same as Fig.~\ref{fig:modest_fits1} for a sunspot group belonging to AR\,11775, when it was located at $\mu\approx0.8$.}
    \label{fig:modest_fits3}
\end{figure*}

\begin{figure*}
    \centering
    \includegraphics[width=.99\textwidth]{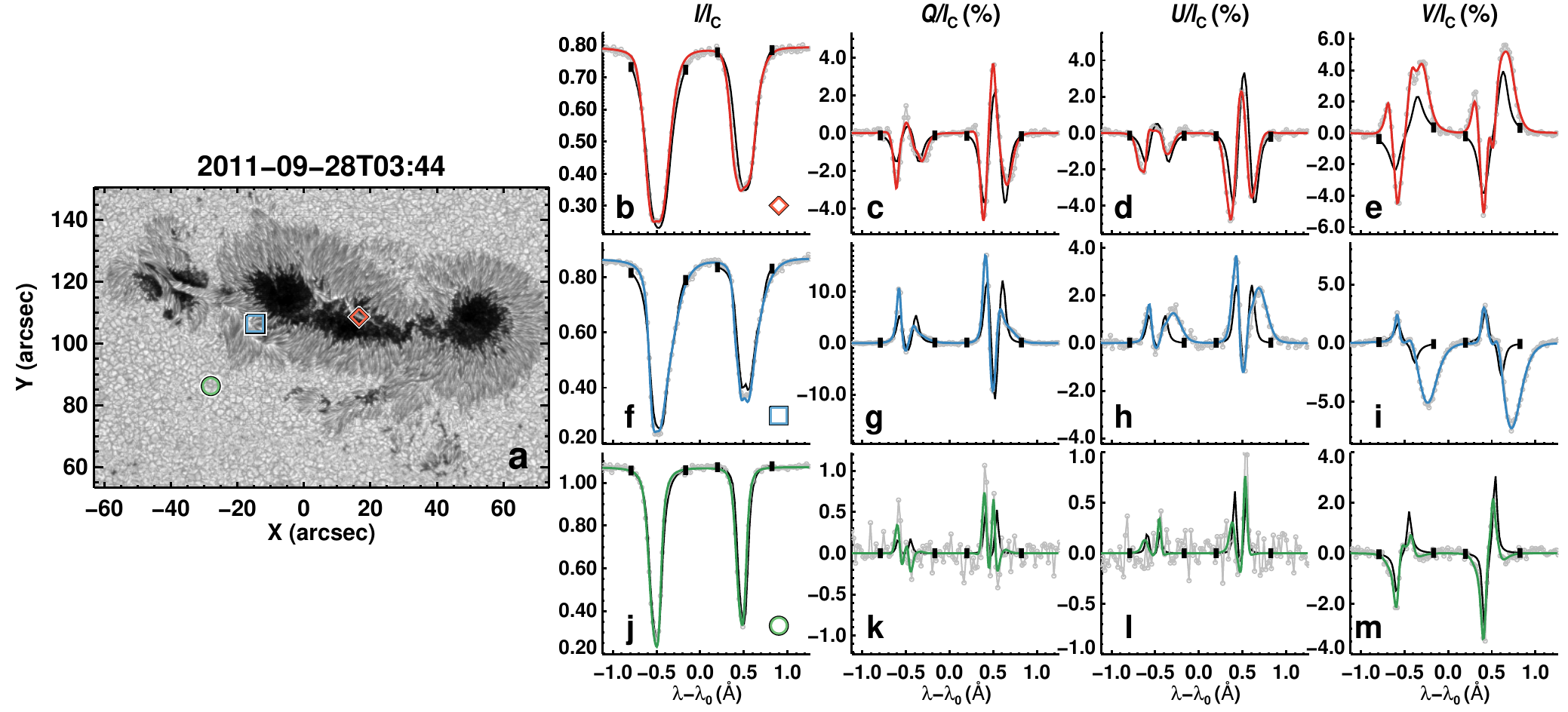}
    \caption{The same as Fig.~\ref{fig:modest_fits1} for a sunspot group belonging to AR\,11302, when it was located at $\mu\approx1.0$.}
    \label{fig:modest_fits5}
\end{figure*}

The node positions were located at $\log\tau_{\rm c}=0.0, -0.8$, and $-2.0$, where $\tau_{\rm c}$ represents the continuum optical depth at 5000\,\AA{}. The position of the top node at $\log\tau_{\rm c}=-2.0$ was chosen following \citet{Danilovic2016A&A...qs..nodes} as large parts of the FOV might be filled with quiet Sun. 
These authors demonstrated that this node position is well-suited under quiet-Sun conditions, but it also does not adversely affect the retrieved atmospheres for complex profiles (see the discussion in Sect.~\ref{sec:modelStokes}; cf. Figs.~\ref{fig:modest_fits1}-\ref{fig:modest_fits5}).
The atmospheric stratification is computed from the coarse three-node grid onto a fine depth grid with a sampling of $\Delta\log(\tau)=0.05$ using a spline approximation \citep[see][]{Frutiger2000PhDT}. 
The atmosphere was extrapolated using a linear function based on the slope of the spline interpolated between the nodes for optical depths ranging between $\log\tau_{\rm c}<-6.0$ and $\log\tau_{\rm c}>1.5$.

\subsection{Inversion strategy}\label{sec:InvStrategyModest}

Selecting a balanced inversion strategy helps to avoid getting stuck in a local minimum in the $\chi^2$-hypersurface. There is the possible existence of extreme cases where any of the retrieved parameters can have an extreme value (e.g., $v_{\rm LOS}>7$\,\kms{} - photosheric sound speed - or $B>4$\,kG). In addition, gradients of the $v_{\rm LOS}$ and the magnetic field, at different heights where the lines are formed, can create complex Stokes profiles \citep{SolankiMotavon1993A&A}. 

To account for these possibilities, each observation was inverted ten times as suggested by \citet{vanNoort2013A&A}. After each inversion run, the output atmosphere for each pixel was smoothed with the surrounding neighbors in case the solution in the pixel reaches a local minimum in the $\chi^2$-hypersurface (see below). Then, that smoothed atmosphere was used as the initial condition for the next inversion run applied to the same set of pixels. In the first inversion run, for the initial conditions we give the average quiet-Sun conditions, weakly magnetized and at rest \citep[e.g.,][]{Danilovic2016A&A...qs..nodes}. We used quiet-Sun conditions to initialize the inversions, as \Percquietsun{} of the pixels inside the FOVs belong to quiet Sun, active region plague or network. 

The number of maximum iterations allow to find the best match between the observed and synthetic Stokes vector increases in each inversion run. The first inversion run has a maximum of 10 iterations and this number is increased after each inversion run until up to 100 iterations \citep{vanNoort2012A&A}.

The allowed ranges (``limits'') of the atmospheric parameters during the first run were chosen to cover typical values that contain quiet sun, penumbral, and umbral regions (see below).  In the subsequent runs, the parameter space was slowly increased. This allowed adequately fitting ever more complex profiles while ensuring that the fits to simple Stokes profiles remained of high quality. In particular, to avoid the inversion reaching extreme values the limits for the magnetic field strength and $v_{\rm LOS}$ were smoothly increased for each inversion run. For the first three runs $B$ and $v_{\rm LOS}$ were limited to $B\leq 3$\,kG and $|v_{\rm LOS}|\leq 5$\,\kms{}. This limit accounts for the subsonic pixels in the quiet sun, penumbrae, and the not-too-dark parts of umbrae \citep[i.e. in umbral dots and the diffuse background, e.g.,][]{SocasNavarro2004ApJ...UDs,  Riethmueller2008ApJL...UD2invs,  Riethmueller2013A&A...UD}. This limit covers the values reached in the majority of the pixels. In case a larger value of either of these quantities would be required to reproduce a particular set of Stokes profiles, the inversion saturates at this limit, but then, the saturated value is used as an input for the next inversion run. 

During inversion runs 4 and 5, the  $v_{\rm LOS}$ limits start covering the supersonic regime to $|v_{\rm LOS}|\leq 8$\,\kms{} and $\leq 10$\,\kms{}, and also the maximum field strength is increased such that $B\leq 5$\,kG. The last 5 inversion runs are devoted to finding the best fits for extreme cases for the supersonic LOS velocities \citep[e.g.][]{delToroIniesta2001ApJL...Evershed,  vanNoort2013A&A, EstebanPozuelo2016ApJ, Siu-Tapia2017A&A} and superstrong magnetic field strengths \citep{vanNoort2013A&A, Okamoto2018ApJ, Siutapia2019, CastellanosDuran2022...Phd}, while further loosening the limits in each run. In the last run, the maximum values for the magnetic field strength and line-of-sight velocity were set to  15\,kG and 30\,\kms{}, respectively.

During a given inversion run, the global minimum of the $\chi^2$-hypersurface can be located inside the allowed parameter space or not, depending on the complexity of the atmosphere. When increasing the limits of the parameters in the next inversion run, the previously fitted atmosphere is only affected if the ``real'' global minimum of the $\chi^2$-hypersurface was not part of the parameter space. Then the next inversion run will improve the solution by increasing the size of the $\chi^2$-hypersurface in each step until the ``real'' global minimum is part of the parameter space. However, by extending limits the global minimum for a given atmosphere does not change the solution if it represents the global minimum also in the increased parameter search space.

In addition, if the previous best-fit atmosphere was stuck in a local  minimum, we help the inversion to go out of it by starting the new inversion run with an offset given by the ``invert-smooth-invert'' approach. This approach assumes that adjacent pixels may have relatively similar atmospheres. Consequently, when using the smoothed atmosphere as an input of the next inversion run, the inversion is likely to  find the global minimum easier, avoiding getting stuck in a local minimum. Furthermore, every time the inversion is restarted, the so-called Marquardt parameter is also reset (see e.g., Eq.~4 of \citet{vanNoort2012A&A}). The Marquardt parameter is used to find the minimum in a  $\chi^2$-hypersurface, since it regulates the amount of linearization assumed by the algorithm. We help the algorithm to leave the local minimum by resetting the Marquardt parameter. The effect of the ``invert-smooth-invert'' approach can be directly seen working in the quiet sun, where the low signal-to-noise tends to retrieve \textit{noisy} maps. Other spatial-regularization schemes are found in the literature \citep[see e.g.,][]{delaCruzRodriguez2019AA...coupled}, but they have not been implemented in the SPINOR code yet. The opted simple spatial-regularization ``invert-smooth-invert'' scheme is applied after each inversion run. 
Therefore, it is straightforwardly applied to any other type of inversion that uses a gradient-based minimization (e.g., SIR-inversions) to fit synthetic Stokes profiles to the observed ones.

As suggested by \citet{vanNoort2013A&A}, data were upsampled before the inversion by a factor of 2 in Fourier space to keep the noise level constant (see Fig.~5 in \citet{vanNoort2013A&A} for illustration). The upsampled data are then inverted by the coupled inversions, which means that the number of pixels inside the same FOV is four times larger than for the original sampling. For normal [fast] maps, the approximate pixel size is 0\farcs16 [0\farcs32]. The inversion is done on upsampled pixels with a size of 0\farcs08 [0\farcs16]. The output of the coupled inversion is later downsampled with the same procedure in Fourier space to return to the nominal scale of the observed FOV, i.e., the final pixels have again a size of 0\farcs16 [0\farcs32]. The upsampled inversions are not presented as a default in the current catalog, but are available on request. The upsampling (and later downsampling) steps are needed to fit substructures that are below the pixel size, but which affect the observed Stokes profiles. As explained by \citet{vanNoort2013A&A}, this approach is similar to having multiple atmospheres within the pixel that are assigned different filling factors. Since these atmospheres or components are coupled by the PSF, this allows the coupled inversions to infer the atmospheric conditions up to the diffraction limit of the telescope when the PSF of the optical system is known and the data are reordered without seeing disturbances. See \citet{vanNoort2017A&A...ImageRestoSpectra} for estimating the PSF and restoration of solar spectra when observations are taken by a ground-based facility.

\subsection{Tiling the FOV}\label{sec:tiling}

During the inversion, a minimization procedure is applied to find the best match between the observed Stokes profiles and the synthetic Stokes profiles that are obtained after the solution of the radiative transfer equation of polarized light for a given atmospheric model. In the classic pixel-to-pixel inversions, every pixel inside the FOV is treated independently. This implies that the size of the matrices needed to be inverted during the minimization procedure is independent of the size of the FOV. Hence, classic inversions are fast and easy to parallelize, and not very demanding in terms of memory usage.  

For coupled inversions, all profiles within the FOV must be inverted simultaneously. The coupling of the solutions between the pixels rapidly increases the size of the matrices needing to be inverted during the minimization procedure with the number of pixels, $N$. The size of the matrix computations roughly increases $\propto\!N^3$, which places greater demands on computer memory with increasing size of the scan \citep[see further details in][]{vanNoort2012A&A}.

The number of pixels in the scans selected for \modest{} ranges from \mbox{$N_{\rm smallest}\sim$8\,500}
($\approx$22\arcsec{}$\times 36\arcsec{}$; fast mode) to $N_{\rm largest}\sim$591\,000 ($\approx$125$\arcsec{}\times110\arcsec{}$; normal mode). In addition, after upsampling the number of pixels increases by a factor of four. Many of the scans are much too large to be handled in one go. One solution is to split the larger scans into sufficiently overlapping tiles. In addition, we were able to use three different computing facilities to perform our inversions. However, this also implied that we had to account for the differences between the facilities such as the type of processors, the available storage, the maximum runtime per job, and the read/write I/O characteristics among others. We chose to divide all scans into tiles of the same size based on all these constraints and for consistency. We obtained good results by dividing each upsampled FOV into tiles of 150$\times$150 spatial pixels resulting in a maximum memory usage of 2\,GBytes per tile (see Fig.~\ref{fig:tiling_modest}). Maps on the left side of Fig.~\ref{fig:tiling_modest} mark the location of each tile within the FOV, where different colors represent each tile. Note the overlap between tiles.

We performed tests by varying the overlapping area between tiles to determine how much overlap is needed to ensure that there are no discontinuities when stitching the atmospheres retrieved in the individual tiles together. These tests allowed us to ensure that the overlap between the tiles was at least twice the width of the core of the PSF that covers most of the power in Fourier space. A conservative value for this overlap was chosen to be 16 pixels. This was done to ensure a homogeneous quality of inversion after all the tiles (now containing the best-fit profiles and resulting atmospheres) are arranged in their original positions (see also Appendix~\ref{sec:effectstiling}).

\begin{table*}[htbp]
\centering
\caption{Features covered by the sample}\label{tab:modest2solarfeatures}
\begin{tabular}{lrrcrrcrr}
\toprule[1.5pt]
\multicolumn{1}{c}{\multirow{2}{*}{Pixels}} & \multicolumn{2}{c}{Fast mode} & & \multicolumn{2}{c}{Normal mode} & &\multicolumn{2}{c}{Total}\\
 \cline{2-3}\cline{5-6}\cline{8-9}
 & \multicolumn{1}{c}{(Num. px)}  & \multicolumn{1}{c}{(Percen.)}  && \multicolumn{1}{c}{(Num. px)} & \multicolumn{1}{c}{(Percen.)}  && \multicolumn{1}{c}{(Num. px)} & \multicolumn{1}{c}{(Percen.)}\\
\midrule[1.5pt]
Quiet Sun/plage/network& 9.1$\times10^{7}$&  68.8\,\%&& 6.2$\times10^{6}$&  60.6\,\%&& 9.7$\times10^{7}$&  68.2\,\%\\
Penumbra& 3.3$\times10^{7}$&  25.1\,\%&& 3.2$\times10^{6}$&  31.8\,\%&& 3.6$\times10^{7}$&  25.6\,\%\\
Warm umbra& 7.1$\times10^{6}$&   5.4\,\%&& 6.6$\times10^{5}$&   6.5\,\%&& 7.7$\times10^{6}$&   5.5\,\%\\
Cold umbra& 9.1$\times10^{5}$&   0.7\,\%&& 1.2$\times10^{5}$&   1.1\,\%&& 1.0$\times10^{6}$&   0.7\,\%\\
Masked& 7.6$\times10^{4}$&  0.06\,\%&& 1.9$\times10^{4}$&  0.18\,\%&& 9.4$\times10^{4}$&  0.07\,\%\\
Total& 1.3$\times10^{8}$& 100.0\,\%&& 1.0$\times10^{7}$& 100.0\,\%&& 1.4$\times10^{8}$& 100.0\,\%\\
\bottomrule[1.5pt]
\end{tabular}
\end{table*}

Maps on the right side of Fig.~\ref{fig:tiling_modest} show each tile within the FOV that then was inverted. After the inversion, tiles were stitched together by removing the outer added edge between neighboring tiles. No boundary or edge effects were visible after putting together the tiles on the large map. No smoothing at the boundaries was required, indicating that the independent inversions of the adjacent tiles found the same global minimum for the pixels located in both tiles, and that the chosen overlap of 16 pixels was sufficiently large.

Appendix~\ref{sec:effectstiling} shows a quantitative test on the effects of tiling the FOV. A scan of AR\,11748  was inverted without tiling the FOV on a so-called ``fat node'' with large memory. This scan contains all types of features (umbrae, penumbrae, and quiet Sun) at the location where the tiles are put together. The inversion was carried out following the same steps as presented in Secs.~\ref{sec:modestcoupledinversions}-\ref{sec:InvStrategyModest}. As the last step, the whole FOV was downsampled to its original size, thus reversing the Fourier upsampling procedure mentioned above. This inversion was then compared with the nominal ``tiled'' \modest{} inversion (Fig.~\ref{fig:tilingeffect}). This test demonstrates that the effects of tiling the FOV are negligible.

\section{Results} \label{sec:results-modest}

The \modest{} catalog contains the atmospheric conditions of sunspot groups that were located roughly within a longitude of $\pm60^{\circ}$  and $\pm25^{\circ}$ in latitude on the solar disk. \modest{} started out focusing on the more complex sunspot groups (Fig.~\ref{fig:sMODESTSample}d). Later, simpler sunspots were added. All ARs with sunspots were observed by Hinode/SOT-SP and span the last part of solar cycle 23, and almost the full cycle 24. Currently, \modest{} is composed of \numinvs{} scans of \numars{} individual ARs with sunspots (or parts of these active regions). The inverted scans frequently show multiple sunspots. The total number of spatial $xy$-pixels is \numpixels{}, and the total number of data points\footnote{One data point is considered to be a monochromatic Stokes profile value. A typical $xy$-pixel has 444 data points, which corresponds to a full Stokes vector at a given wavelength $(I, Q, U, V)(\lambda)$ times $\sim$111 spectral wavelength positions covered by each Stokes profile. The number of wavelength positions may vary depending on the observing mode.} that was fitted is \numdatapoints{}.

In December 2006, i.e. still in the early stages of the mission, a hardware failure in the X-band at 8.4\,GHz reduced the downlink transmission bandwidth between Hinode and ground stations considerably. In terms of data volume, fast mode scans with a spatial sampling of 0\farcs32 are considerably smaller than normal scans with a spatial sampling of 0\farcs16. Due to the technical problem with the X-band, the majority of the Hinode/SOT-SP scans are therefore taken in fast mode (the fifth column in Table~\ref{tab:sample-ARS-MODEST} lists whether the scan was taken in the fast or normal mode).

\begin{figure*}[htbp]
    \centering
    
   \includegraphics[width=.85\textwidth]{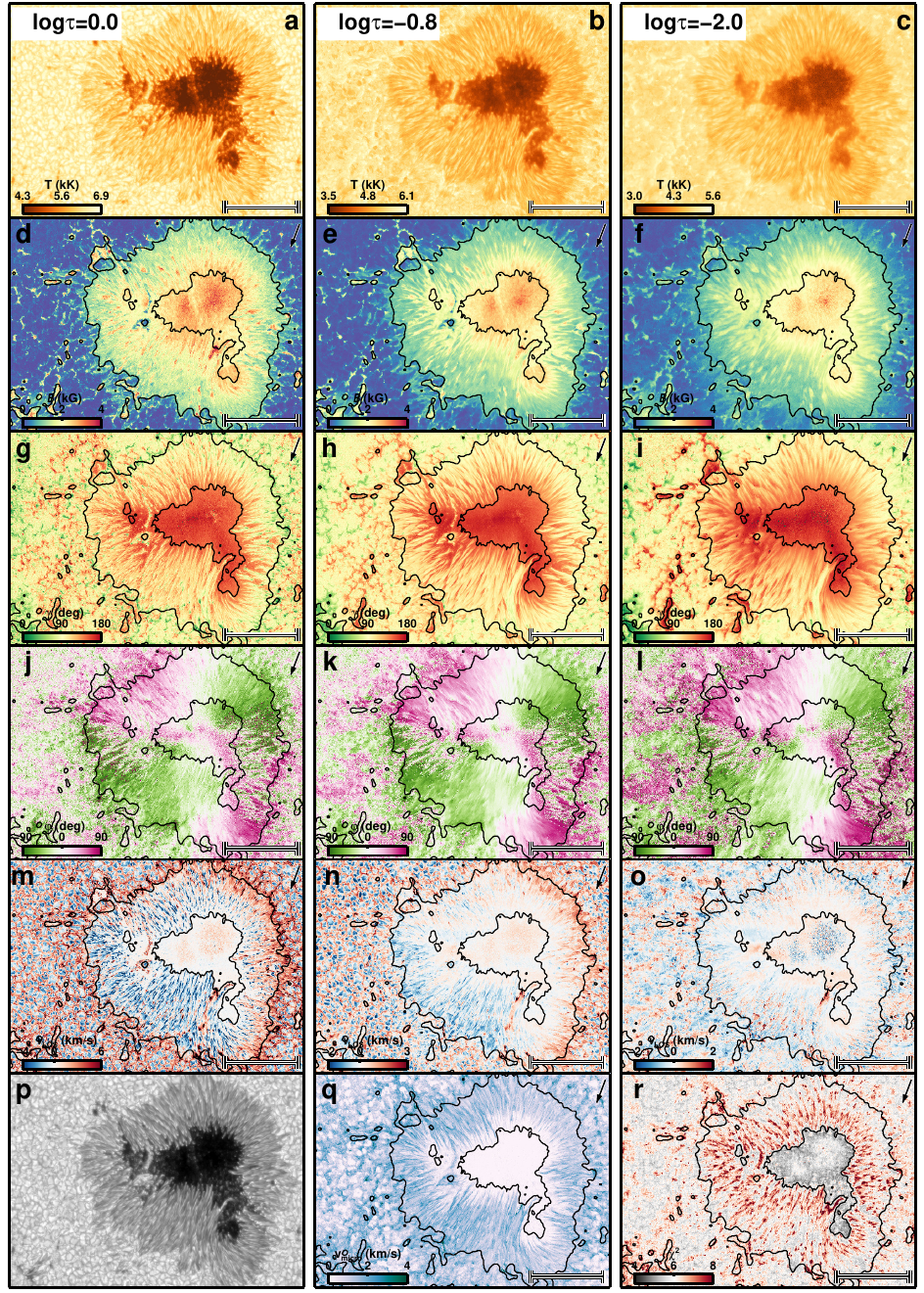}
           \caption{Depth-dependent atmospheric conditions retrieved by the coupled inversions of AR\,10953 (2007 May 1). Rows from top to bottom show maps of temperature, magnetic field strength, inclination, azimuth, and line-of-sight velocity, while explanations for the quantities plotted in the bottom row are given below. In the first five rows, the columns show from left to right these quantities retrieved at $\log\tau =0$, $-0.8$ and $-2.0$, i.e. at the bottom, middle, and top nodes. Panels (p)-(r) display the best-fit continuum, micro-turbulence, and  $\chi^2_{\rm reg}$ maps.  This scan was observed in the normal mode with a pixel size of 0\farcs16. The FOV is composed of $8\times10$ tiles.
           }
    \label{fig:modesttratified1}
\end{figure*}

\begin{figure*}[!htbp]
    \centering
   \includegraphics[width=.9\textwidth]{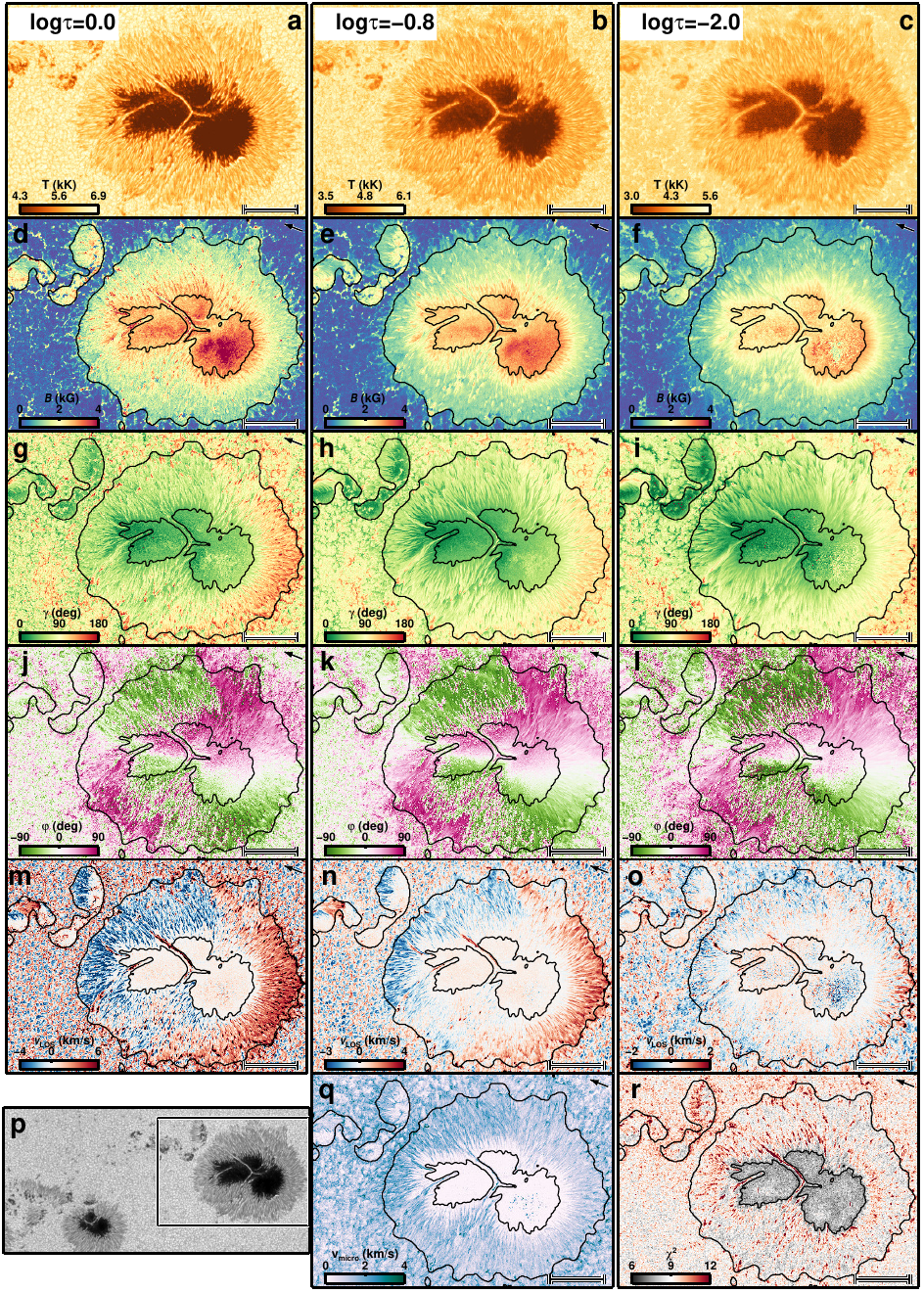}
   \caption{Same as Fig.~\ref{fig:modesttratified1}, but for of part of AR\,11944 (2014 January 9). This scan was observed in the fast mode with a pixel size of 0\farcs32. The black rectangle inlaid in panel (p) marks the FOV covered by the other panels. The full FOV is composed of $13 \times 6$ tiles.} 
    \label{fig:modesttratified2}
\end{figure*}

Table~\ref{tab:modest2solarfeatures} summarizes the total number of $xy$-pixels covering different solar features depending on whether they were observed in the fast or normal mode, and the combined total. Roughly 69\% of the areas of the scans observe quiet Sun, active region plage, and network regions surrounding sunspot groups, while 25\% of the pixels cover sunspot penumbrae and 5\% belong to sunspots umbrae. In addition, Figs.~\ref{fig:MODESTcontinuum000} to \ref{fig:MODESTcontinuum043} show a mosaic of the best-fit continuum images as well as the  line-of-sight magnetograms of the current sample of \modest{} (see also summary Table~\ref{tab:sample-ARS-MODEST}).

The following sub-sections present examples taken from the catalog. They include examples of typical fits to the observed Stokes vectors at different $\mu-$values of different solar structures, the retrieved atmospheric maps for different ARs, as well as the intrinsic limitations of the \ion{Fe}{I} pair at 6302\,\AA{}, comparison with the standard Level-2 data product and the 6302\,\AA{} lines when observed in cold umbrae.

\subsection{Modeling of the observed Stokes vector}\label{sec:modelStokes}

The spatially coupled inversions apply a one-component depth-depended atmospheric model to each pixel of an FOV, irrespective of the type of (photospheric) solar features observed. Figures~\ref{fig:modest_fits1} to \ref{fig:modest_fits5} show three scans at $\mu$-values of 0.6, 0.8 and 1.0, respectively. Panel (a) shows the best-fit continuum image of the scan, while columns 2 to 5 present three pixels within the FOV selected on the basis of the complexity of their observed Stokes spectra. The observed Stokes profiles are shown by the gray-dotted lines, while the best fits obtained by the inversions are presented by the colored lines. Pixels in the bottom row belong to the quiet Sun (green line), while the middle and top rows show pixels taken from penumbrae (blue and green lines, sometimes these pixels are located in light bridges). For comparison, the standard MERLIN inversions, that are a part of the Hinode/SOT-SP pipeline (Level-2 data product), are shown by the black lines. The vertical bars mark the reduced spectral window used by the standard Level-2 inversions. Figures~\ref{fig:modest_fits1} to \ref{fig:modest_fits5} corroborate that, regardless of the $\mu$-value of the scan and the photospheric feature, the spatially coupled inversions can retrieve fairly good fits to the sometimes very complex observed Stokes vector using a single atmospheric model. Further examples of typical good fits to complex atmospheres obtained by coupled inversions have been presented by \citet{vanNoort2013A&A, Siu-Tapia2017A&A, Siutapia2019, CastellanosDuran2020, CastellanosDuran2023...ejectionCEFs}.

\subsection{Physical parameters of the atmospheres}
In this section, we present examples of the maps of atmospheric physical parameters contained in the \modest{} catalog, with an inverted normal mode scan being plotted in Fig.~\ref{fig:modesttratified1} and a fast mode scan in Fig.~\ref{fig:modesttratified2}. From top to bottom, the first five rows show temperature, magnetic field strength, line-of-sight inclination and azimuth of the magnetic field, and line-of-sight velocity, respectively. Each column depicts the atmospheric quantities at $\log(\tau)=0$ (left),  $\log(\tau)=-0.8$ (middle) and $\log(\tau)=-2.0$ (right). The bottom row displays the best-fit continuum (panel (p)), micro-turbulence (panel (q)), and the $\chi^2$-map (panel (r)).

The magnetic field is in the line-of-sight reference frame. To transform the magnetic vector to the local reference frame, the intrinsic ambiguity of the magnetic field's azimuth must be taken into account. Tests have been performed with currently available techniques \citep{Leka2009}, however, the disambiguation is currently not part of the \modest{} pipeline (see Section~\ref{sec:modest180disambiguation}). 

Atmospheric maps show emblematic features observed in the solar photosphere. For example: 
outside the sunspot, the upflows in the granules and downflows in the intergranular lanes are well visible (panel (m)) as is the concentration of the magnetic field in the intergranular lanes. Also, the well-known drop of the magnetic field strength of the sunspots from inside the umbra towards the outer parts of the penumbra is clearly visible. The magnetic field decreases with height and the magnetic field canopy is observed when comparing panels (d) and (f).  The temperature variation between penumbral spines and intraspines is clearly visible \citep[panels (a)-(c), see also e.g.,][]{Schmidt1992A&A...penumbra, Lites1993ApJSpinesInPenumbra, Title1993ApJ...EF, Langhans2005A&AL}, which is associated with changes in the magnetic field inclination \citep[panels (g)-(h), cf.][]{Tiwari2013}. The normal Evershed flow pattern is seen as a blueshift in the center-side penumbra and as a redshift in the limb-side penumbra \citep[panel (\ref{fig:modesttratified2}m) in Fig.~\ref{fig:modesttratified2}, cf.][]{Evershed1909}. The same panel (\ref{fig:modesttratified2}m) also shows elongated counter Evershed flows associated with a filamentary light bridge \citep[cf.][]{CastellanosDuran2021...rareCEFs}, as well as small-scale penumbral downflows \citep[cf.][]{Katsukawa2010A&A...Penumbradownflows, JurcakKatsukawa2010A&A...penumbraldownflows, vanNoort2013A&A}.

\section{Discussion}
 
\begin{figure*}[htbp]
    \centering
       \includegraphics[width=.99\textwidth]{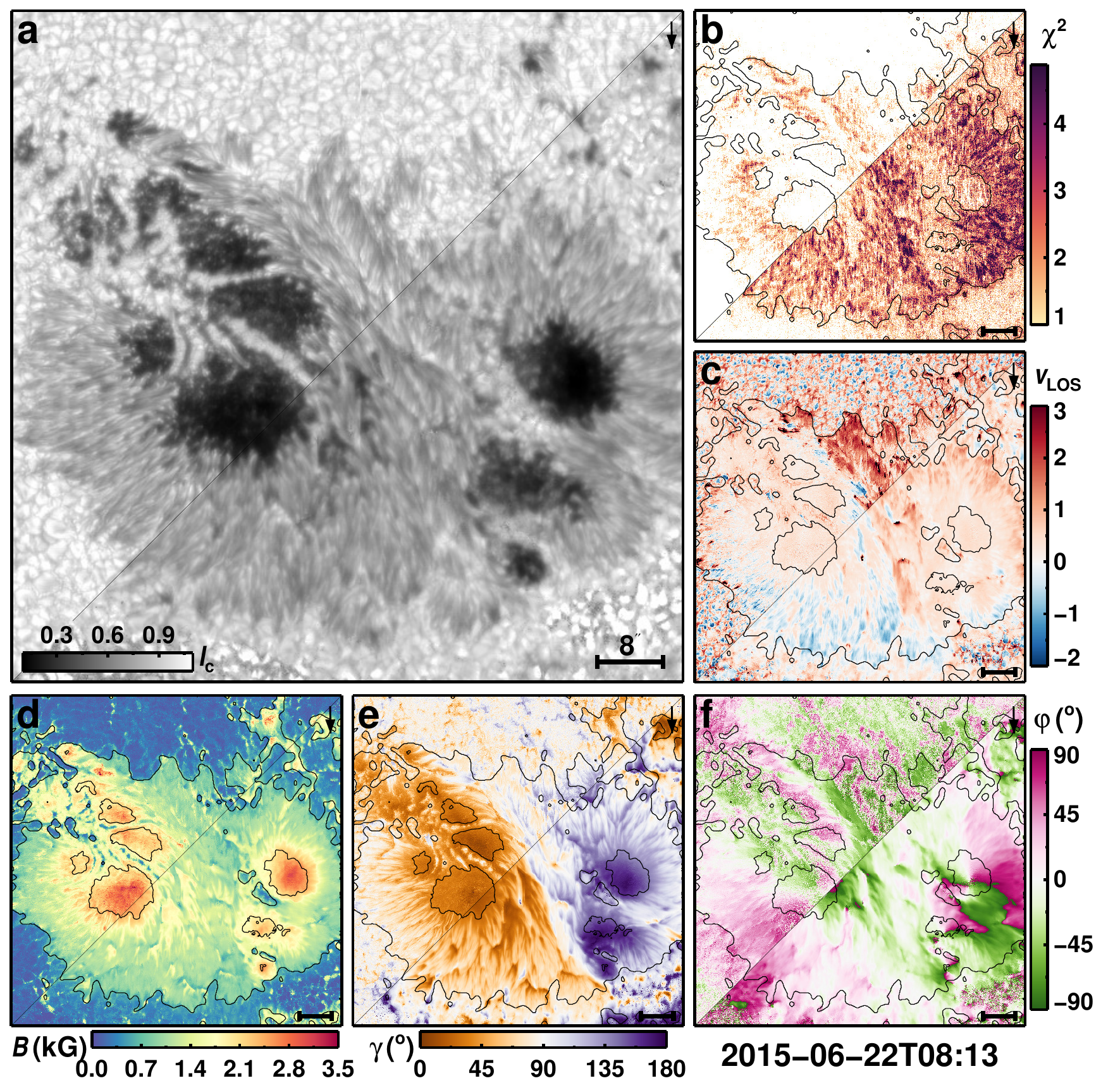}
    \caption{Atmospheric maps of AR\,12371 obtained with the spatially coupled inversions at the middle node ($\log\tau=-0.8$) (top-left half of each panel) and MERLIN (Level-2; bottom-right half). Panel (a) shows the best-fit continuum map. The atmospheric parameters retrieved by both inversion schemes are shown in the panels: (b) $\chi^2$-map, (c) line-of-sight velocity, (d) magnetic field strength, (e) magnetic field inclination relative to the line-of-sight, (f) azimuth relative to the line-of-sight. Atmospheric maps at the middle node were used for the coupled inversions.  Horizontal bars on the bottom-right mark 8\arcsec{}. This scan was observed in the normal mode with a pixel size of 0\farcs16.
    } 
    \label{fig:maps-modest1}
\end{figure*}

\begin{figure*}[htbp]
    \centering
     
   \includegraphics[width=.99\textwidth]{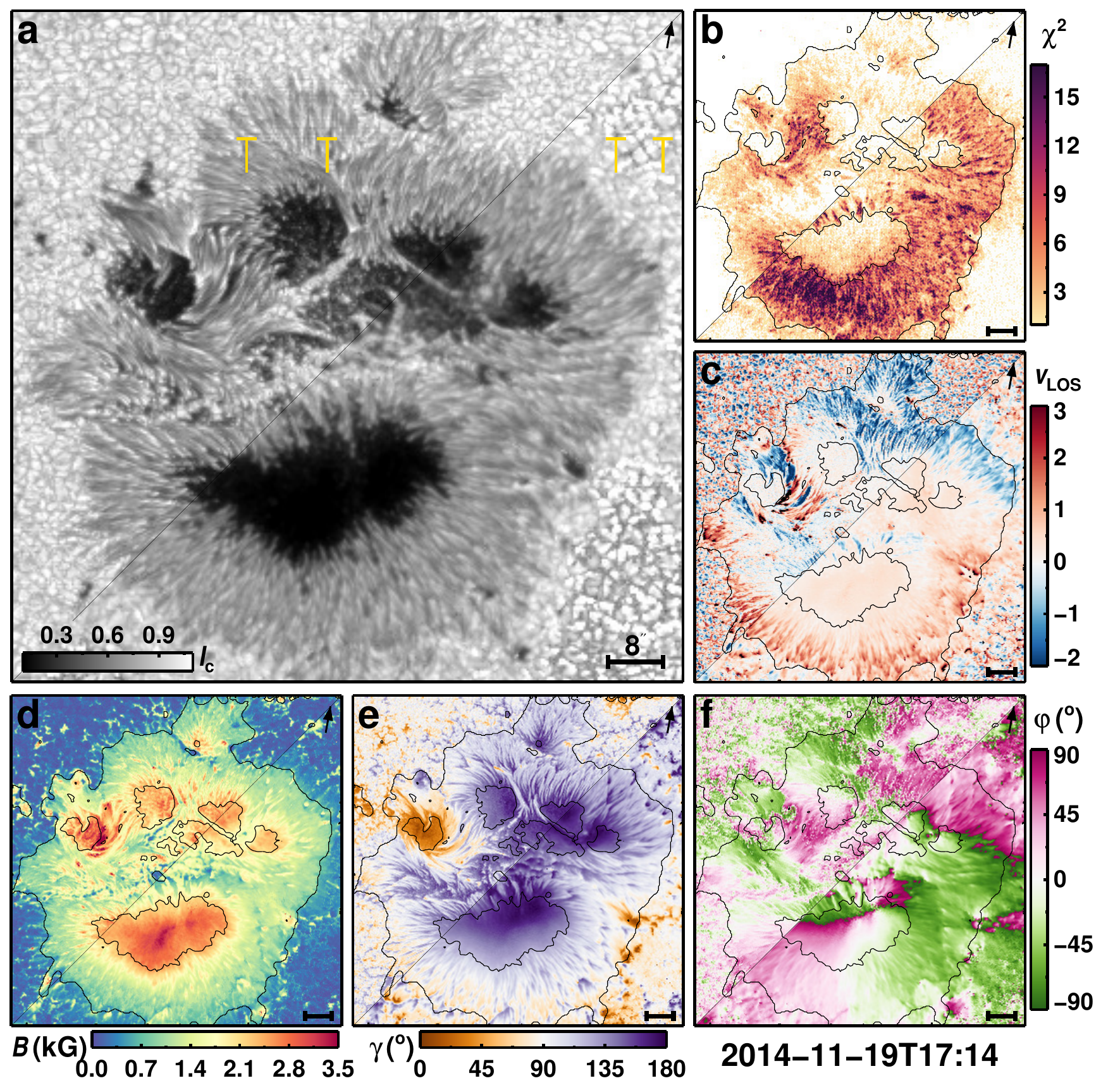}
   \caption{Same as Fig.~\ref{fig:maps-modest1}, but for of part of AR\,12209. This scan was observed in the fast mode with a pixel size of 0\farcs32.}
    \label{fig:maps-modest4}
\end{figure*}

\subsection{Comparison with the standard Level-2 inversions}

We present two examples of the depth-dependent atmospheric conditions retrieved using coupled inversions and compare them with the standard Level-2 ME-inversions that are a part of the Hinode/SOT-SP pipeline. Figure~\ref{fig:maps-modest1} is a normal-mode scan, while Fig.~\ref{fig:maps-modest4} presents a fast-mode scan. The panels in Figs.~\ref{fig:maps-modest1} and \ref{fig:maps-modest4} show the best-fit continuum image (a), $\chi^2$ (b), magnetic field strength (c), line-of-sight velocity (d), magnetic field inclination (e) and the magnetic field azimuth (f). Each panel is diagonally split into two halves, with the top-left half showing the results of the 2D coupled inversions presented in this paper. All these maps show the results for the middle node ($\log\tau_{c}=-0.8$), which is best suited for comparison with ME results  \citep[cf.][]{OrozcoSuarez2010A&A...height..ME}. The bottom-right halves of the same panels show the same atmospheric quantities returned by the Level-2 ME-inversions of the complementary part of the same dataset. The continuum images in panels (a) have slightly different brightnesses in the two halves, likely because the SPINOR and MERLIN codes normalize  the observed Stokes profiles differently during the inversion. This should not affect the rest of the parameters, however.

Not seen in the examples shown here is that coupled inversions provide depth-dependent information that a ME-inversion cannot provide by design. This is a qualitative difference between the atmospheres provided in the \modest{} database and those obtained by inversions with MERLIN. 

Equally important is the relative goodness of fit, or $\chi^2$. For direct comparison of the standard Level-2 with the coupled inversion, the $\chi^2$-maps were estimated by adopting the generic form of the merit function

\begin{equation}
    \chi^2=\frac{1}{4N_{\lambda}-\varrho}\sum_{i=1}^{N_{\lambda}}\sum_{j=1}^4
    \frac{w_{j}^2}{\sigma_j^2}\left(I_{ij}^{\rm obs}-I_{ij}^{\rm syn}\right)^2,
\end{equation}

\noindent where $j$ stands for the four Stokes profiles, $i$ runs over the number of the observed wavelength points ($N_{\lambda}$), $\varrho$ represents the number of free parameters of each inversion\footnote{The number of free parameters of the stratified-\modest{} inversions and the ME-Level-2 inversions are $\varrho_{\rm MODEST}=16$ (Sect.~\ref{sec:freeparameters}) and $\varrho_{\rm MERLIN}=11$.}, $w_{j}$ is the weight applied to each Stokes profile, $\sigma_j$ is the noise in the observed Stokes profile, and $I_{ij}^{\rm syn}$ and $I_{ij}^{\rm obs}$ are synthetic and observed Stokes profiles.  In addition, $\chi^2$-values were estimated only within the two spectral windows that MERLIN fits (black lines in Figs.~\ref{fig:modest_fits1}--\ref{fig:modest_fits5}), and we use the same weights that MERLIN assumes. Even though these assumptions may favor the MERLIN inversions, the $\chi^2$-values obtained by coupled inversion are significantly smaller (see panel (b) in Figs.~\ref{fig:maps-modest1} and \ref{fig:maps-modest4}).

The superior visibility of the fine-scale structure from the SPINOR coupled inversions is evident. This difference is best visible in fast-mode scans (e.g. Fig.~\ref{fig:maps-modest4}), but is also present for the normal-mode maps. One sign of the removal of the spatial degradation by the PSF by the coupled inversion is the enhanced contrast in the best-fit continuum map when comparing both inversions (panels (a)), as can be seen from Figs.~\ref{fig:maps-modest1} and \ref{fig:maps-modest4}. In addition, in the outer part of the penumbra, the Level-2 continuum maps show a ``halo'' in the outer penumbra boundary that extends into the surrounding granulation. This spatial contamination between different solar features is absent in the coupled inversions. This is a direct consequence of the spatially coupled inversions that account for the smearing by the PSF. Other atmospheric maps show comparable results at large scales, but on the smaller scales, the coupled inversions retrieve finer structures. 

This smearing by the telescope PSF makes the Level-2 inversions look spatially smoother over scales of the PSF width. Inverting these smeared maps with Level-2 inversions maintains this smearing and results in apparently smoother maps. The 2D inversions act like a deconvolution of these smeared maps, bringing back the original fine structure with higher contrast in all physical quantities \cite[see for example Fig.~11 of][and compared Figs.~3 and 4 of \citet{CastellanosDuran2020}]{vanNoort2012A&A}.

\subsection{Molecular blends in the coolest parts of sunspot umbrae}\label{sec:modestmolecular}

In the comparatively cool environment of sunspot umbrae, diatomic molecules can form. Due to the large number of possible molecular transitions, molecular spectral lines appear as bands in many spectral regions \citep[e.g.,][]{Berdyugina2002A&A...MolI,AsensioRamos2004ApJ603L}. Of particular importance in the spectral window around 6302\,\AA{} are transitions of the CaH and TiO diatomic molecules \citep[][see also \citet{Berdyugina2003A&A...MolII,Berdyugina2005A&A...MolIII}]{Berdyugina2011ASPC..mol6302}.

\begin{figure*}[htbp]
    \centering
    \includegraphics[width=.99\textwidth]{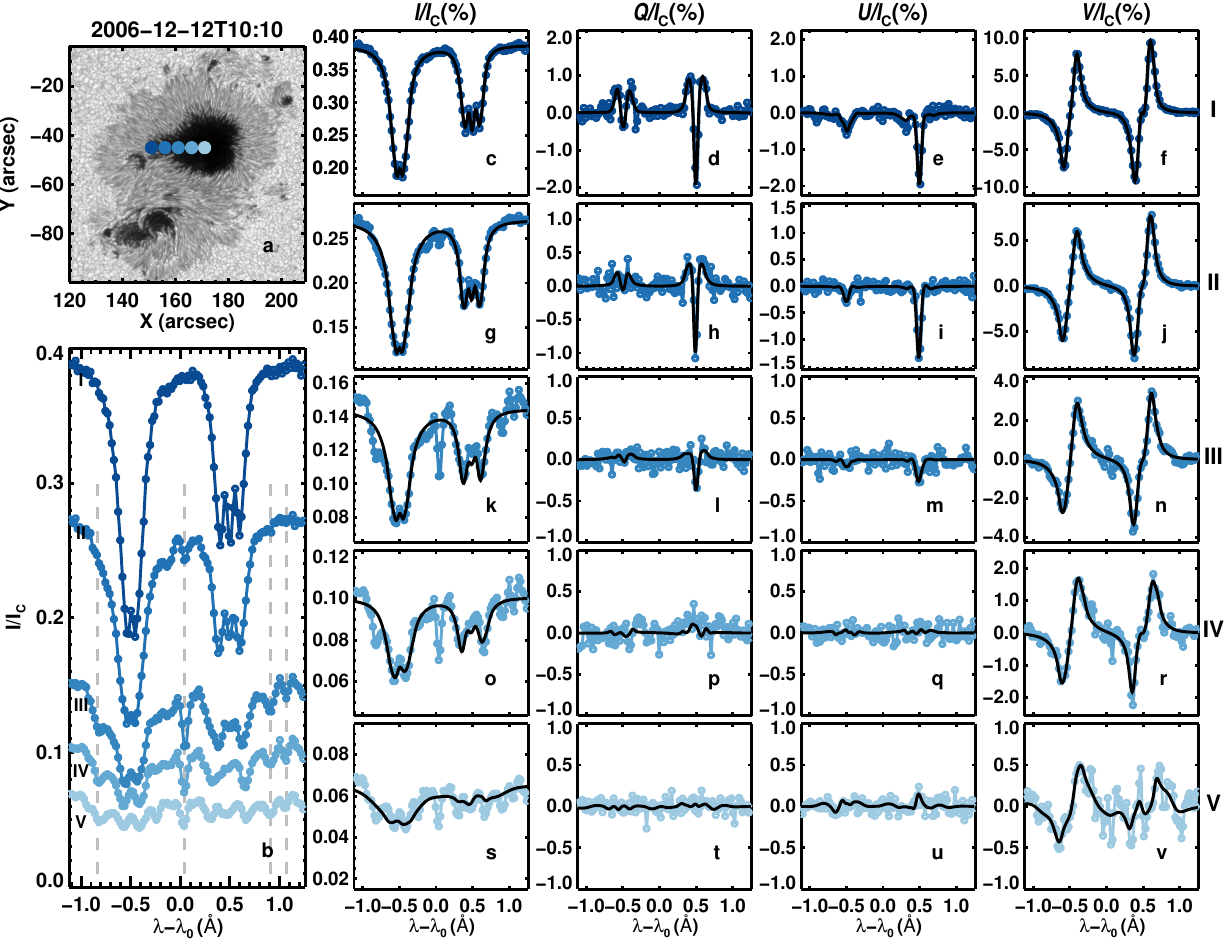}
    \caption{Hinode/SOT-SP observations of AR\,10930 observed almost at disk-center ($\mu=0.99)$. Panel (a) shows the continuum image. Blue circles overplotted on the continuum image mark the locations inside the umbra of the spectra plotted in the remaining panels, color-coded according to the region they were observed in. Lighter blue represents darker and cooler umbral regions, where blends with molecular lines appear. Panel (b) shows the five normalized Stokes $I(\lambda)$ profiles. The spectra are displayed on the same intensity scale for comparison (i.e. all spectra are normalized to the average quiet Sun continuum intensity). Dashed-gray vertical lines mark four prominent molecular blends identified at 6301.2\,\AA{}, 6302.0\,\AA{}, 6302.9\,\AA{}, and 6303.1\,\AA{}. 
    Columns 2 to 5 (panels c-v) show the full observed Stokes vectors emerging at the same spatial pixels and the best fits found by the coupled inversions (black lines).}
    \label{fig:hinode_profiles-modest}
\end{figure*}

Larger sunspots harbor darker and hence cooler  umbrae \citep[e.g.,][]{Mathew2007A&A...spotcontrast}. In large dark umbrae the molecular blends are the strongest. Figure~\ref{fig:hinode_profiles-modest} shows five selected Stokes profiles arising along a line from the edge to the center of the umbra of a sunspot (AR\,10930) that was located near the disk center.  The main sunspot of AR\,10930 is one of the largest spots in the sample.
The pixels with the plotted profiles were chosen to sample umbra from warm ($I_{\rm c}\approx0.4\,I_{\rm qs}$; marked with the darkest blue color) to cold ($I_{\rm c}\lesssim0.15\,I_{\rm qs}$; marked with the lightest blue color). Vertical dashed lines in panel (\ref{fig:hinode_profiles-modest}b) mark the most prominent (but by far not all the) molecular blends. At the lowest umbral temperatures, the \ion{Fe}{I} line pair is of similar strength as the molecular lines, making it difficult for the inversion code to find the correct solution. Columns 2 to 5 show the observed Stokes vector and the obtained best fits. The blending molecular lines can make the identification of the correct position and splitting of the \ion{Fe}{I} line pair at 6302\,\AA{} difficult for the inversion code, potentially leading to errors in the field strength or the otherwise very robustly determined velocity.  
In Fig.~\ref{fig:hinode_profiles-modest}v, the Stokes $V$ of the \ion{Fe}{I} line at 6302\,\AA{} shows a ``wiggle'' that the code sometimes fits by adding complexity to the atmosphere (e.g., unrealistically large line-of-sight velocities in umbrae). This ``wiggle'' is thought to have a molecular origin, as it is not seen in the \ion{Fe}{I} line at 6301\,\AA{}. Also, the continuum level is suppressed to an unknown value due to a molecular blend that appears at $\approx$6300.3\,\AA{}. In addition to the molecular blends,
the \ion{Fe}{I} line pair 6302\,\AA{} weaken in the coldest parts of umbrae due to their relatively high excitation potentials, making these lines not very sensitive to low temperatures \citep{Smitha2021...Ti22micron}. The issues presented by the appearance of molecular lines must be taken into account when analyzing the cold regions in sunspots. The user of the catalog must be aware of this issue and should use the inversion results from the regions where the spectra are affected by molecular lines with caution, or use masking techniques to avoid them completely.

Fortunately, umbral regions cold enough to form strong molecular lines do not appear too often on the Sun. In fact, pixels with low intensities  $I_{\rm c}<0.15\,I_{\rm qs}$, where molecular blends are sufficiently prominent to start affecting the inversion results,  just represent \PercColdumbraWRTtotalumbra{} of all umbral pixels ($I_{\rm c}\leq0.5\,I_{\rm qs}$), and less than \Perccoldumbra{} of all pixels that are part of the \modest{} catalog. All the same, these 11\%\ of the umbrae are also the coldest and hence likely the parts with the strongest field strengths (see the relationship between temperature and field strength studied by, e.g., \citet{KoppRabin1992SoPh..Bfield..Temp..spots, Solanki1993A&A...IRlines..wilson, Mathew2004A&A...wilson}), so that the properties of the strongest-field umbral regions may not be so reliable in \modest{} as other parts of the sunspots.

\subsection{180$^{\circ}$ disambiguation of the magnetic field}\label{sec:modest180disambiguation}

Measurements of the magnetic field based on the Zeeman effect have an intrinsic 180$^{\circ}$ ambiguity in the azimuthal angle of the component of the magnetic field transverse to the light-of-sight. Single-height-disambiguation techniques exist to solve the 180$^{\circ}$ ambiguity through minimizing a given parameter of the system \citep[e.g., electric currents, free energy, etc. For a review see][]{Leka2009}. These single-height techniques are currently, to some extent, considered to be straightforward and they have been already automated for the ME-based inversions part of the SDO/HMI and Hinode/SOT-SP standard data products. These techniques work especially well in simple configurations but tend to struggle in more complex active regions. For example, the disambiguation techniques undergo problems in highly sheared regions and complex polarity inversion lines, which are present in many active regions. 

Another more recently developed technique combines observations taken with different view-angles by two instruments (for example SO/PHI \citep[][]{Solanki2020} onboard Solar Orbiter \citep{Muller2020A&A...SOLARORBITER} and Hinode/SOT-SP or SDO/HMI). This technique solves the 180$^{\circ}$ ambiguity by geometrical considerations without the need of minimizing any parameter \citep{Valori2022SoPh..disambiguation, Valori2023A&A...disambiguityPHI}. However, this technique also solves the 180$^{\circ}$ ambiguity at a single optical depth only, because of the limited number of wavelength points recorded by SO/PHI. 

There is currently no standard technique available to properly disambiguate height-stratified inversions. Hence, current methods find the minimum of a given parameter for one height but these techniques do not find the global minimum for a stratified inversion. To avoid this problem, it is customary to disambiguate the data node by node, but that can easily result in inconsistent disambiguation for the different layers. If the sunspot is located close to the disk center and it is a relatively regular sunspot,  the disambiguation can be applied to the middle node and the same solutions can then be consistently applied to the other nodes. This choice assumes that the magnetic field does not change dramatically with height, which cannot be assumed in general. For example, it is expected that this approach will fail at the edges of facular magnetic concentrations, where the opposite polarity weak field lies under the expanding fields of the flux concentrations \citep[see, e.g.,][]{Buehler2015A&A...plage}. Similar problems are expected under the canopies surrounding sunspots.

Due to the current lack of an available disambiguation technique that fully accounts for the variation of the magnetic field with height, we opt, in this early phase of \modest{}, to provide the magnetic field vector directly as it is obtained by the inversion, with the intrinsic 180$^{\circ}$ ambiguity. This also means that all quantities are given in the line-of-sight reference frame. Disambiguation codes can be easily found and the user of \modest{} can select a map and apply a method of choice, bearing in mind the caveats given above.

\subsection{Quality masks}\label{sec:qualitymasks}

After the inversion of a scan is fully converged, we create quality masks for that scan. Three different types of pixels are marked. These masks are not related to the quality of the fit to the observed Stokes vector, but rather to the integrity of each Hinode/SOT-SP scan.  We define three quality levels:

\begin{itemize}
    \item[(I)] Good-quality pixels
    \item[(II)] Medium-quality pixels.  In rare events, there are sharp cuts found in the Hinode/SOT-SP data. These pixels in most cases are well fitted on both parts of the sharp cut. However, because of the coupled nature of the inversions, with the signal in nearby pixels being coupled to each other by the point spread function, we mark these places to warn the user that the results may be less reliable than for good-quality pixels.
    An example is shown on Fig.~\ref{fig:modest_fits_medium} in Appendix~\ref{sec:lowqualitypixels}
    \item[(III)] Low-quality pixels. In a very small fraction of pixels, there are significant spikes in one of the Stokes profiles. These most likely come from cosmic rays or readout problems. As the inversion still finds the best possible fit to these pixels, including to the spike, unrealistic atmospheric conditions are retrieved in the pixel(s) and, due to the spatial coupling, their surroundings. All these types of pixels must be excluded from any type of analysis. An example is shown on Fig.~\ref{fig:modest_fits_low} in Appendix~\ref{sec:lowqualitypixels}.
\end{itemize}

Together, pixels of types (II) and (III) make up \Percbadpixels{} of all pixels within the catalog.

\subsection{Residuals in the atmospheric maps}

In some scans we observe a slight discontinuity of the atmospheric parameters for a certain pixel of the slit, resulting in spurious horizontal lines in the maps. These spurious residuals are best seen in the top node of the temperature maps, but they are also visible in the continuum maps. These spurious lines are most often located in the top part of the FOV \citep[cf. Fig.~1 of][]{Tiwari2013}. For some scans, these spurious lines can all be observed directly on the continuum maps of both, the coupled inversions and the standard Level-2 data. In Fig.~\ref{fig:maps-modest4} they are pointed out by the yellow marks on panel (\ref{fig:maps-modest4}a). The dispersion of these residuals in the temperature maps is of the order of a few Kelvin, but it is difficult to separate this variation from the observed feature. This pattern highly resembles the Hinode/SOT-SP dark images \citep[see Fig.~2 in][]{Lites2013SoPh}, and is seen more often in scans taken after $\sim$2014. This might be an indication of some sort of degradation, but further analysis is needed. The magnitude of the residuals is small, so they could not be detected before the inversions. For the output of the inversion, we avoid removing these residuals ad-hoc and leave it up to the user of the catalog to either leave them as they are, or to apply a filter for their removal.

\section{Summary and conclusions}\label{sec:conclusion-modest}

In this paper, we introduced the \modest{} catalog consisting of height-dependent maps of atmospheric parameters to study a wide range of phenomena in the photosphere of the Sun. The \modest{} catalog encompasses all types of photospheric features, from umbral dots to network fields and quiet Sun regions. In addition, all types of sunspot groups from $\alpha$-spots to complex $\delta$ ARs are covered by this catalog (Fig.~\ref{fig:sMODESTSample}). The catalog currently contains the outputs of inversion of \numinvs{} spatial scans by Hinode/SP of \numars{} individual sunspot groups,  making a total of \numpixels{} spatial pixels with retrieved height-dependent physical conditions of the solar photosphere.

The large variety of scans, features, types of ARs, and locations on the solar disk that are contained in the \modest{} catalog can be used to perform a range of studies, some of which are listed below. Thanks to the large number of inverted sunspots with height-dependent information, it may be possible to extend on existing studies.

\begin{itemize} 
    \item The photospheric structure of the penumbra \citep[e.g.,][]{Tiwari2013, Tiwari2015A&A...penumbra}, its formation and temporal evolution on timescales of hours-days \citep[e.g.,][]{Scharmer2008ApJL...evershed, Schlichenmaier2010...formationPenumbra, BelloGonzalez2019ASPC}, as well as penumbral features such as penumbral grains \citep[e.g.,][]{Muller1973SoPh...penumbralgrains.a, Muller1973SoPh...penumbralgrains.b, Sobotka1999A&A...penumbralgrains, Sobotka..Puschmann2022A&A...PGs,Sobotka2023A&A...PGs} or orphan penumbrae \citep[e.g.,][]{Zirin..Wang1991AdSpR..OrphanPenumbra, Loeptien2023A&A...orphanpenumbra}
    \item Statistics of granular  \citep{Vazquez1973SoPh...LB, Lites1990ApJ...penumbrafinestruc, Lites1991,Sobotka1993, RouppevanderVoort2010ApJ...LB, Lagg2014A&A...LB, Schlichenmaier2016A&A...LB, GrionMarin2021...LB} and filamentary light bridges \citep{Adjabshirzadeh1980A&A...UD, Rimmele2008ApJ...UD, Katsukawa2007PASJ...LB,Guglielmino2017ApJ}
    \item Bipolar light bridges and the superstrong magnetic fields they harbor \citep{Zirin1993b, Livingston2006SoPh, Okamoto2018ApJ, Wang2018RNAAS, CastellanosDuran2020}
    \item The 3-D structure and evolution (on hours-days timescale) of the normal Evershed flow \citep{Evershed1909, RimmeleMarino2006ApJ...EF} and counter Evershed flow \citep[e.g,][]{Schlichenmaier2011IAUS..CEF, Kleint2012ApJ...Cflare2CEF, Kleint2013ApJ, Louis2014A&A...CEF, Siu-Tapia2017A&A, CastellanosDuran2021...rareCEFs, CastellanosDuran2023...ejectionCEFs}
    \item Dependence of properties of sunspots on their size, shape, complexity, etc \citep[e.g.,][]{Collados1994A&A...spots, Mathew2007A&A...spotcontrast, Rezaei2012A&A...sunspotProperties, Rezaei2015A&A...SunspotProperties}. The physical structure of the umbra-penumbra boundary as a function of various sunspot parameters  \citep{Jurcak2011A&A...JurcakCriterion, Jurcak2018A&A...JurcakCriterion, Lindner2020A&A...JurcakCriterion, Loeptien2020A&A...umbrapenumbra,  GarciaRivas2021A&A...JurcakCriterion}
    \item Properties of the granulation (velocities, contrast, sizes, etc. as a function of height) \citep[e.g.,][]{Danilovic2008A&AL...contrast, Hirzberger2010ApJ...QSContrasts, Ishikawa2020ApJ...convection}, and of small-scale  magnetic features \citep[e.g.,][]{Buehler2015A&A...plage, Buehler2019A&A...PlageNetwork,  Kahil2017ApJS..Brightness, Kahil2019A&A...Icontrast}. 
    \item Study changes in sunspot or granulation properties, or the amount of magnetic flux over the $\sim$1.5 solar cycles that are currently covered by \modest{} \citep[see e.g.,][]{Muller1984SoPh...NetworkVariability, Mathew2007A&A...spotcontrast, Livingston2012ApJ...Sunspotfield, Buehler2013A&A...localdynamo, Lites2014PASJ...NetworkCycleVariation, Kiess2014A&A...umbra} 
    \item Use the large sample of ARs at different $\mu-$values, for example, to train artificial neural networks \citep{Carroll2001A&A...ANNinversions, Socas-Navarro2005ApJ...ANNinversions, AsensioRamos2019A&A...ANNinversions, Milic2020A&A...ANNinversions,  Liu2020ApJ...ANNinversions, Socas-Navarro2021...inversionANA, Centeno2022...ANNinversions}
    \item The connection between flares and the detailed properties of underlying sunspots \citep[e.g.,][]{Shimizu2014PASJ...Flowsflares, Toriumi2019LRSP, Yardley2022...FlareMagneticEnvironment}.
\end{itemize}

The spatially coupled inversions effectively retrieve excellent fits to the observed Stokes profiles for all kinds of solar features, almost irrespective of the complexity of the measured Stokes profile. Only in the very cold umbra are the inversions of lower quality, caused by the intrinsic problem of molecular lines appearing in the spectral window observed by Hinode/SOT-SP. In addition, recent works have demonstrated the significance of considering non-LTE effects on the formation of Iron lines in the photosphere \citep{Smitha2020A&A......NLTEFeI, Smitha2021A&A...NLTEFeI, Smitha2022....NLTE6173}. Compared to LTE-treated Stokes profiles, non-LTE Stokes profiles showed significant differences. In order to fully assert how non-LTE effects impact the retrieved atmospheric conditions, additional work is required. 

Besides inverting the Hinode/SOT-SP scans of a larger number of active regions, additional improvements are conceivable. For example, modeling molecular lines and implementing them into the inversion procedure could improve the results obtained for the darkest parts of umbrae. The development and application of a disambiguation technique that covers the height dependence would open up new applications.

\textit{Observatory --} Hinode (SOT-SP)\par
\textit{Computing facilities --} HPC-GWDG (G\"ottingen, Germany), HPC-MPS (G\"ottingen, Germany), and HPC-MPCDF (Garching, Germany).

\begin{acknowledgements}
We are grateful to Rebecca Centeno Elliott for providing the MERLIN input files that were used to obtain synthetic Stokes profiles of standard Level-2 inversions. We thank Stephan Thoma (MPS) for his continuous support and development of a web-application to access the catalog. J.~S. {Castellanos~Dur\'an} was funded by the Deutscher Akademischer Austauschdienst (DAAD) and the International Max Planck Research School (IMPRS) for Solar System Science at the University of G\"ottingen. This project has received funding from the European Research Council (ERC) under the European Union’s Horizon 2020 research and innovation program (grant agreement No. 695075).  Hinode is a Japanese mission developed and launched by ISAS/JAXA, with NAOJ as domestic partner and NASA and UKSA as international partners. It is operated by these agencies in cooperation with ESA and NSC (Norway).
\end{acknowledgements}

\bibliographystyle{aa}
\bibliography{references}

\onecolumn

\appendix

\section{Spectral stray light and spurious polarization}\label{sec:greystrayModest}

We tested the influence of the spectral stray light on the atmospheric parameters retrieved by the inversions.  We used Hinode/SOT-SP data of the quiet sun taken at disk center and compared it to the National Solar Observatory/Kitt Peak Fourier Transform Spectrometer atlas \citep[FTS;][]{Brault1976JOSA...FTS,Brault1978fsoo...FTS}. We found that the spectral stray light is up to $\zeta\sim$3\%, which is consistent with the nominal values found by \cite{Lites2013SoPh}. We chose a test FOV that contains quiet Sun, a small pore, part penumbra and umbra. We calibrated this FOV using the nominal Hinode/SOT-SP data, and created two datacubes: one with the nominal calibration, and another with $\zeta=3\%$ spectral stray light  subtracted. To apply this subtraction, we followed the approach described in \cite{Bianda1998A&A}. The corrected Stokes $I(\lambda)$ is given by 

\begin{equation}\label{eq:Icorr}
    \left(\frac{I(\lambda)}{I_{c}}\right)=(1+\zeta)\frac{I(\lambda)^{{\rm corr}}}{I_c}-\zeta,
\end{equation}
where $I_c$ is the continuum intensity. The corresponding Stokes profiles for the linear ($F_{Q,U}(\lambda)$) and circular polarization ($F_V(\lambda)$) are
\begin{equation}\label{eq:Scorr}
    F_{Q,U,V}(\lambda)= F_{Q,U,V}^{{\rm corr}}(\lambda) +k_{Q,U,V}I(\lambda),
\end{equation}

\noindent $k_{Q,U,V}$ is the spurious polarization. $k_{Q,U,V}$ is usually well corrected by the \texttt{sp\_prep} routines \citep{Lites2013SoPh}, however, it has been  shown that $k_{Q,U,V}$ are not negligible in extreme cases \citep{Okamoto2018ApJ}. Combining Eqs.~\eqref{eq:Icorr}-\eqref{eq:Scorr}, the corrected Stokes profiles are

 \begin{figure*}[hb]
 \begin{center}
 \includegraphics[width=.85\textwidth]{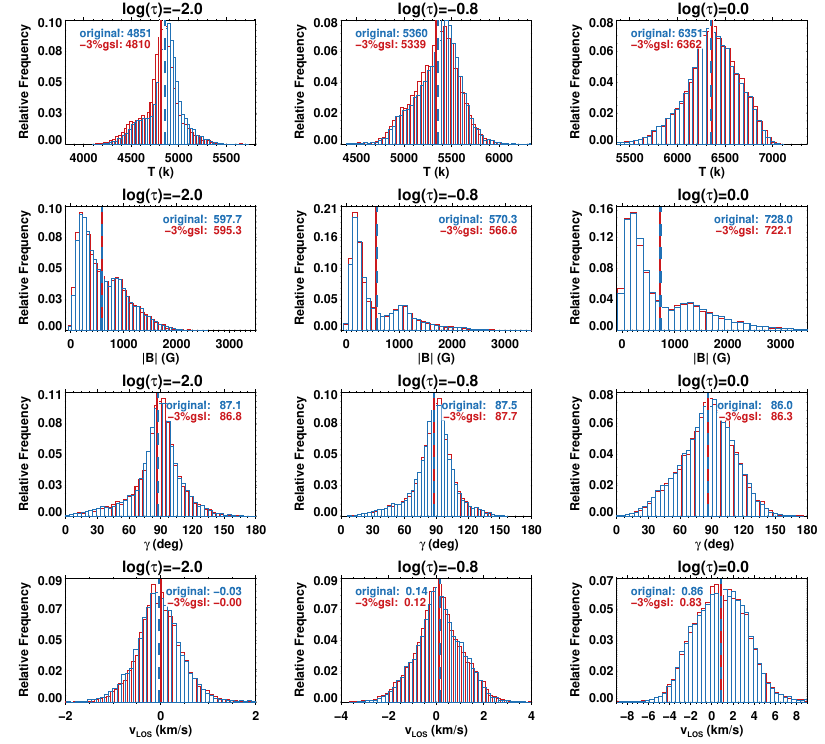}
 \caption{Effects of the gray stray light. Each column shows the retrieved atmospheric conditions for each node.  Plotted from top to bottom rows are: the retrieved temperature, $B$, $\gamma$, and $v_{\rm LOS}$.  Results obtained from the original Hinode/SOT-SP data are in blue, and data corrected with 3\% -level of gray stray light are red.   }\label{fig:greyhito}
 \end{center}
 \end{figure*}

\begin{equation}
    \left(\frac{F_{Q,U,V}(\lambda)}{I_c}\right)^{{\rm corr}}=\frac{F_{Q,U,V}(\lambda)}{I_c\left(1-\frac{\zeta}{1+\zeta} \right)}.
\end{equation}

Figure~\ref{fig:greyhito} shows the results of this test, where labels ``original'' and ``-3\%gsl'' mark the two datacubes. Panels in Fig.~\ref{fig:greyhito} show the histograms of the same atmospheric quantity. No strong variation was observed with and without the 3\% gray stray light. Based on these findings and as the gray stray light might change over time, no correction of this type was applied. 

\section{Effects of tiling the FOV}\label{sec:effectstiling}

Splitting the FOV into multiple tiles was necessary to invert large Hinode/SOT-SP scans because of the high-computational demand of the coupled inversions, mainly in terms of memory usage. To mitigate boundary effects, the tiles need to overlap sufficiently. We chose this overlap region to be 3 times the FWHM of the PSF. In this appendix, we demonstrate that the tiling with this large spatial overlap does not produce any artifacts to the inversion results. 

\begin{figure*}[htbp]
 \begin{center}
 \sidecaption
 \includegraphics[width=.99\textwidth]{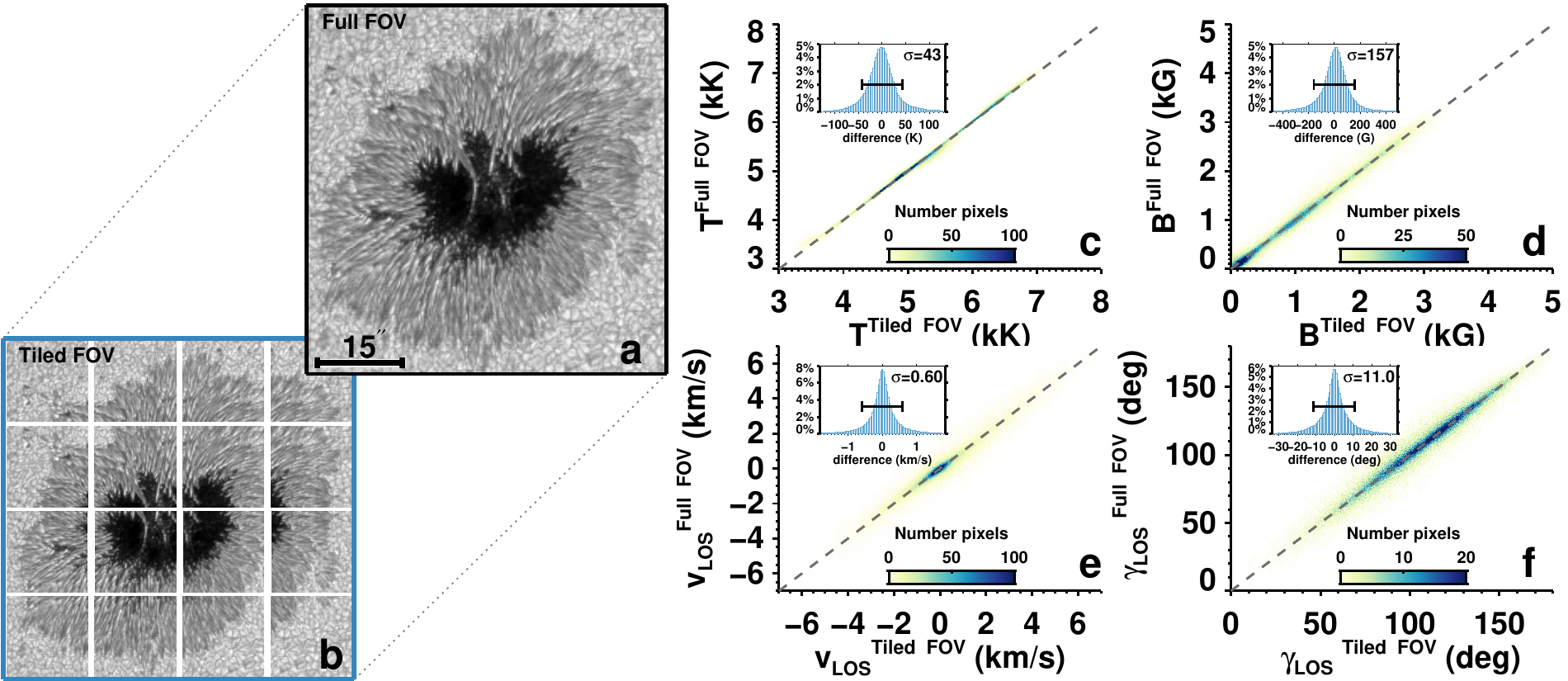}
 \caption{Effects of tiling the FOV. Panel (a) shows the full untiled continuum image of the AR\,11748 when it was located close to disk center. Panel (b) shows the FOV broken up into overlapping tiles. The edges of each tile are marked by the white lines FOV. Scatter plots of atmospheric parameters obtained with or without the tiling are displayed in Panels (c)-(d). Data points from all three optical depth nodes are plotted in each panel. The individual panels display the temperature (c), magnetic field strength (d), line-of-sight velocity (e), and line-of-sight inclination of the magnetic field (f). The $x$-axis displays the inversions performed with the tiled FOV, while the $y$-axis shows the inversions for the whole FOV inverted in one go for the three-node height. The insets inside panels (c)-(f) show the histograms of differences between the two inversions, where the black bars indicate 1$\sigma$.  }\label{fig:tilingeffect}
 \end{center}
 \end{figure*}

Figure~\ref{fig:tilingeffect} shows an experiment where the same scan was inverted using two different approaches. The first inversion was performed for the full FOV without any tiling (\ref{fig:tilingeffect}a). The second inversion used the same tiling scheme as in \modest{}, and the FOV was divided into 16 tiles (\ref{fig:tilingeffect}b). This scan was chosen because all types of features are present (the quiet Sun, penumbrae, and umbrae) in the area where the tiles are stitched together. These inversions were carried out following the same steps as in Secs.~\ref{sec:modestcoupledinversions}-\ref{sec:InvStrategyModest}.

The results of these inversions are shown in Fig.~\ref{fig:tilingeffect}. Panels (\ref{fig:tilingeffect}c) to (\ref{fig:tilingeffect}f) depict the scatter plots of the retrieved atmospheric conditions for the three nodes combined. The abscissa results of the inversion where the FOV was divided, while the ordinate displays the inversion of the non-tiled FOV. Insets within panels (c)-(f) show the scatter of the difference between the atmospheric quantities from the two inversions for the three-node heights. The differences are Gaussian with a 1$\sigma$-level of 43\,K for the temperature, 157\,G for the magnetic field strength 0.6\,\kms{} for the line-of-sight velocity, and 11\,deg for the line-of-sight inclination of the magnetic field. As expected, the magnetic-field-inclination scatter is the largest. The match between the two inversions reveals that the effect of tiling the FOV is small.

\section{Medium- and low-quality pixels}\label{sec:lowqualitypixels}

\begin{figure*}
    \centering
    \includegraphics[width=.95\textwidth]{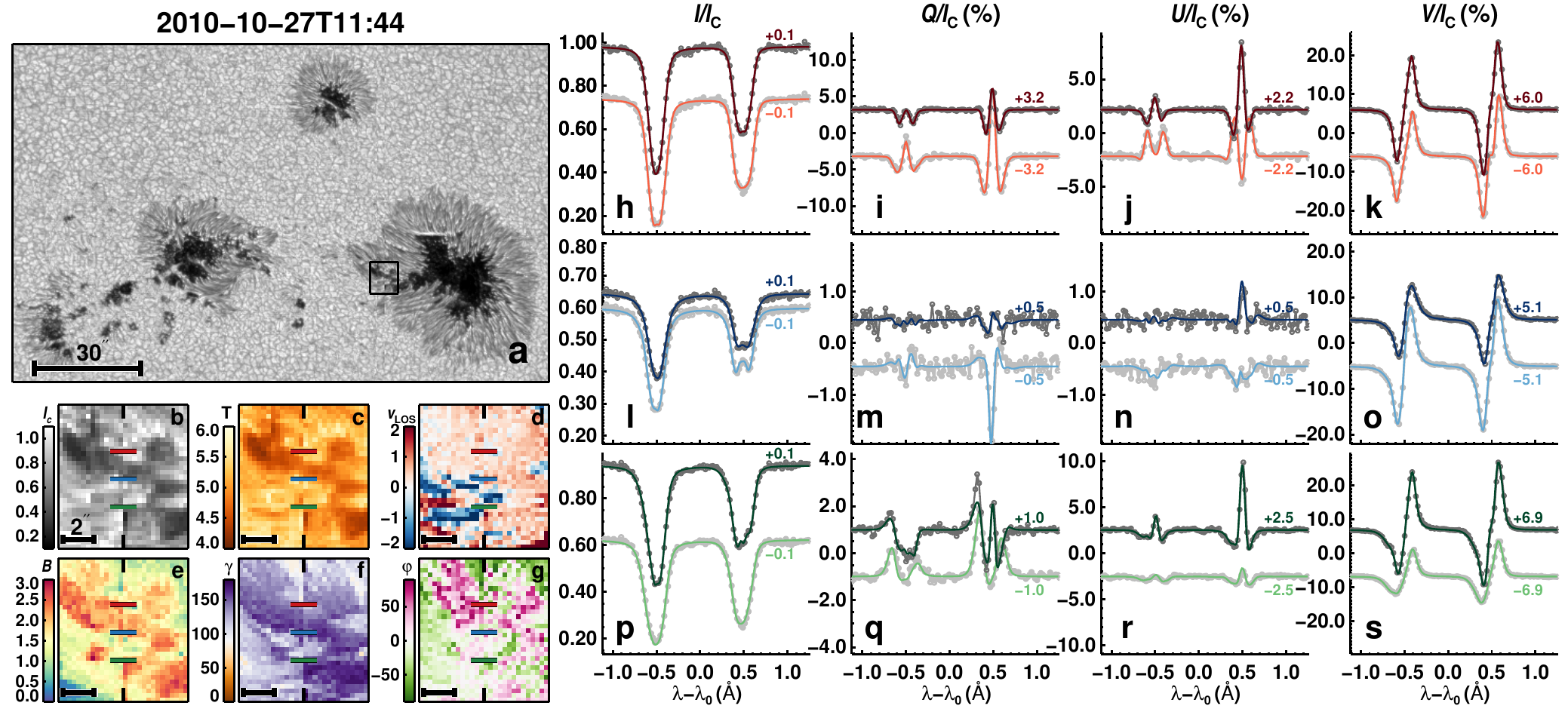}
    \caption[Medium-quality pixels in \modest{}]{
    Fits obtained by the coupled inversions to the observed Stokes profiles for medium-quality pixels. Panel (a) shows the continuum image of the AR\,11117 when it was located at $\mu\approx0.9$. Maps in panels (b) to (g) display the continuum intensity (b), temperature (c), $v_{\rm LOS}$ (d), magnetic field strength (e), inclination (f), and azimuth (g). These maps are from the region outlined by the black rectangle in panel (a).
    Columns 2 to 5 show the observed Stokes profiles (gray dotted lines) and the best fits obtained by the coupled inversions (colored lines). A vertical shift was added to the Stokes profiles for better representation. The shift values  are displayed on the right side of each profile.  Each row shows two Stokes vectors coming from neighboring pixels lying immediately to the left  (darker lines) and to the right (lighter lines) of a sharp discontinuity in the data (indicated by black vertical tickmarks in (b)--(g)). Colored horizontal lines in panels (b) to (g) mark the locations of the profiles. $\lambda_0=6302$\,\AA{}. }
    
    \label{fig:modest_fits_medium}
\end{figure*}

\begin{figure*}
    \centering
    \includegraphics[width=.95\textwidth]{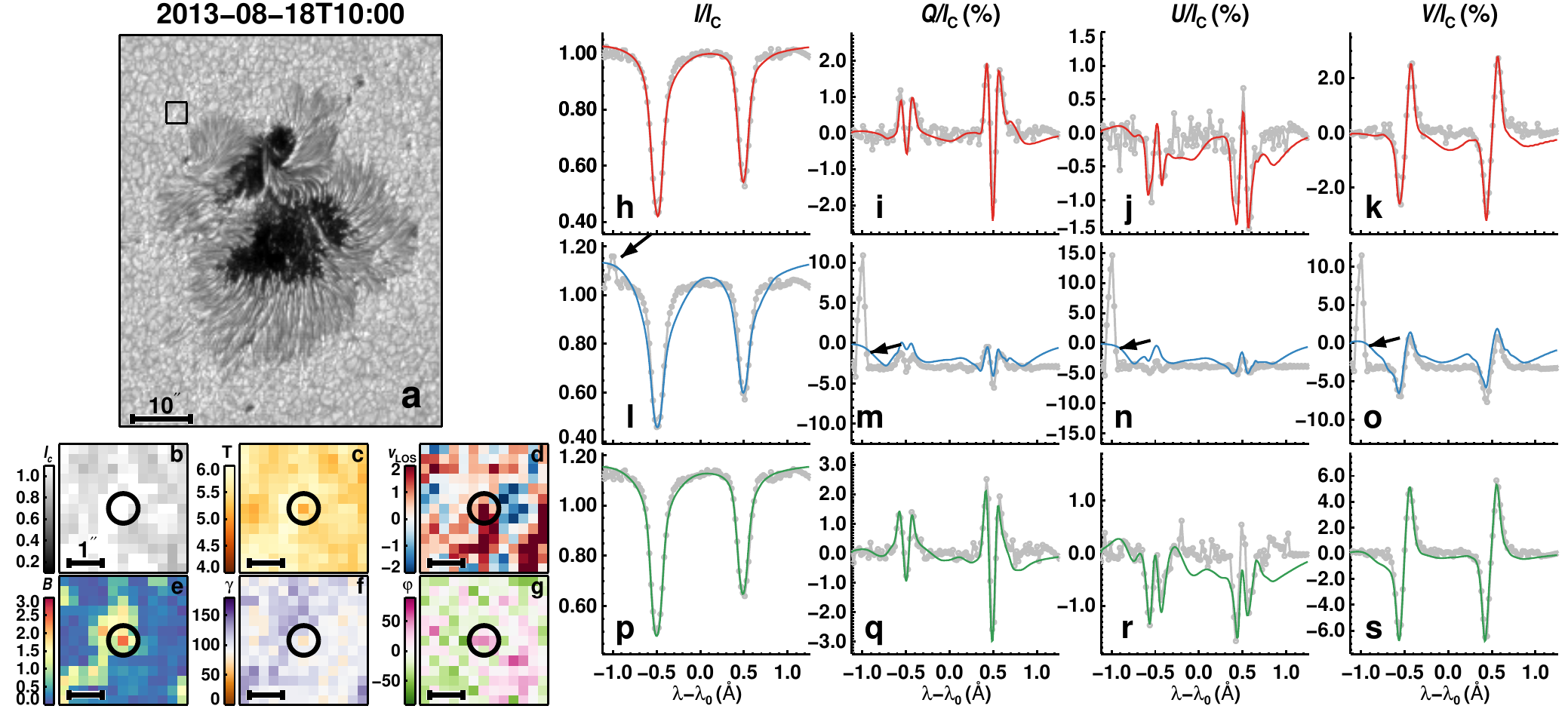}
    \caption[Low-quality pixels in \modest{}]{
    A low-quality pixel. Same as Fig.~\ref{fig:modest_fits_medium} for a sunspot group belonging to AR\,11818, when it was located at $\mu\approx0.76$. The open circles mark the location of the low-quality pixel. Black arrows in panels (l) to (o) mark an artifact likely produced by a cosmic ray which strongly influences the fits.}
    
    \label{fig:modest_fits_low}
\end{figure*}

Figures~\ref{fig:modest_fits_medium} and \ref{fig:modest_fits_low} show examples of pixels, which are likely affected by instrumental or data reduction artifacts, making them of only medium or low quality (see Sect.~\ref{sec:qualitymasks}). The medium-quality pixels can easily be identified in maps of the Stokes parameters, and also in the corresponding maps of the inverted parameters, as sharp discontinuities in the vertical or horizontal direction. 

Figure~\ref{fig:modest_fits_medium} shows the atmospheric maps and the fits on both sides of one such discontinuity. The Stokes profiles are well-fitted as is common for medium-quality pixels. However, the fit values are likely not as reliable as for normal pixels due to the coupling of the pixels on both sides of the divide by the PSF.

Figure~\ref{fig:modest_fits_low} shows an example of a low-quality pixel. There are spikes in the Stokes profiles (black arrows in panels \ref{fig:modest_fits_low}l -- \ref{fig:modest_fits_low}o). Especially in the polarized Stokes profiles the spike is much larger than the actual solar signals.  The inversion still tries to invert this pixel and its surroundings (red and green profiles), but poor fits and unrealistic atmospheric conditions are retrieved in these pixels and in its surroundings (the latter again due to the coupling via the PSF).

\newpage
\section{Sample of sunspot groups observed by Hinode/SOT-SP part of the \modest{} catalog}

\begin{figure*}[htbp]
\begin{center}
\includegraphics[trim={0 2.55cm 0 0},clip,width=1\textwidth]{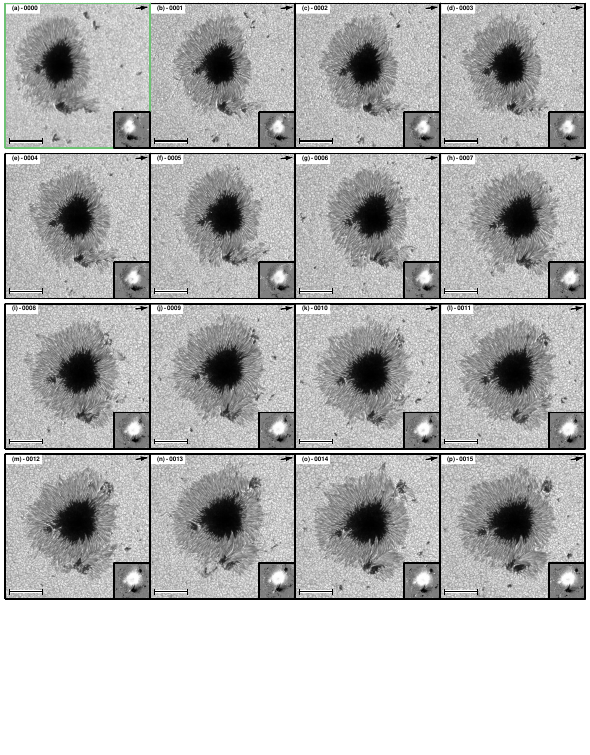}
\caption{Best-fit continuum maps and magnetograms of the sample of sunspots in the \modest{} catalog. Numbers on the upper-left part of each panel are the \textsc{INV\_ID} of each inversion. \textsc{INV\_ID}s can be found on the second column of Table~\ref{tab:sample-ARS-MODEST}. Bars on the bottom-left corner have a length of 20$^{\prime\prime}$, and arrows on the top-right corner point toward disk center. Insets on the lower-right show the magnetogram ($B\cos(\gamma_{\rm LOS})$) in the LOS reference frame clipped at 1.5\,kG obtained at the middle node. Axes colors display whether the scan was taken in fast (black) or normal mode (green).}\label{fig:MODESTcontinuum000}
\end{center}
\end{figure*}

\begin{figure*}[htbp]
\begin{center}
\includegraphics[width=1\textwidth]{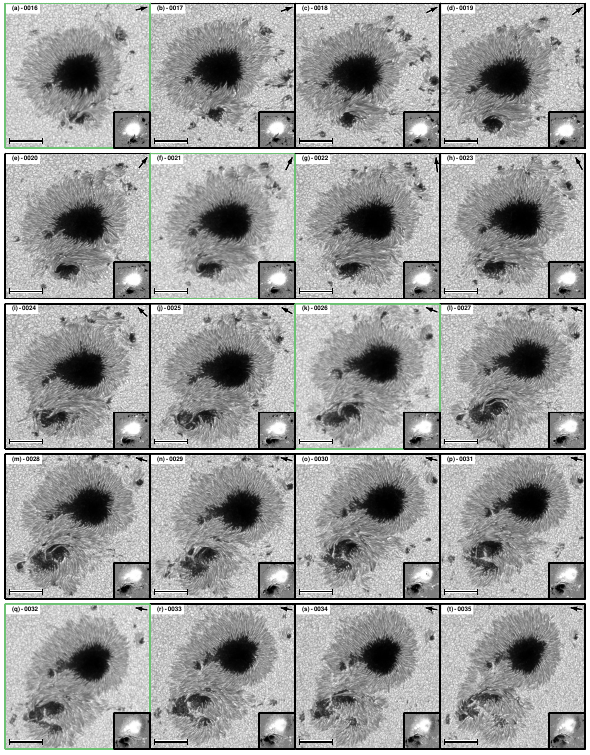}
\caption{Same as Fig.~\vref{fig:MODESTcontinuum000}. Axes colors display whether the scan was taken in fast mode (black) or normal mode (green).}\label{fig:MODESTcontinuum001}
\end{center}
\end{figure*}

\begin{figure*}[htbp]
\begin{center}
\includegraphics[width=1\textwidth]{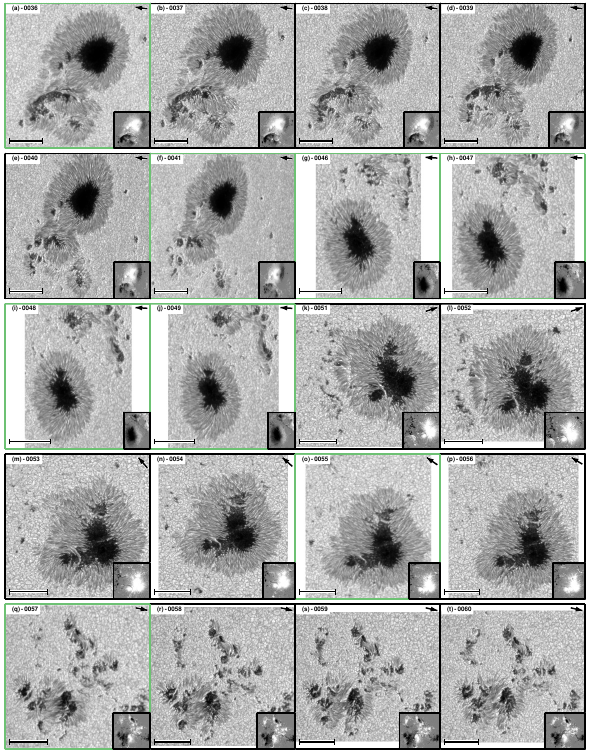}
\caption{Same as Fig.~\vref{fig:MODESTcontinuum000}. Axes colors display whether the scan was taken in fast mode (black) or normal mode (green).}\label{fig:MODESTcontinuum002}
\end{center}
\end{figure*}

\begin{figure*}[htbp]
\begin{center}
\includegraphics[width=1\textwidth]{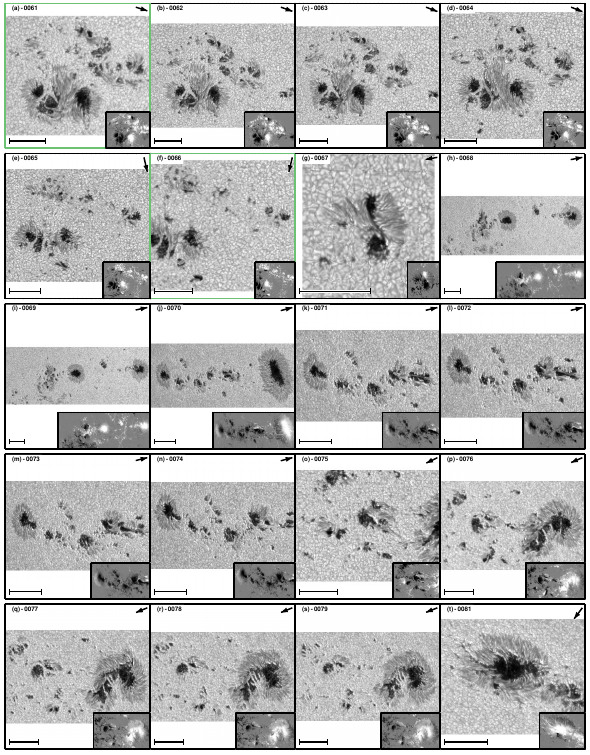}
\caption{Same as Fig.~\vref{fig:MODESTcontinuum000}. Axes colors display whether the scan was taken in fast mode (black) or normal mode (green).}\label{fig:MODESTcontinuum003}
\end{center}
\end{figure*}

\begin{figure*}[htbp]
\begin{center}
\includegraphics[width=1\textwidth]{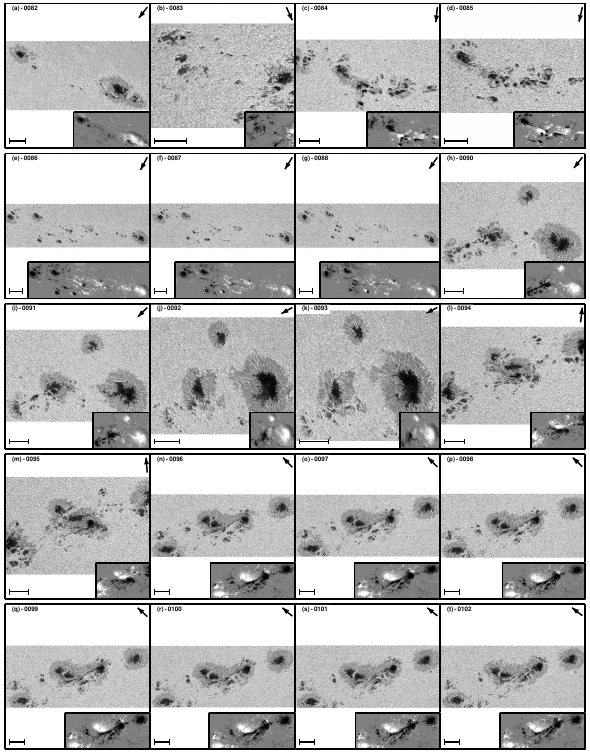}
\caption{Same as Fig.~\vref{fig:MODESTcontinuum000}. Axes colors display whether the scan was taken in fast mode (black) or normal mode (green).}\label{fig:MODESTcontinuum004}
\end{center}
\end{figure*}

\begin{figure*}[htbp]
\begin{center}
\includegraphics[width=1\textwidth]{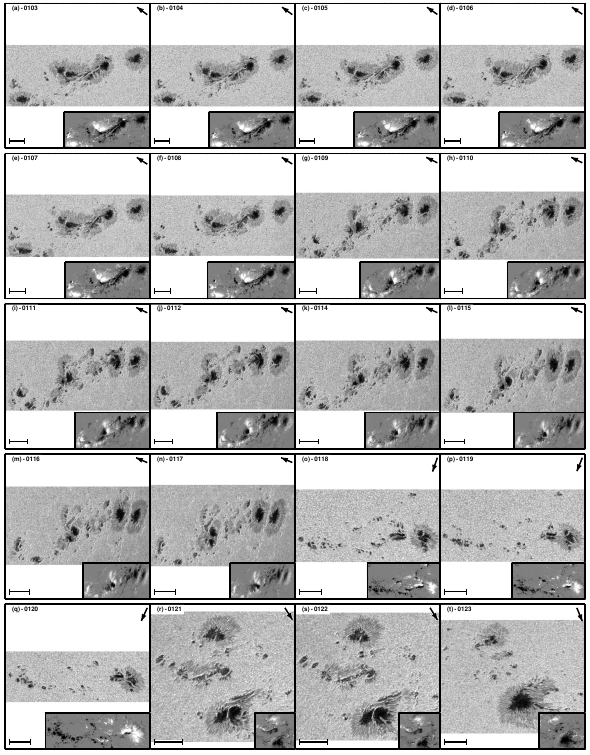}
\caption{Same as Fig.~\vref{fig:MODESTcontinuum000}. Axes colors display whether the scan was taken in fast mode (black) or normal mode (green).}\label{fig:MODESTcontinuum005}
\end{center}
\end{figure*}

\begin{figure*}[htbp]
\begin{center}
\includegraphics[width=1\textwidth]{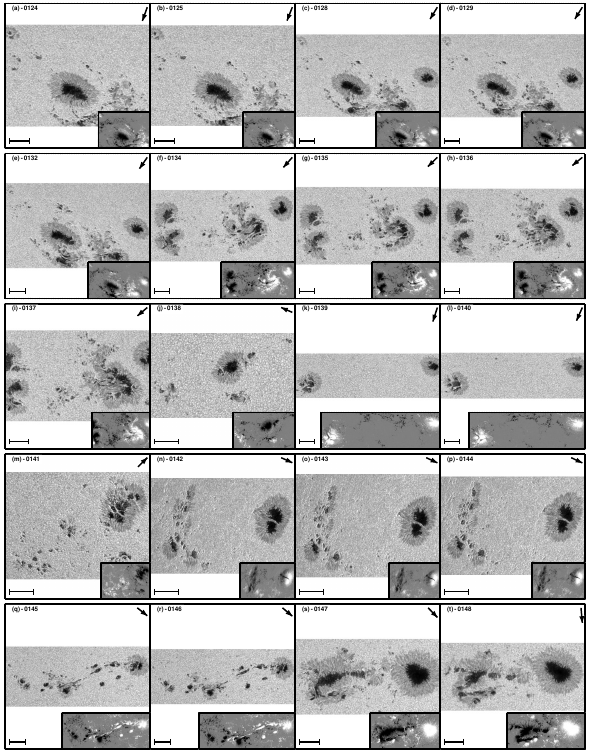}
\caption{Same as Fig.~\vref{fig:MODESTcontinuum000}. Axes colors display whether the scan was taken in fast mode (black) or normal mode (green).}\label{fig:MODESTcontinuum006}
\end{center}
\end{figure*}

\begin{figure*}[htbp]
\begin{center}
\includegraphics[width=1\textwidth]{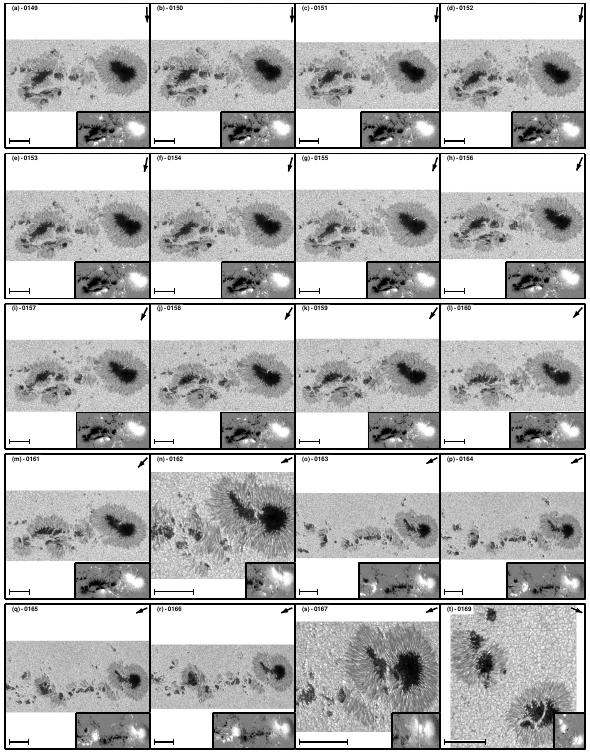}
\caption{Same as Fig.~\vref{fig:MODESTcontinuum000}. Axes colors display whether the scan was taken in fast mode (black) or normal mode (green).}\label{fig:MODESTcontinuum007}
\end{center}
\end{figure*}

\begin{figure*}[htbp]
\begin{center}
\includegraphics[width=1\textwidth]{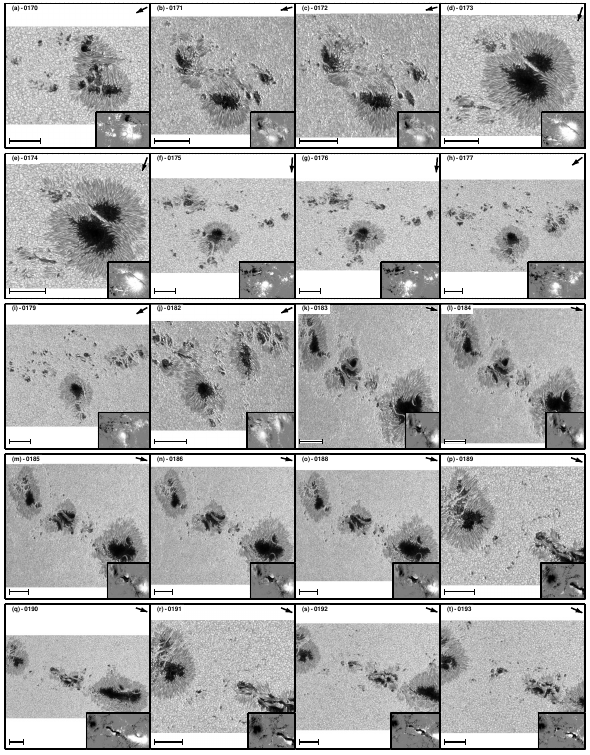}
\caption{Same as Fig.~\vref{fig:MODESTcontinuum000}. Axes colors display whether the scan was taken in fast mode (black) or normal mode (green).}\label{fig:MODESTcontinuum008}
\end{center}
\end{figure*}

\begin{figure*}[htbp]
\begin{center}
\includegraphics[width=1\textwidth]{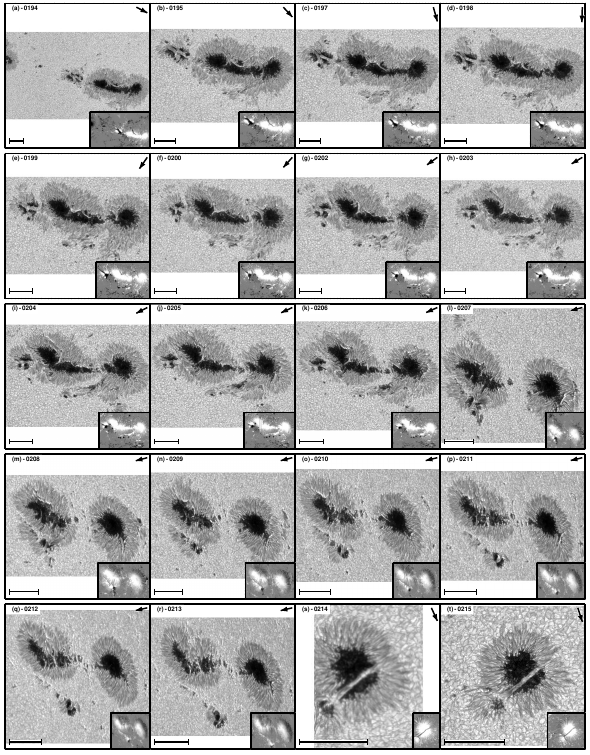}
\caption{Same as Fig.~\vref{fig:MODESTcontinuum000}. Axes colors display whether the scan was taken in fast mode (black) or normal mode (green).}\label{fig:MODESTcontinuum009}
\end{center}
\end{figure*}

\begin{figure*}[htbp]
\begin{center}
\includegraphics[width=1\textwidth]{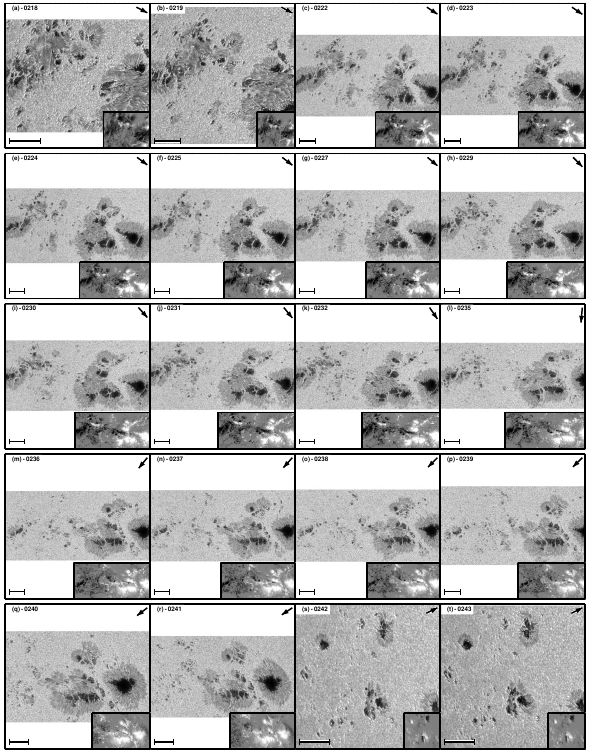}
\caption{Same as Fig.~\vref{fig:MODESTcontinuum000}. Axes colors display whether the scan was taken in fast mode (black) or normal mode (green).}\label{fig:MODESTcontinuum010}
\end{center}
\end{figure*}

\begin{figure*}[htbp]
\begin{center}
\includegraphics[width=1\textwidth]{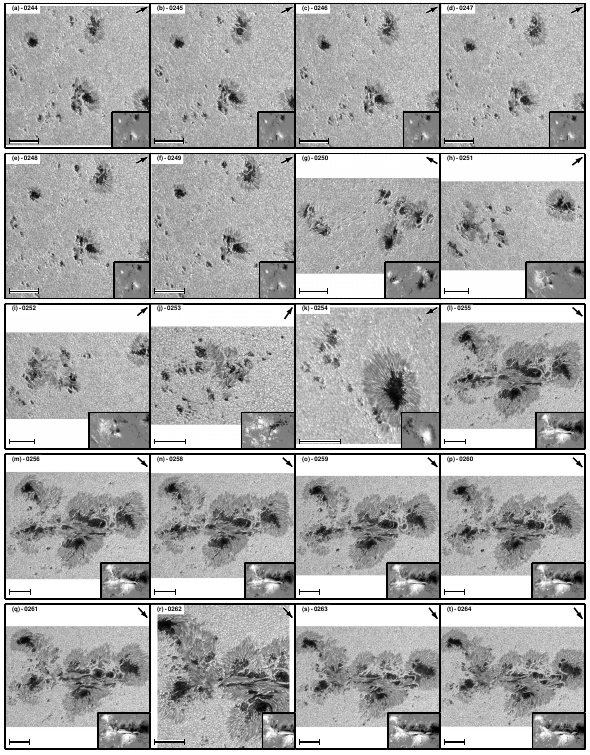}
\caption{Same as Fig.~\vref{fig:MODESTcontinuum000}. Axes colors display whether the scan was taken in fast mode (black) or normal mode (green).}\label{fig:MODESTcontinuum011}
\end{center}
\end{figure*}

\begin{figure*}[htbp]
\begin{center}
\includegraphics[width=1\textwidth]{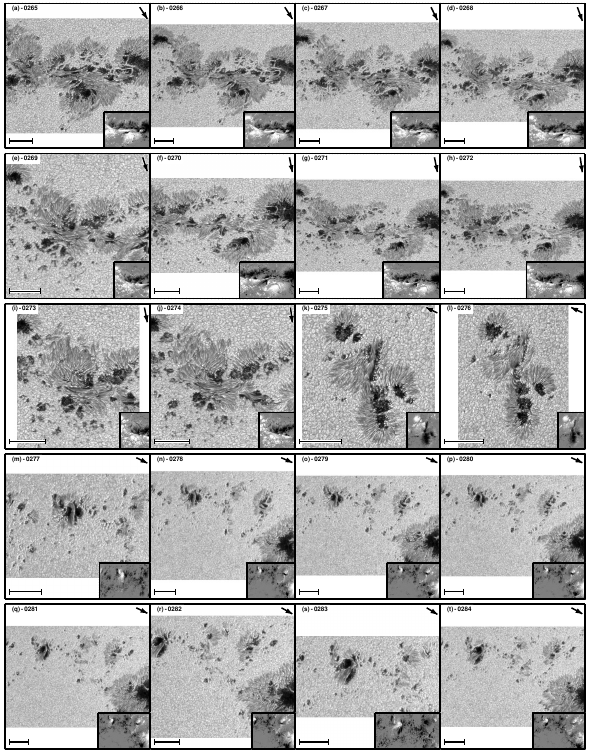}
\caption{Same as Fig.~\vref{fig:MODESTcontinuum000}. Axes colors display whether the scan was taken in fast mode (black) or normal mode (green).}\label{fig:MODESTcontinuum012}
\end{center}
\end{figure*}

\begin{figure*}[htbp]
\begin{center}
\includegraphics[width=1\textwidth]{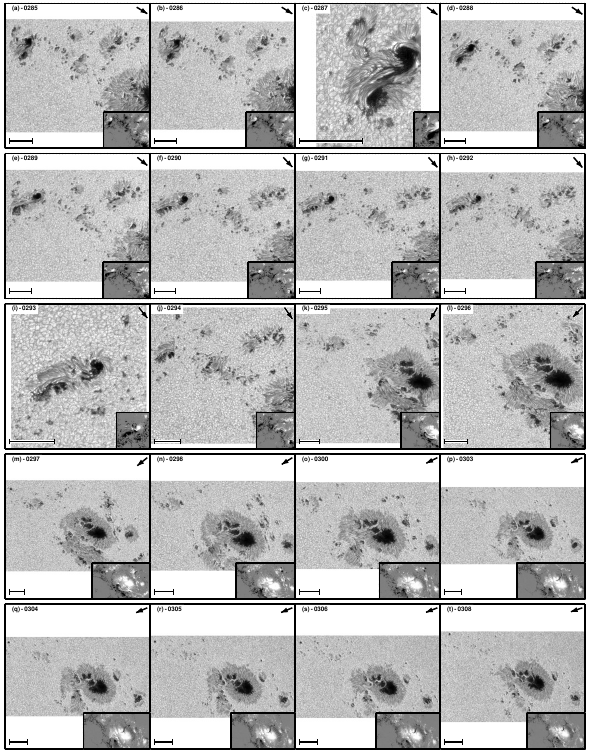}
\caption{Same as Fig.~\vref{fig:MODESTcontinuum000}. Axes colors display whether the scan was taken in fast mode (black) or normal mode (green).}\label{fig:MODESTcontinuum013}
\end{center}
\end{figure*}

\begin{figure*}[htbp]
\begin{center}
\includegraphics[width=1\textwidth]{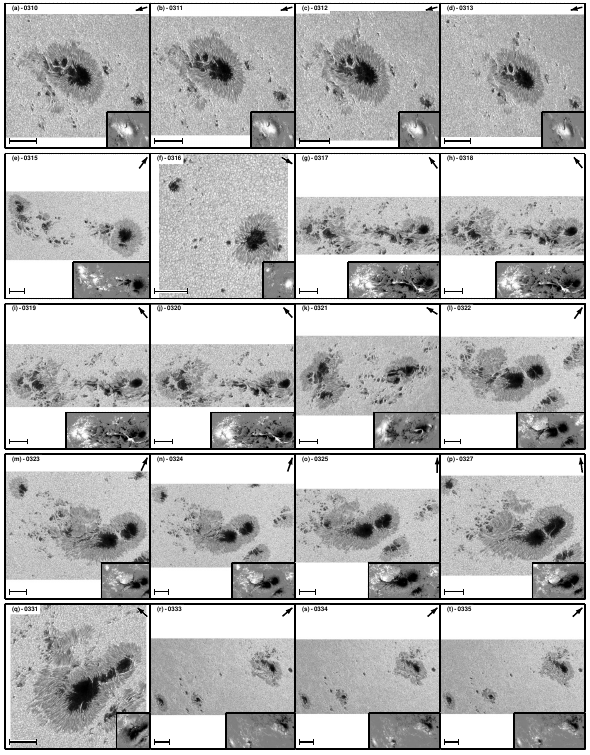}
\caption{Same as Fig.~\vref{fig:MODESTcontinuum000}. Axes colors display whether the scan was taken in fast mode (black) or normal mode (green).}\label{fig:MODESTcontinuum014}
\end{center}
\end{figure*}

\begin{figure*}[htbp]
\begin{center}
\includegraphics[width=1\textwidth]{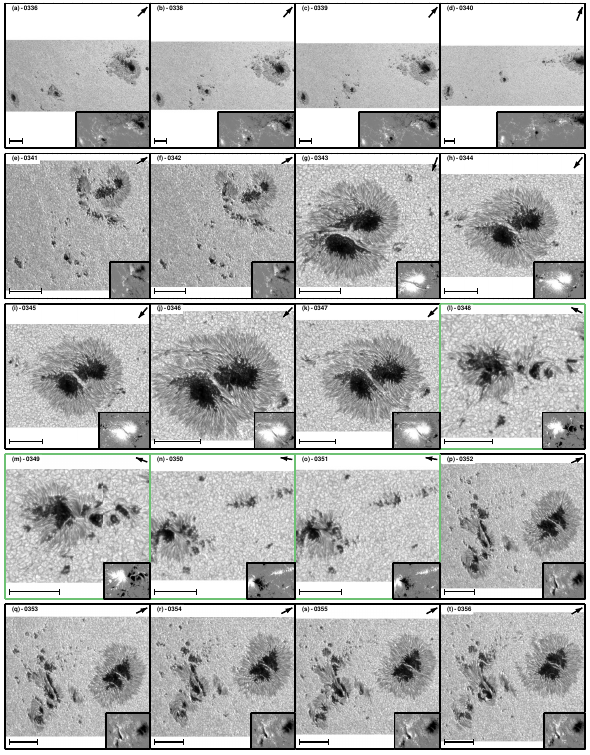}
\caption{Same as Fig.~\vref{fig:MODESTcontinuum000}. Axes colors display whether the scan was taken in fast mode (black) or normal mode (green).}\label{fig:MODESTcontinuum015}
\end{center}
\end{figure*}

\begin{figure*}[htbp]
\begin{center}
\includegraphics[width=1\textwidth]{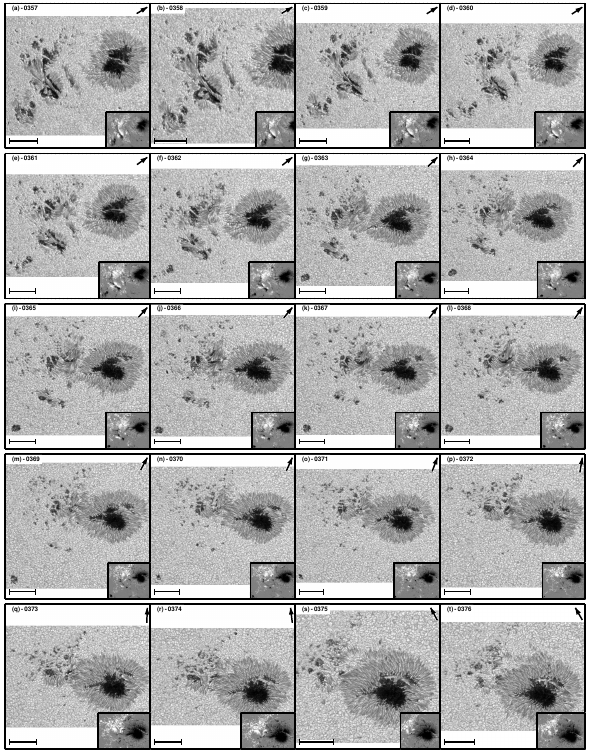}
\caption{Same as Fig.~\vref{fig:MODESTcontinuum000}. Axes colors display whether the scan was taken in fast mode (black) or normal mode (green).}\label{fig:MODESTcontinuum016}
\end{center}
\end{figure*}

\begin{figure*}[htbp]
\begin{center}
\includegraphics[width=1\textwidth]{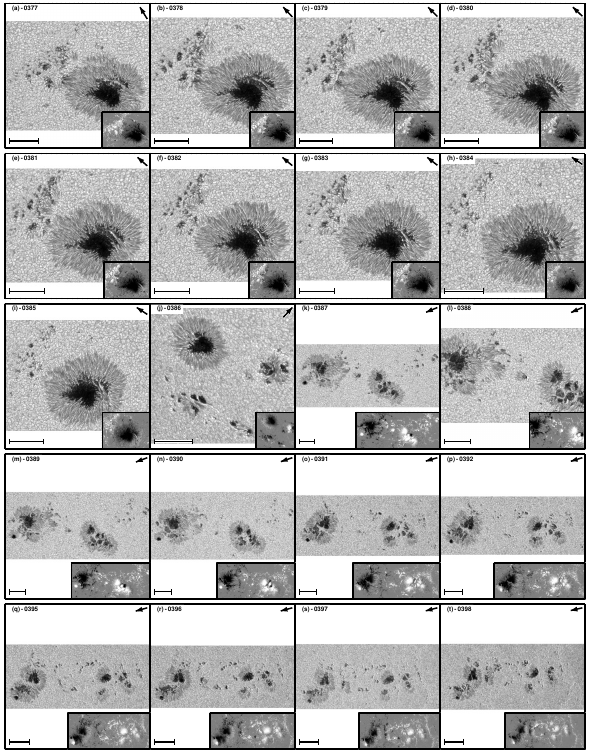}
\caption{Same as Fig.~\vref{fig:MODESTcontinuum000}. Axes colors display whether the scan was taken in fast mode (black) or normal mode (green).}\label{fig:MODESTcontinuum017}
\end{center}
\end{figure*}

\begin{figure*}[htbp]
\begin{center}
\includegraphics[width=1\textwidth]{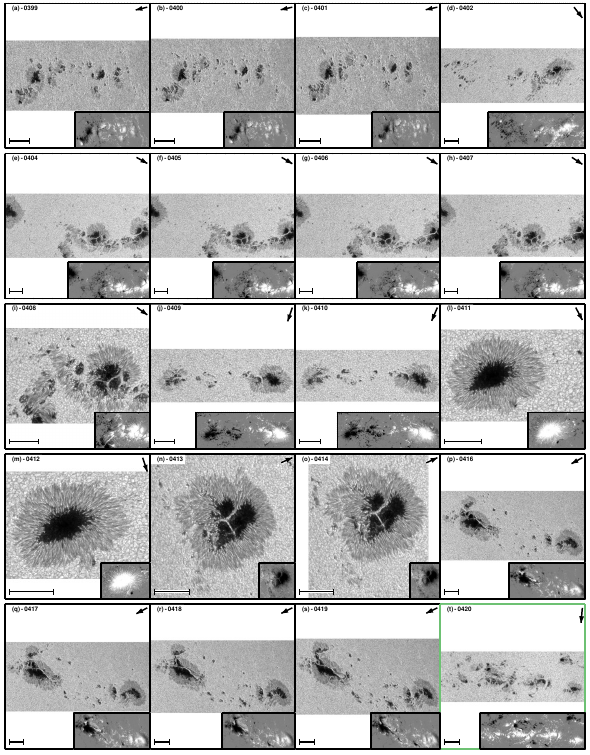}
\caption{Same as Fig.~\vref{fig:MODESTcontinuum000}. Axes colors display whether the scan was taken in fast mode (black) or normal mode (green).}\label{fig:MODESTcontinuum018}
\end{center}
\end{figure*}

\begin{figure*}[htbp]
\begin{center}
\includegraphics[width=1\textwidth]{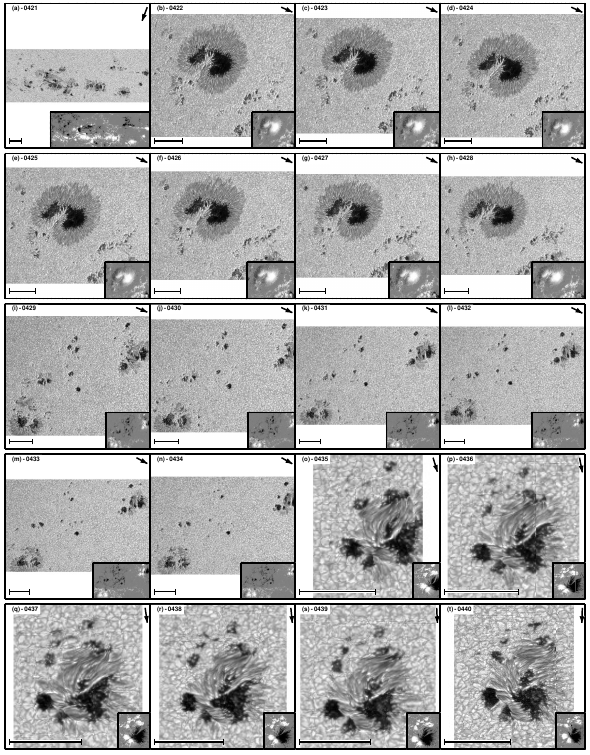}
\caption{Same as Fig.~\vref{fig:MODESTcontinuum000}. Axes colors display whether the scan was taken in fast mode (black) or normal mode (green).}\label{fig:MODESTcontinuum019}
\end{center}
\end{figure*}

\begin{figure*}[htbp]
\begin{center}
\includegraphics[width=1\textwidth]{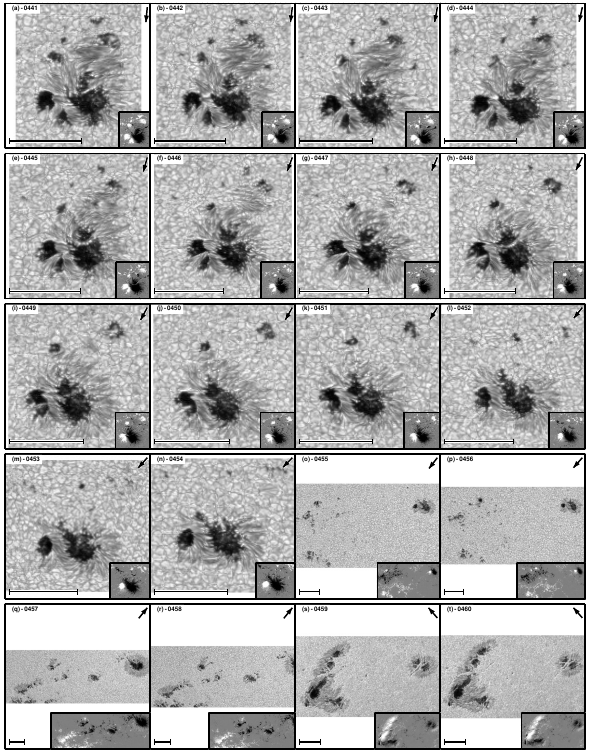}
\caption{Same as Fig.~\vref{fig:MODESTcontinuum000}. Axes colors display whether the scan was taken in fast mode (black) or normal mode (green).}\label{fig:MODESTcontinuum020}
\end{center}
\end{figure*}

\begin{figure*}[htbp]
\begin{center}
\includegraphics[width=1\textwidth]{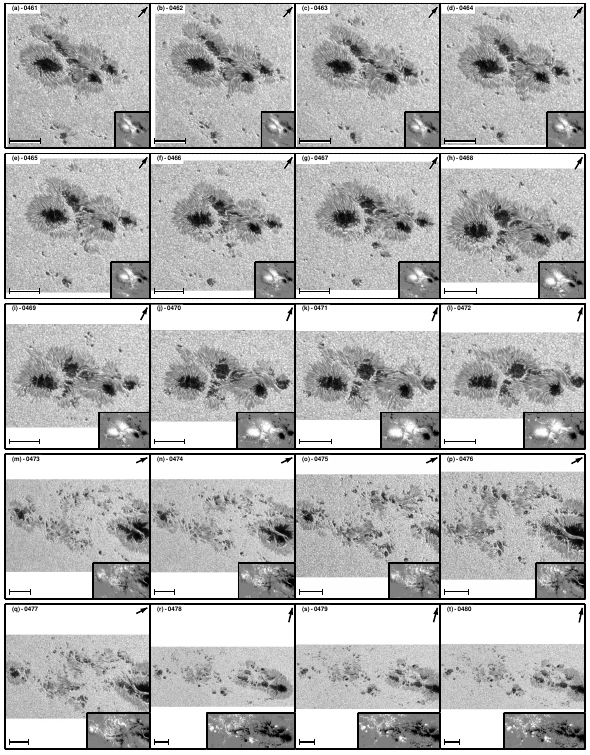}
\caption{Same as Fig.~\vref{fig:MODESTcontinuum000}. Axes colors display whether the scan was taken in fast mode (black) or normal mode (green).}\label{fig:MODESTcontinuum021}
\end{center}
\end{figure*}

\begin{figure*}[htbp]
\begin{center}
\includegraphics[width=1\textwidth]{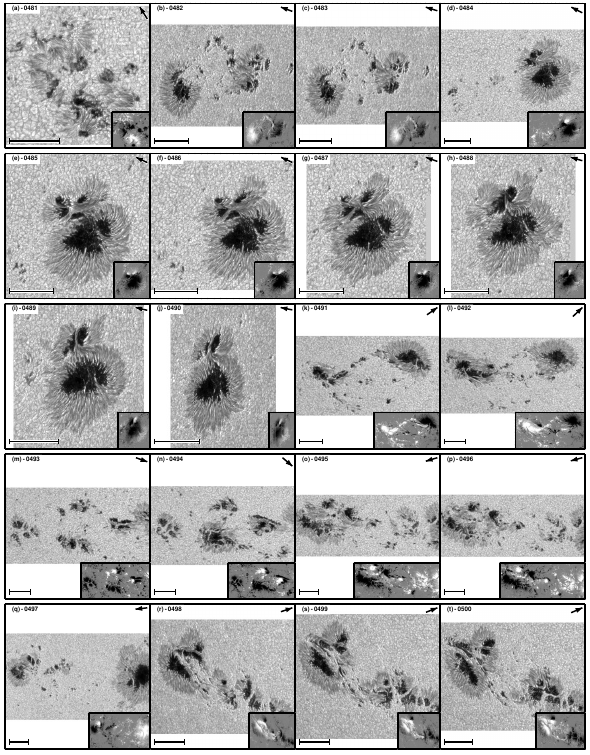}
\caption{Same as Fig.~\vref{fig:MODESTcontinuum000}. Axes colors display whether the scan was taken in fast mode (black) or normal mode (green).}\label{fig:MODESTcontinuum022}
\end{center}
\end{figure*}

\begin{figure*}[htbp]
\begin{center}
\includegraphics[width=1\textwidth]{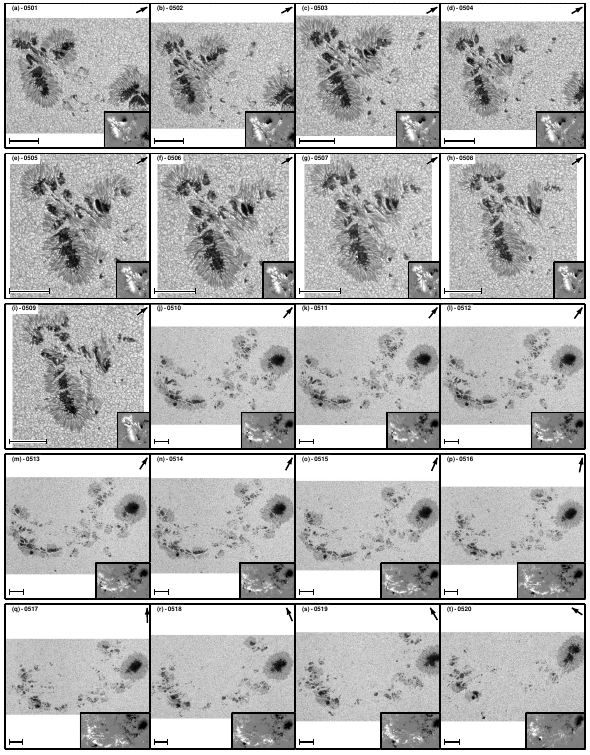}
\caption{Same as Fig.~\vref{fig:MODESTcontinuum000}. Axes colors display whether the scan was taken in fast mode (black) or normal mode (green).}\label{fig:MODESTcontinuum023}
\end{center}
\end{figure*}

\begin{figure*}[htbp]
\begin{center}
\includegraphics[width=1\textwidth]{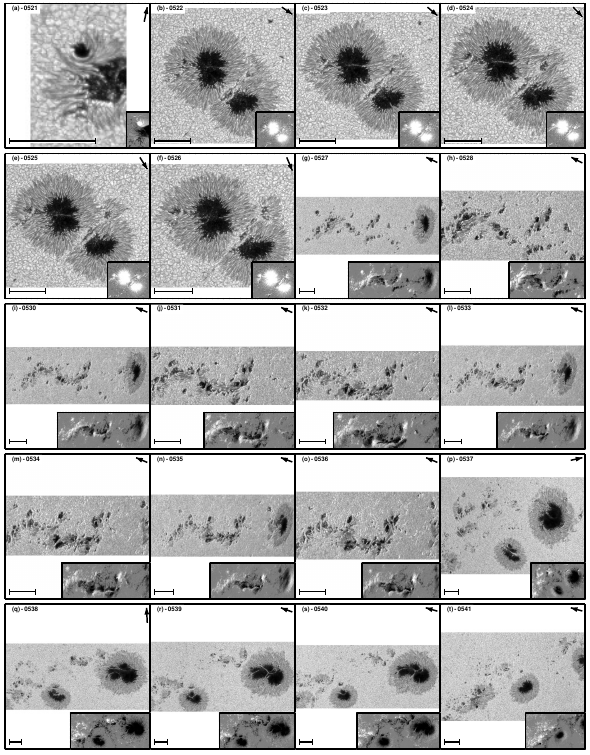}
\caption{Same as Fig.~\vref{fig:MODESTcontinuum000}. Axes colors display whether the scan was taken in fast mode (black) or normal mode (green).}\label{fig:MODESTcontinuum024}
\end{center}
\end{figure*}

\begin{figure*}[htbp]
\begin{center}
\includegraphics[width=1\textwidth]{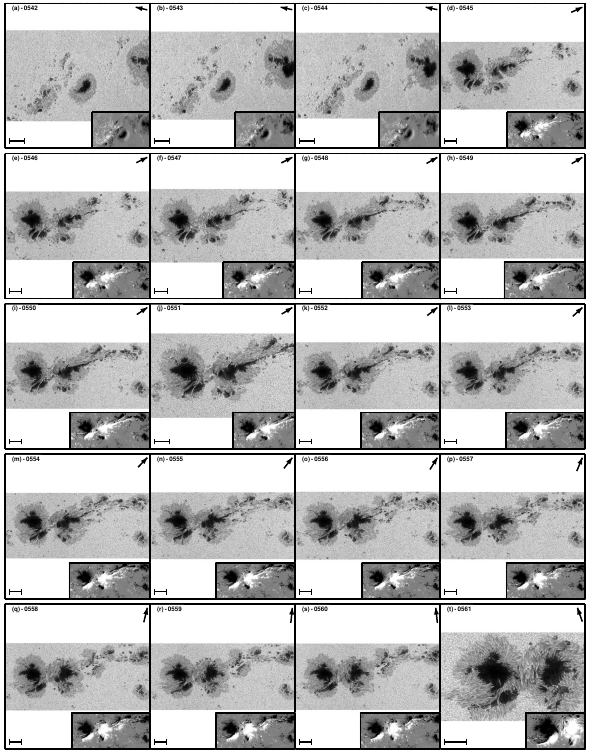}
\caption{Same as Fig.~\vref{fig:MODESTcontinuum000}. Axes colors display whether the scan was taken in fast mode (black) or normal mode (green).}\label{fig:MODESTcontinuum025}
\end{center}
\end{figure*}

\begin{figure*}[htbp]
\begin{center}
\includegraphics[width=1\textwidth]{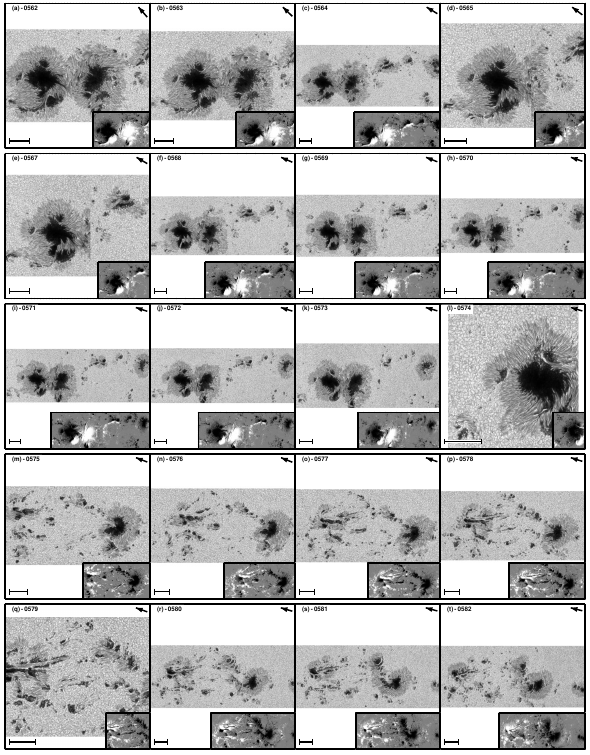}
\caption{Same as Fig.~\vref{fig:MODESTcontinuum000}. Axes colors display whether the scan was taken in fast mode (black) or normal mode (green).}\label{fig:MODESTcontinuum026}
\end{center}
\end{figure*}

\begin{figure*}[htbp]
\begin{center}
\includegraphics[width=1\textwidth]{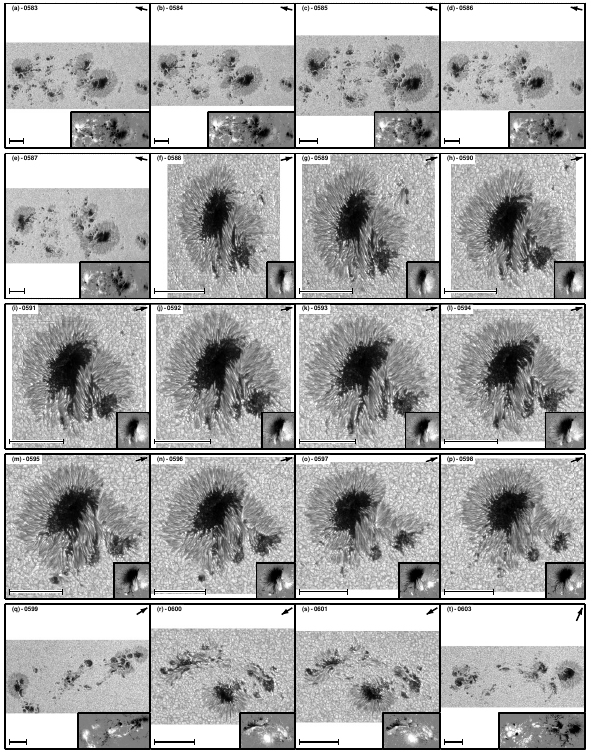}
\caption{Same as Fig.~\vref{fig:MODESTcontinuum000}. Axes colors display whether the scan was taken in fast mode (black) or normal mode (green).}\label{fig:MODESTcontinuum027}
\end{center}
\end{figure*}

\begin{figure*}[htbp]
\begin{center}
\includegraphics[width=1\textwidth]{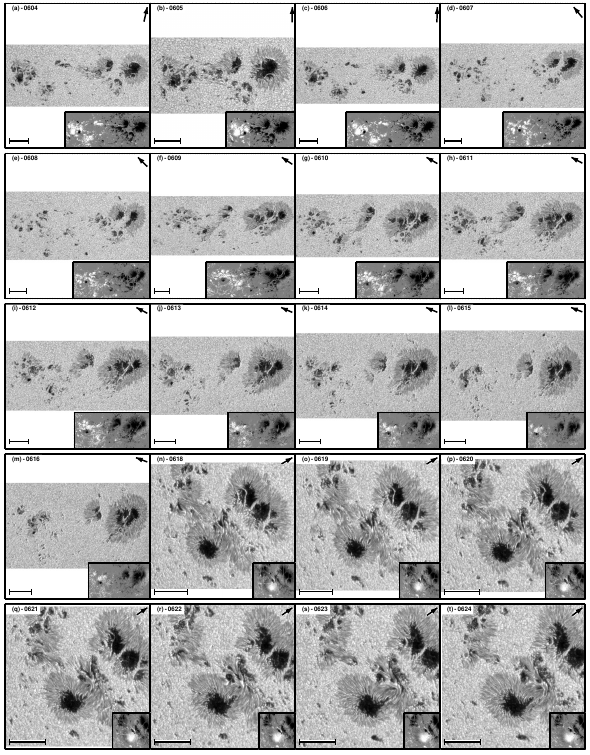}
\caption{Same as Fig.~\vref{fig:MODESTcontinuum000}. Axes colors display whether the scan was taken in fast mode (black) or normal mode (green).}\label{fig:MODESTcontinuum028}
\end{center}
\end{figure*}

\begin{figure*}[htbp]
\begin{center}
\includegraphics[width=1\textwidth]{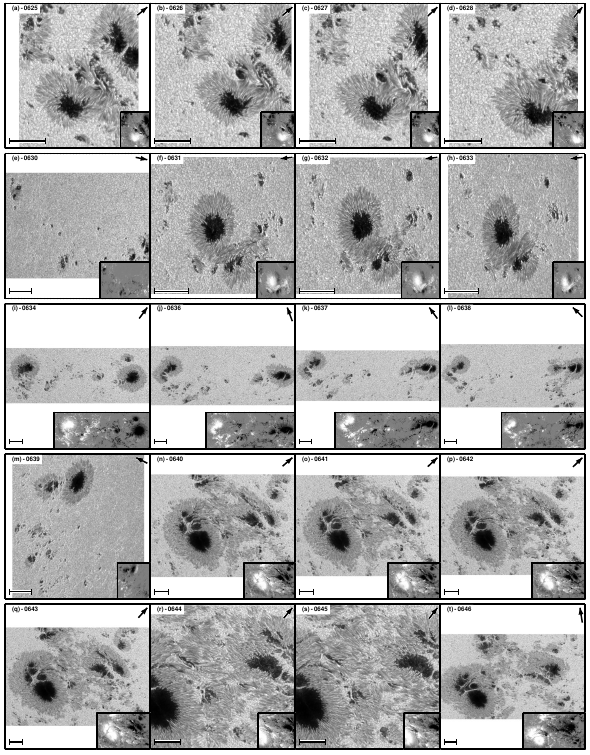}
\caption{Same as Fig.~\vref{fig:MODESTcontinuum000}. Axes colors display whether the scan was taken in fast mode (black) or normal mode (green).}\label{fig:MODESTcontinuum029}
\end{center}
\end{figure*}

\begin{figure*}[htbp]
\begin{center}
\includegraphics[width=1\textwidth]{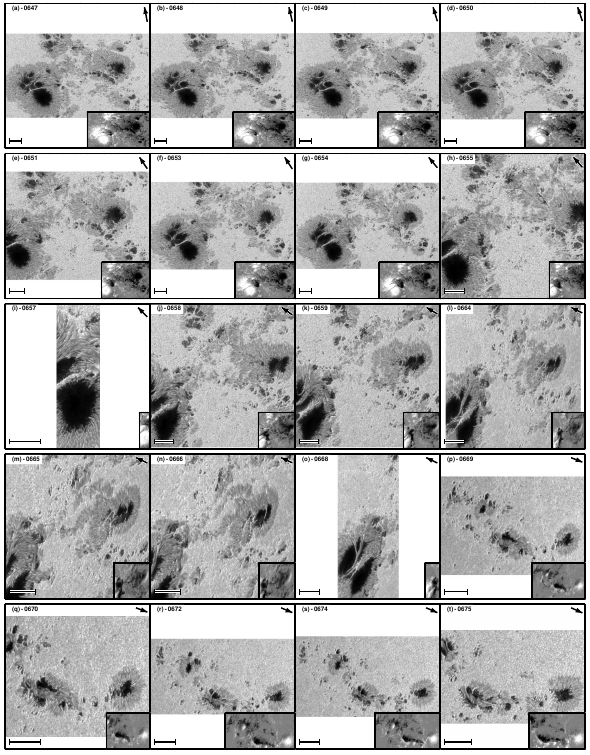}
\caption{Same as Fig.~\vref{fig:MODESTcontinuum000}. Axes colors display whether the scan was taken in fast mode (black) or normal mode (green).}\label{fig:MODESTcontinuum030}
\end{center}
\end{figure*}

\begin{figure*}[htbp]
\begin{center}
\includegraphics[width=1\textwidth]{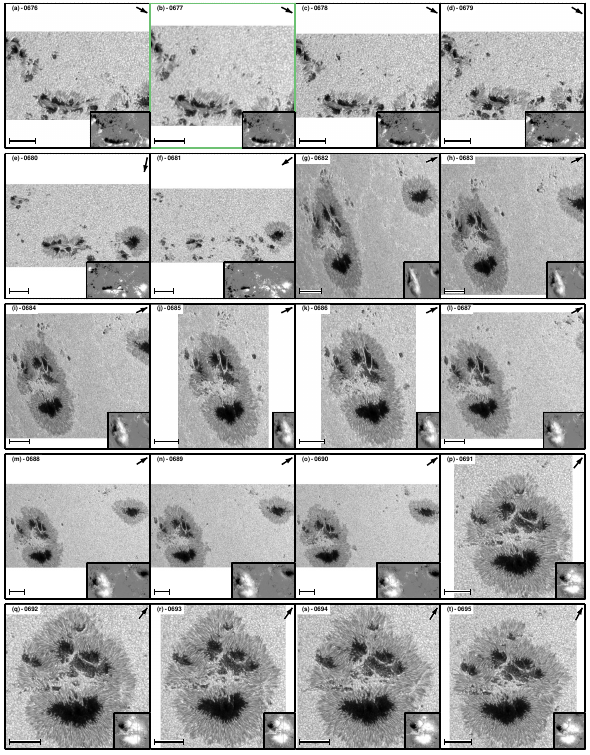}
\caption{Same as Fig.~\vref{fig:MODESTcontinuum000}. Axes colors display whether the scan was taken in fast mode (black) or normal mode (green).}\label{fig:MODESTcontinuum031}
\end{center}
\end{figure*}

\begin{figure*}[htbp]
\begin{center}
\includegraphics[width=1\textwidth]{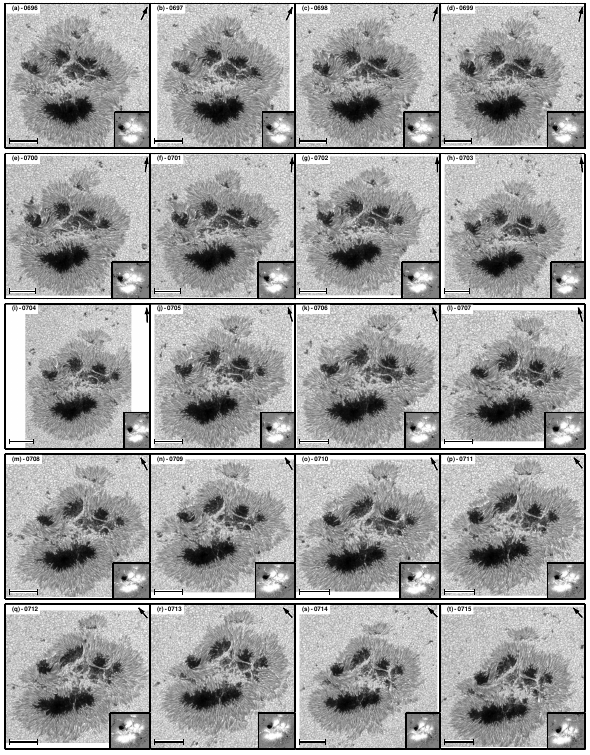}
\caption{Same as Fig.~\vref{fig:MODESTcontinuum000}. Axes colors display whether the scan was taken in fast mode (black) or normal mode (green).}\label{fig:MODESTcontinuum032}
\end{center}
\end{figure*}

\begin{figure*}[htbp]
\begin{center}
\includegraphics[width=1\textwidth]{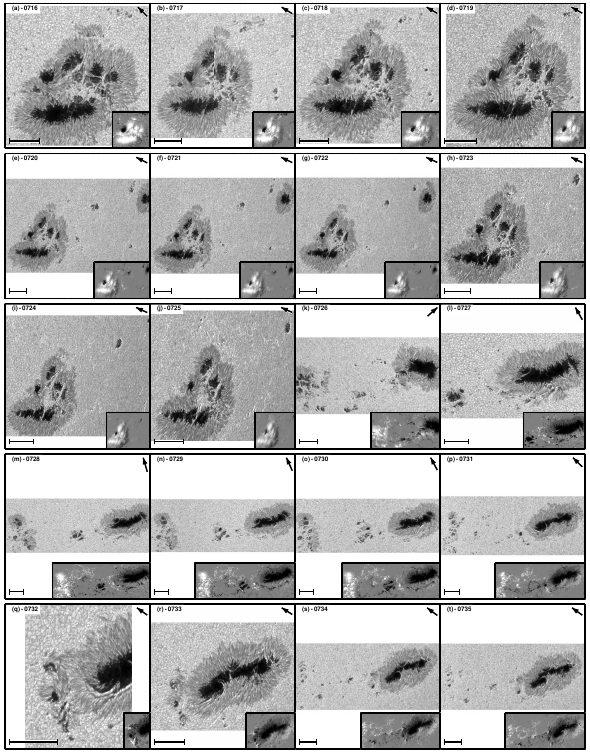}
\caption{Same as Fig.~\vref{fig:MODESTcontinuum000}. Axes colors display whether the scan was taken in fast mode (black) or normal mode (green).}\label{fig:MODESTcontinuum033}
\end{center}
\end{figure*}

\begin{figure*}[htbp]
\begin{center}
\includegraphics[width=1\textwidth]{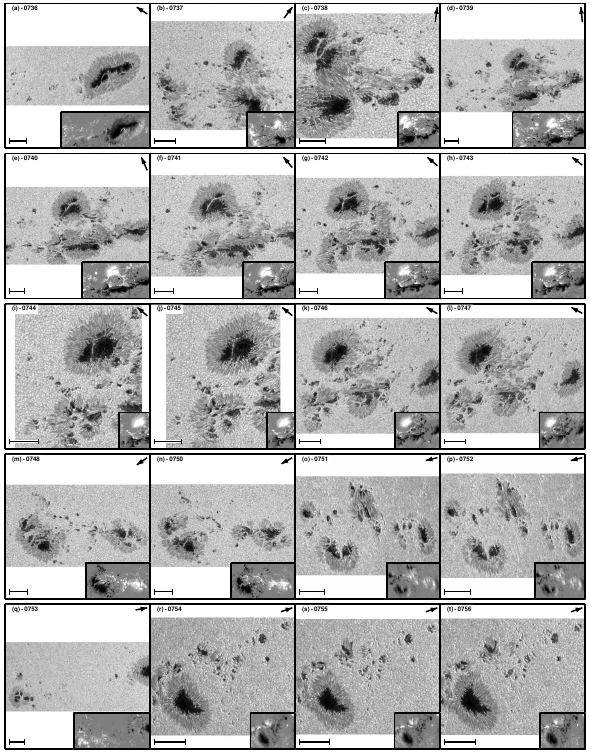}
\caption{Same as Fig.~\vref{fig:MODESTcontinuum000}. Axes colors display whether the scan was taken in fast mode (black) or normal mode (green).}\label{fig:MODESTcontinuum034}
\end{center}
\end{figure*}

\begin{figure*}[htbp]
\begin{center}
\includegraphics[width=1\textwidth]{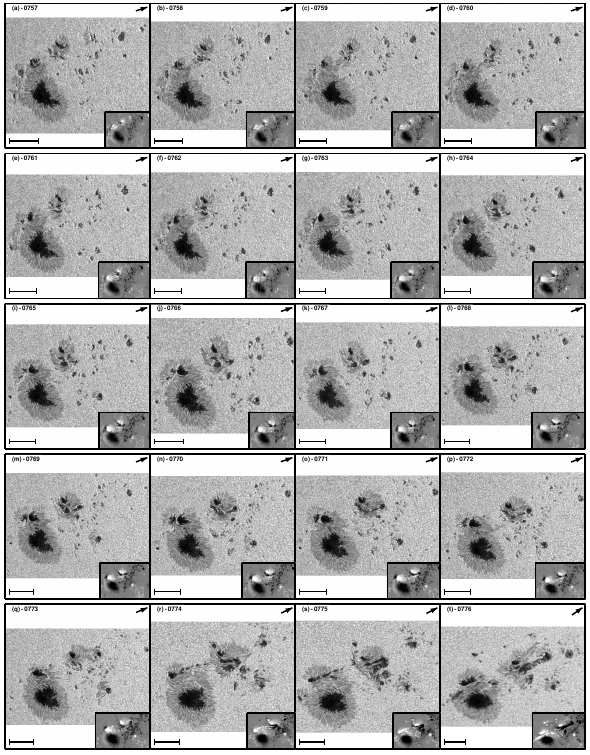}
\caption{Same as Fig.~\vref{fig:MODESTcontinuum000}. Axes colors display whether the scan was taken in fast mode (black) or normal mode (green).}\label{fig:MODESTcontinuum035}
\end{center}
\end{figure*}

\begin{figure*}[htbp]
\begin{center}
\includegraphics[width=1\textwidth]{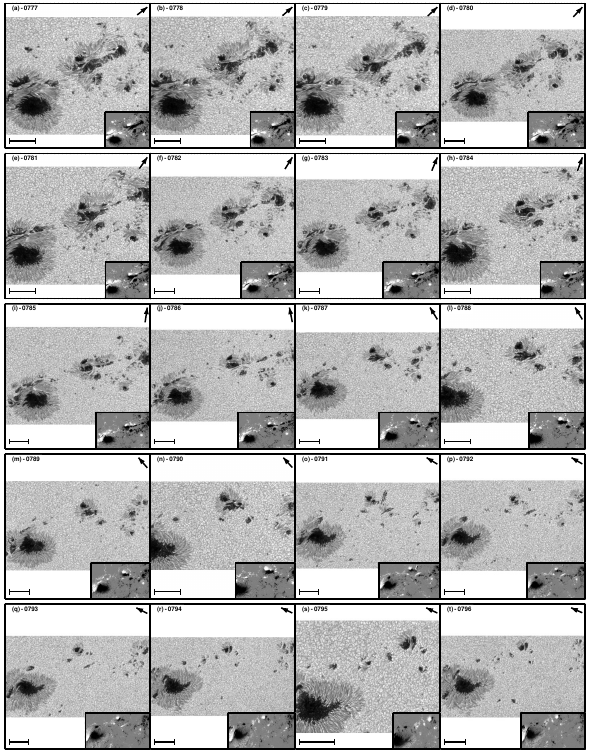}
\caption{Same as Fig.~\vref{fig:MODESTcontinuum000}. Axes colors display whether the scan was taken in fast mode (black) or normal mode (green).}\label{fig:MODESTcontinuum036}
\end{center}
\end{figure*}

\begin{figure*}[htbp]
\begin{center}
\includegraphics[width=1\textwidth]{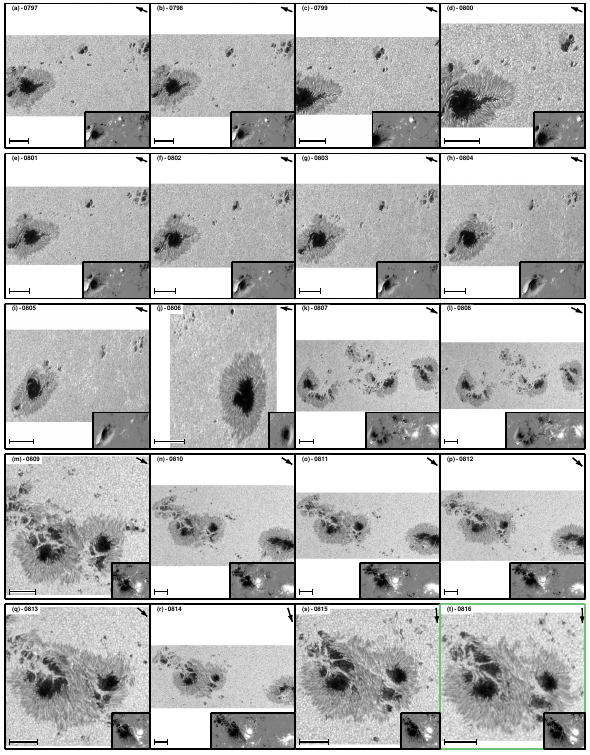}
\caption{Same as Fig.~\vref{fig:MODESTcontinuum000}. Axes colors display whether the scan was taken in fast mode (black) or normal mode (green).}\label{fig:MODESTcontinuum037}
\end{center}
\end{figure*}

\begin{figure*}[htbp]
\begin{center}
\includegraphics[width=1\textwidth]{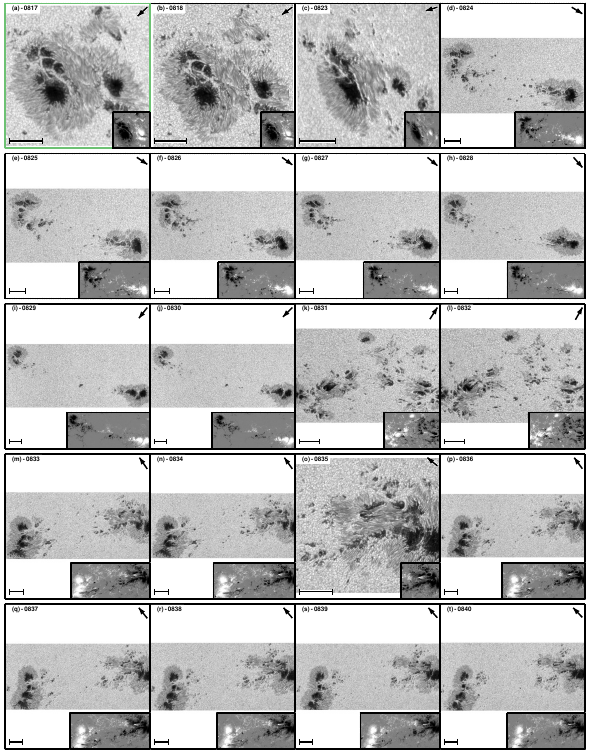}
\caption{Same as Fig.~\vref{fig:MODESTcontinuum000}. Axes colors display whether the scan was taken in fast mode (black) or normal mode (green).}\label{fig:MODESTcontinuum038}
\end{center}
\end{figure*}

\begin{figure*}[htbp]
\begin{center}
\includegraphics[width=1\textwidth]{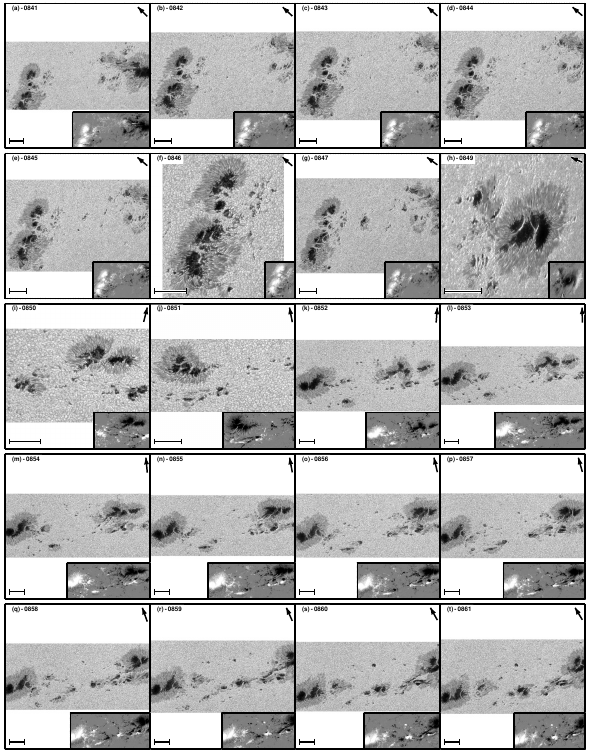}
\caption{Same as Fig.~\vref{fig:MODESTcontinuum000}. Axes colors display whether the scan was taken in fast mode (black) or normal mode (green).}\label{fig:MODESTcontinuum039}
\end{center}
\end{figure*}

\begin{figure*}[htbp]
\begin{center}
\includegraphics[width=1\textwidth]{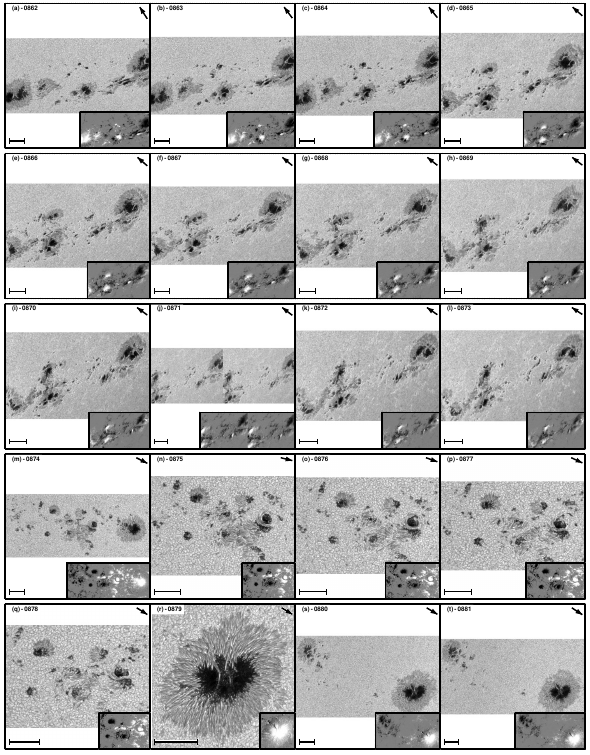}
\caption{Same as Fig.~\vref{fig:MODESTcontinuum000}. Axes colors display whether the scan was taken in fast mode (black) or normal mode (green).}\label{fig:MODESTcontinuum040}
\end{center}
\end{figure*}

\begin{figure*}[htbp]
\begin{center}
\includegraphics[width=1\textwidth]{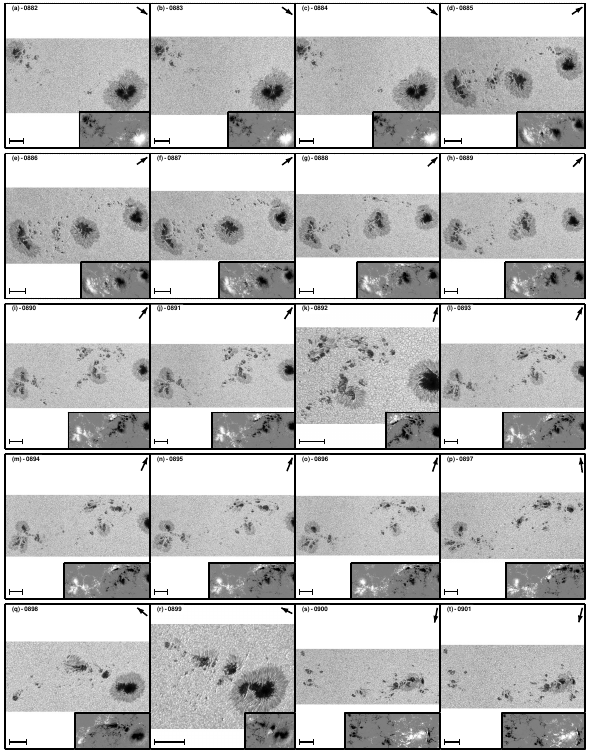}
\caption{Same as Fig.~\vref{fig:MODESTcontinuum000}. Axes colors display whether the scan was taken in fast mode (black) or normal mode (green).}\label{fig:MODESTcontinuum041}
\end{center}
\end{figure*}

\begin{figure*}[htbp]
\begin{center}
\includegraphics[width=1\textwidth]{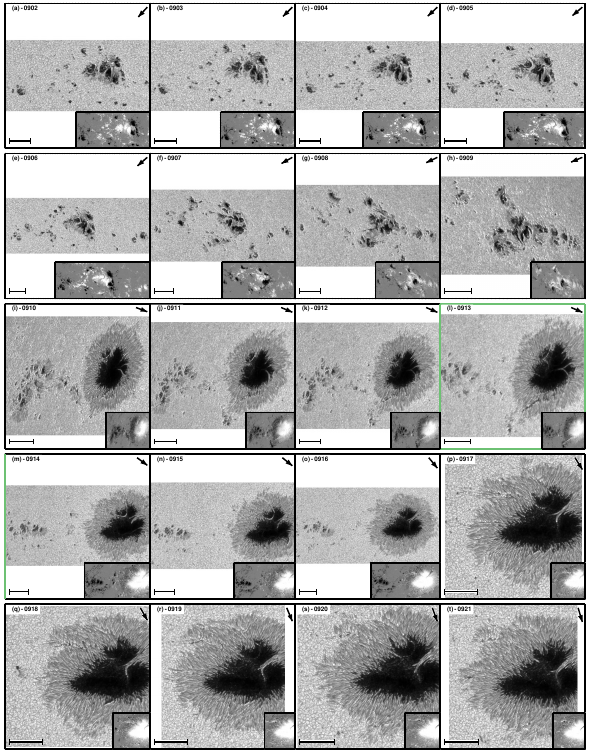}
\caption{Same as Fig.~\vref{fig:MODESTcontinuum000}. Axes colors display whether the scan was taken in fast mode (black) or normal mode (green).}\label{fig:MODESTcontinuum042}
\end{center}
\end{figure*}

\begin{figure*}[htbp]
\begin{center}
\includegraphics[width=1\textwidth]{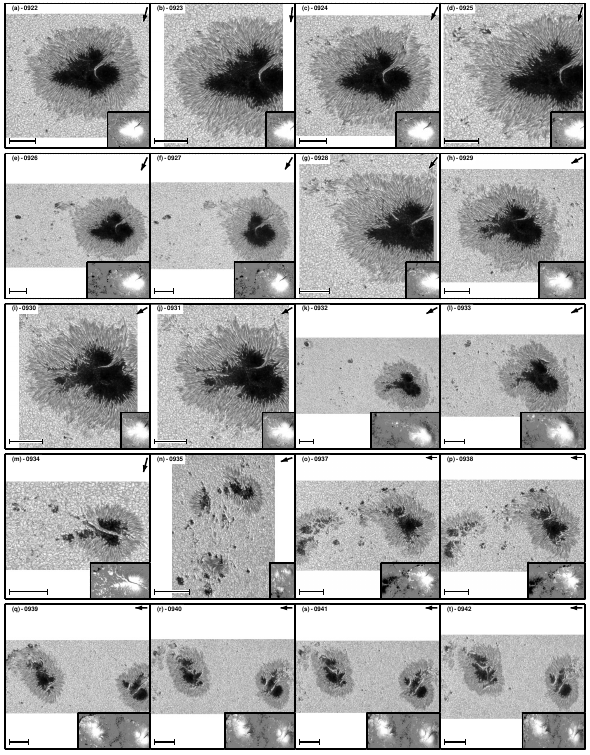}
\caption{Same as Fig.~\vref{fig:MODESTcontinuum000}. Axes colors display whether the scan was taken in fast mode (black) or normal mode (green).}\label{fig:MODESTcontinuum043}
\end{center}
\end{figure*}

\begin{figure*}[htbp]
\begin{center}
\includegraphics[width=1\textwidth]{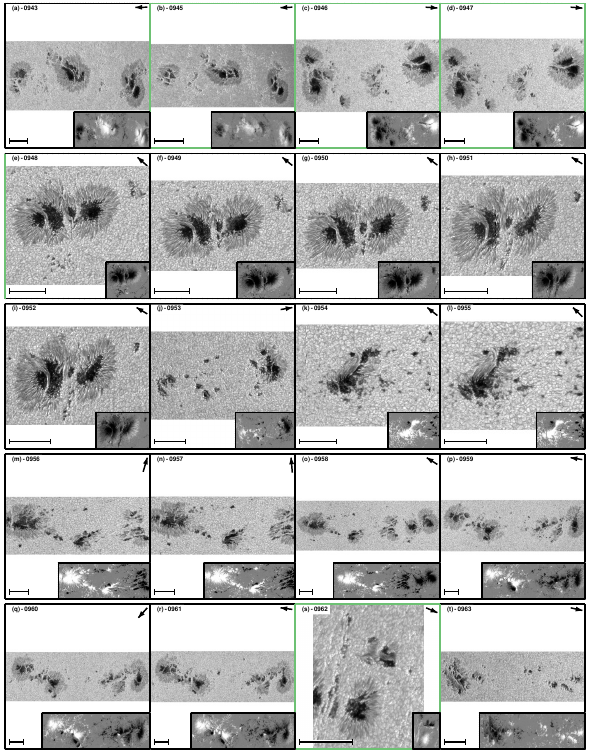}
\caption{Same as Fig.~\vref{fig:MODESTcontinuum000}. Axes colors display whether the scan was taken in fast mode (black) or normal mode (green).}\label{fig:MODESTcontinuum044}
\end{center}
\end{figure*}

\begin{figure*}[htbp]
\begin{center}
\includegraphics[width=1\textwidth]{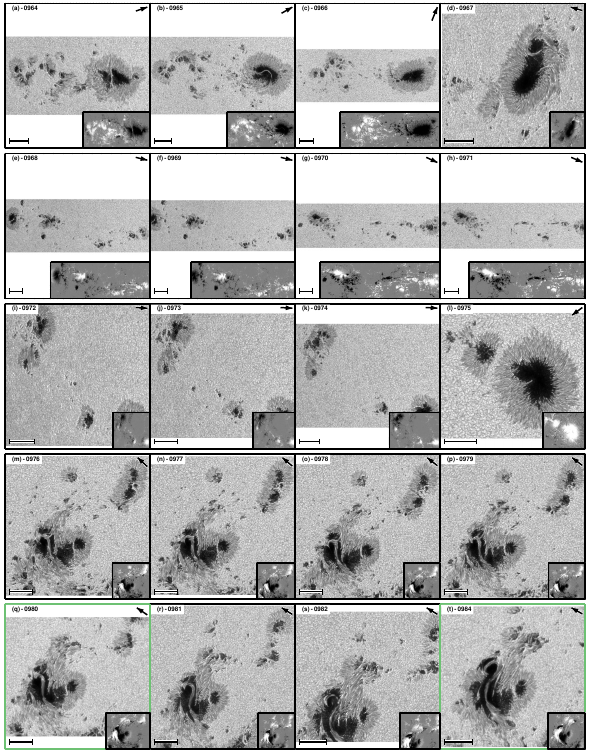}
\caption{Same as Fig.~\vref{fig:MODESTcontinuum000}. Axes colors display whether the scan was taken in fast mode (black) or normal mode (green).}\label{fig:MODESTcontinuum045}
\end{center}
\end{figure*}

\begin{figure*}[htbp]
\begin{center}
\includegraphics[width=1\textwidth]{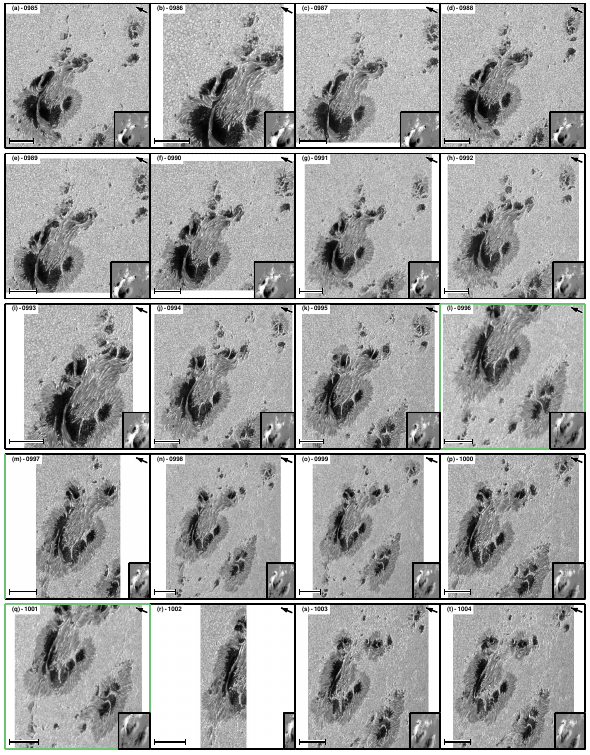}
\caption{Same as Fig.~\vref{fig:MODESTcontinuum000}. Axes colors display whether the scan was taken in fast mode (black) or normal mode (green).}\label{fig:MODESTcontinuum046}
\end{center}
\end{figure*}

\begin{figure*}[htbp]
\begin{center}
\includegraphics[trim={0 7.65cm 0 0},width=1\textwidth]{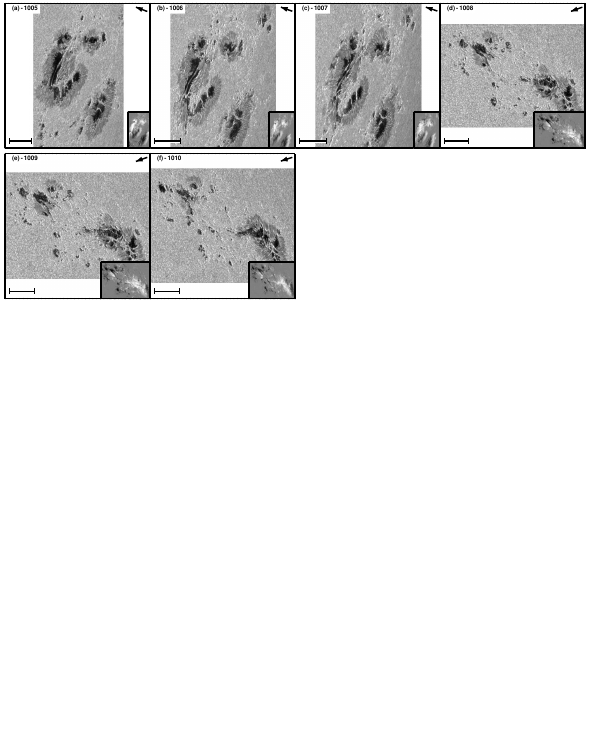}
\caption{Same as Fig.~\vref{fig:MODESTcontinuum000}. Axes colors display whether the scan was taken in fast mode (black) or normal mode (green).}\label{fig:MODESTcontinuum047}
\end{center}
\end{figure*}

{\footnotesize
\begin{longtable}{lccccccccccccl}
\caption{Sample of sunspot groups part of the \modest{} catalog}\label{tab:sample-ARS-MODEST}\\
\toprule[1.5pt]
\multirow{2}{*}{index}& \multicolumn{1}{c}{\multirow{2}{*}{inv\_id}}& \multicolumn{1}{c}{obs\_id}& NOAA&type &$x_{\rm c}$&$y_{\rm c}$&\multicolumn{1}{c}{\multirow{2}{*}{$\mu$}}& num &size&area&Z\"urich&magnetic&\multicolumn{1}{c}{Fig.}\\
&        &  \multicolumn{1}{c}{(date)\_(time)}  & AR num   & scan  & ($\arcsec{}$)  & ($\arcsec{}$)  &   & tiles& FOV &(MSH)&class& class& \multicolumn{1}{c}{(panel)}\\
\toprule[1.5pt]
\endfirsthead
\captionsetup{list=no}\caption[]{continued.}\\
\toprule[1.5pt]
\multirow{2}{*}{index}& \multicolumn{1}{c}{\multirow{2}{*}{inv\_id}}& \multicolumn{1}{c}{obs\_id}& NOAA&type &$x_{\rm c}$&$y_{\rm c}$&\multicolumn{1}{c}{\multirow{2}{*}{$\mu$}}& num &size&area&Z\"urich&magnetic&\multicolumn{1}{c}{Fig.}\\
&        &  \multicolumn{1}{c}{(date)\_(time)}  & AR num   & scan  & ($\arcsec{}$)  & ($\arcsec{}$)  &   & tiles& FOV &(MSH)&class& class& \multicolumn{1}{c}{(panel)}\\
\toprule[1.5pt]
\endhead
\toprule[1.5pt]
\endfoot
   1 & 0000 & 20061208\_061114 & 10930 & Nor. & -797 &  -72 & 0.56 & 100 & $ 89\arcsec{}\times 95\arcsec{}$ & 430 & Dki &  $\beta\gamma\delta $ & \vref{fig:MODESTcontinuum000}(a) \\
   2 & 0001 & 20061208\_105104 & 10930 & Fast & -670 &  -58 & 0.60 &  25 & $ 89\arcsec{}\times 95\arcsec{}$ & 430 & Dki &  $\beta\gamma\delta $ & \vref{fig:MODESTcontinuum000}(b) \\
   3 & 0002 & 20061208\_121357 & 10930 & Fast & -661 &  -58 & 0.61 &  25 & $ 89\arcsec{}\times 95\arcsec{}$ & 430 & Dki &  $\beta\gamma\delta $ & \vref{fig:MODESTcontinuum000}(c) \\
   4 & 0003 & 20061208\_131320 & 10930 & Fast & -634 &  -57 & 0.62 &  25 & $ 89\arcsec{}\times 95\arcsec{}$ & 430 & Dki &  $\beta\gamma\delta $ & \vref{fig:MODESTcontinuum000}(d) \\
   5 & 0004 & 20061208\_145005 & 10930 & Fast & -643 &  -53 & 0.63 &  25 & $ 89\arcsec{}\times 95\arcsec{}$ & 430 & Dki &  $\beta\gamma\delta $ & \vref{fig:MODESTcontinuum000}(e) \\
   6 & 0005 & 20061208\_180506 & 10930 & Fast & -621 &  -54 & 0.66 &  25 & $ 89\arcsec{}\times 95\arcsec{}$ & 430 & Dki &  $\beta\gamma\delta $ & \vref{fig:MODESTcontinuum000}(f) \\
   7 & 0006 & 20061208\_212005 & 10930 & Fast & -598 &  -54 & 0.69 &  25 & $ 89\arcsec{}\times 95\arcsec{}$ & 430 & Dki &  $\beta\gamma\delta $ & \vref{fig:MODESTcontinuum000}(g) \\
   8 & 0007 & 20061209\_034005 & 10930 & Fast & -548 &  -53 & 0.74 &  25 & $ 89\arcsec{}\times 95\arcsec{}$ & 430 & Dki &  $\beta\gamma\delta $ & \vref{fig:MODESTcontinuum000}(h) \\
\end{longtable}
}

\end{document}